\documentclass{iopart}
\usepackage{graphicx}
\usepackage{iopams}

\usepackage[sort&compress]{natbib}

\providecommand{\newblock}{\hskip .11em \@plus.33em \@minus.07em}

\usepackage[colorlinks=true,citecolor=blue]{hyperref}

\newcommand{\GF}{G_{\mathrm{F}}}
\newcommand{\NA}{N_{\mathrm{A}}}
\newcommand{\diag}{\mathrm{diag}}
\newcommand{\sgn}{\mathrm{sgn}}

\newcommand{\vac}{\mathrm{vac}}
\newcommand{\pr}{\mathrm{pr}}
\newcommand{\matt}{\mathrm{matt}}
\newcommand{\eff}{\mathrm{eff}}
\newcommand{\tot}{\mathrm{tot}}
\newcommand{\thetav}{\theta_\mathrm{v}}
\newcommand{\Es}{E^\mathrm{s}_\nu}

\newcommand{\vP}{\vec{P}}
\newcommand{\vs}{\vec{s}}

\newcommand{\ve}{\vec{e}}
\newcommand{\vte}{\vec{\tilde{e}}}
\newcommand{\vH}{\vec{H}}
\newcommand{\vtH}{\vec{\tilde{H}}}
\newcommand{\nf}{\tilde{n}}
\newcommand{\ff}{\tilde{f}}

\newcommand{\bsf}{\mathrm{(f)}}
\newcommand{\bsv}{\mathrm{(v)}}

\newcommand{\bir}{\bi{r}}
\newcommand{\bip}{\bi{p}}

\newcommand{\rmm}{\mathrm{m}}
\newcommand{\rmH}{\mathrm{H}}

\newcommand{\rmC}{\mathrm{C}}

\newcommand{\taup}{\tau^\prime}

\begin{document}

%\preprint{INT PUB 08-27}

\topical{Neutrino flavour transformation in supernovae}
\author{H Duan$^1$ and J P Kneller$^{2,3}$}

\address{$^1$ Institute for Nuclear Theory,University of Washington,Seattle, WA 98195, USA}
\address{$^2$ School of Physics and Astronomy,University of Minnesota,Minneapolis, MN 55455, USA}
\address{$^3$ Institut de Physique Nucl\'eaire, F-91406 Orsay cedex, CNRS/IN2P3 and University of Paris-XI, France}

\ead{\mailto{huaiyu.duan@mailaps.org}, \mailto{kneller@ipno.in2p3.fr}}

\begin{abstract}
Rapid progress has been made during recent years in the understanding
of the flavour oscillations that occur as neutrinos traverse through
supernova.  The previous paradigm has given way and it is now clear
that the neutrino signals we shall receive from future Galactic
supernovae will allow us both to peer inside these extraordinary
cosmic events and to probe some of the fundamental properties of these
elusive particles.  In this review we aim to distill the progress that
has been made focusing upon the effects of the dynamic density profile
and the emergence of collective flavour oscillations due to neutrino
self-interactions.
\end{abstract}

\maketitle

\section{Introduction\label{sec:introduction}}

The detection of neutrinos from the Sun
\citep{Davis:PhysRevLett.20.1205,Davis:1984ts} and Supernova 1987A
\citep{Hirata:1987hu,Bionta:1987qt,Alekseev:1987JETPL..45..589A} have
laid the foundation for a new field, neutrino astronomy. These
discoveries not only opened a new window upon the Universe through the
observation of cosmic neutrinos but also established a new way to
probe fundamental properties of these elusive particles in the grand
``laboratories'' of stars or even the entire Universe.  One such
experiment was the Sudbury Neutrino Observatory (SNO)
\citep{Ahmad:PhysRevLett.89.011301} which showed that neutrinos other
than $\nu_e$, the only neutrino species produced in the Sun, were
present in the solar neutrino flux at Earth.  This transformation to
other flavours and the concomitant deficit in $\nu_e$ --- the solar
neutrino problem \citep{Bahcall:1972ax} --- can be elegantly explained
by the Mikheyev-Smirnov-Wolfenstein (MSW) mechanism
\citep{Wolfenstein:1977ue,Mikheyev:1985aa}.  If neutrinos have masses
and the flavour states are not the mass eigenstates, then the electron
neutrinos emitted in the high density at the core of the Sun
experience resonant flavour conversion as they propagate to the vacuum
thereby altering the composition of the flux.  SNO, other solar
neutrino experiments, atmospheric neutrino experiments and also
terrestrial experiments have now 
compiled compelling evidence that indeed neutrinos of different
flavours can mutate into each other and this phenomenon is known as
``neutrino flavour transformation'' or ``neutrino (flavour)
oscillation'' (see, e.g.\ \citealp{Amsler:2008zzb} for a review).

But the focus of this review is core-collapse supernova, which we will
refer to simply as supernova from now on.  The explosion marks the
death of a star with mass greater than $\sim 8 M_\odot$ (see,
e.g.\ \citealp{Woosley:2002zz} for a review) initiated as the iron
core of the star collapses when it is incapable of generating
sufficient supporting thermal energy through further nuclear
fusion \citep{Weaver:1978ApJ...225.1021W}. A proto neutron star (PNS)
\citep{Baade:1934PNAS...20..259B} forms at the centre of the supernova
if the collapse can be halted after supra nuclear density is reached. 
Most of the gravitational binding energy ($\sim
3\times10^{53}\,\mathrm{erg}$) of the PNS is released in the
form of neutrinos of all species with energies $\sim 10\,\mathrm{MeV}$
over a period of 10 seconds or so.  Supernovae are such luminous
sources of neutrinos that a supernova located $10\,\mathrm{kpc}$ from
Earth would flood a neutrino detector here with a flux of $\sim
10^{11}\,\mathrm{cm}^{-2}\mathrm{s}^{-1}$. Over the duration of the
burst this flux is more or less the same as that from the Sun but with
much larger average energy.  With current neutrino detector technology
we should observe at least several thousands from the next supernova
in our Galaxy
\citep{Ahrens:2001tz,Ikeda:2007sa,2002PhRvD..66a3012S,Cadonati:2000kq,2008arXiv0810.1416D,Beacom:2002hs}.

Because supernova energetics are dominated by neutrinos, it was
immediately realized after the discovery of the MSW mechanism that any
neutrino flavour transformation in supernovae could have important
effects \citep{Fuller:1987aa}. There are many flavour transformation
scenarios that could occur in the supernovae but for this review we
shall consider just the flavour transformation of active neutrinos
outside the neutrino sphere where neutrinos start free streaming. In
this regard the flavour transformation is similar to that in the Sun
but there are three key aspects of neutrino flavour transformation in
a supernova that also make it very different from the solar
environment: (a) the matter density in a supernova is much higher
($\rho\sim 10^{12}\,\mathrm{g}\,\mathrm{cm}^{-3}$ at the neutrino
sphere) so that oscillations can be affected by both neutrino
mass-squared differences; (b) the density profile of a supernova is
rich with prominent features which evolve over the duration of the
neutrino emission; and (c) the neutrino flux near the neutrino sphere
can be so large that neutrino self-interaction must be taken into
account.  The vast majority of the neutrinos emitted by the PNS
propagate through the overlying material unimpeded; only a small
percentage will be absorbed by nucleons/nuclei. But due to neutrino
oscillations what is detected here on Earth is not the same as that
emitted by the PNS.  The neutrino evolves as it propagates and for
practical purposes the evolution to Earth is solved in a sequence of
steps: first, we calculate the state at the surface of the supernova;
second, we add on the effect of propagating through the vacuum to
Earth; and third, an ``Earth matter'' effect
\citep{Dighe:1999bi} is sometimes
included to account for the possibility that the neutrinos may have to
pass through the Earth before reaching a detector. Of the three steps
in calculating the neutrino signal at Earth it is the first step, the
calculation to the surface of the supernova, that is by far the most
difficult and which forms the focus of this review. We should mention
here that at the present time neutrino oscillations are not part of
supernova simulations or are treated only approximately. So to date
neutrinos are usually sent through the supernova in a post-processing
stage, i.e.\ the supernova is an input.

%%%%%%%%%%%%%%%%%%%%%%%%%%%%%%%%%%%%%%%%%%%%%%%%%%%%%%%%%%%%%%%%%%%%%%%%%%%%%%%%%%%

\subsection{MSW flavour transformation with dynamic density profiles}

The first studies of the electron dominated/pure MSW effect in
supernovae employed purely static density profiles. Temporal evolution
of the profile was ignored and any time dependence of the neutrino
signal was limited to the variation of the neutrino spectra emitted by
the core. But in recent years this has changed and the density
profiles used now have become increasingly sophisticated.

The first, generic, evolutionary feature of any core-collapse
supernova is the forward shock.  That a shock is formed in the
supernova was discovered in the first simulations by
\citet{Colgate:1960PhRvL...5..235C} and the feature is generated by
the rebound of the PNS during the implosion
\citep{Colgate:1961AJ.....66S.280C}.  Initially it was thought that
the shock would then proceed to shed the mantle of the supernova, the
so-called prompt explosion mechanism, but as the physics input into
the simulations became more accurate it was eventually realised that
actually the prompt explosion mechanism fails
\citep{Hillebrandt:1981A&A...103..147H,vanRiper:1982ApJ...257..793V}. The
outward motion of the forward shock stalls at a distance $\sim
200\;{\rm km}$ from the PNS chiefly due to nuclear dissociation
\citep{Mazurek:1982ApJ...259L..13M,Arnett:1982ApJ...263L..55A}.  If
the star is to explode the shock must be revived from its stalled
position and this can only occur by heating the material behind the
shock. Due to the prodigious amounts of energy they convey, the
neutrinos streaming out from the PNS were invoked by
\citet{Bethe:1985ApJ...295...14B} as the transporters of energy from
the hot core to the region behind the shock. This current paradigm of
a stalled shock that is later revived is termed the delayed explosion
mechanism.  Revival of a shock due to neutrino heating does appear to
work in the limited case of stars in the mass range 8--11 $M_{\odot}$
which have O-Ne-Mg cores
\citep{Kitaura:2006A&A...450..345K,Dessart:2006ApJ...644.1063D}. The
explosion energies of the simulations are smaller than the typical
observed energies of supernova by around an order of magnitude but
such weak supernovae do seem to match the energetics of some faint,
\textsuperscript{56}Ni deficient, Type II P supernovae
\citep{Chugai:2000A&A...354..557C,Pastorello:2007Natur.449....1P}.
But for more massive stars where the core is composed of iron all
spherically symmetric simulations of the supernova using Boltzmann
neutrino transport do not explode
\citep{Rampp:2000ApJ...539L..33R,Mezzacappa:2001PhRvL..86.1935M,Liebendorfer:2001PhRvD..63j3004L,Thompson:2003ApJ...592..434T,Hix:PhysRevLett.91.201102}. Any
simulation that does explode uses ``gray'' neutrino transport
\citep{Mezzacappa:2005ARNPS..55..467M}.  While less satisfying than an
explosion based upon first principles such simulations are,
nonetheless, useful because we can learn from them which features we
should expect to find when truly successful supernova simulations are
performed. \citet{Schirato:2002tg} --- and later
\citet{Lunardini:2003eh}, \citet{Takahashi:2003APh....20..189T}, and
\citet{Fogli:2003PhRvD..68c3005F} --- were the first to consider the
effect of the dynamic profile upon the neutrinos. They demonstrated
that the forward shock would race out through the mantle of the star
and reach the H resonance region --- a position we shall clarify in
the next section --- within the first couple of seconds. Upon arrival
its presence in this region alters the adiabaticity of the neutrinos
as they propagate through the supernova and thus the flavour content
of the neutrino signal at Earth.  This discovery that the evolution of
the density profile alters the flavour composition of the neutrino
signal is the reason that one now needs to use time dependent density
profiles in order to make accurate predictions.

After \citet{Schirato:2002tg} additional features in the profile were
also found to influence the neutrinos.  The energy deposition does not
switch off once the forward shock is revived and has begun moving
outwards again. The material close to the PNS continues to be heated
and the effect is to create a wind with a velocity that increases with
distance from the core \citep{Duncan:1986ApJ...309..141D}. This wind
pushes against the slower moving material in front of it creating a
bubble or cavity. If the post shock-revival energy deposition is even
more vigorous the wind strength grows to the point where its velocity
may supersede the local sound speed. This leads to the formation of a
new shock in the profile that faces the PNS rather than the exterior
\citep{Janka:1995ApJ...448L.109J,Burrows:1995ApJ...450..830B},
i.e.\ it is ``reversed''. Compared to the forward shock, the reverse
shock is much more skittish. The reverse shock can be so strong that
it penetrates to densities below those in front of the forward shock
and then later it can diminish in size as the wind abates and even
stall and return to the core
\citep{Tomas:2004JCAP...09..015T,Arcones:2007A&A...467.1227A,Kneller:2008PhRvD..77d5023K}.
The effect upon the neutrinos of the presence of the reverse shock in
the the profile was first considered by
\citet{Tomas:2004JCAP...09..015T}. Unfortunately these authors
unintentionally omitted the consequences of the correlation of the
wavefunction between the multiple H resonances now present in the
profile which can lead to phase effects. That multiple resonances can
lead to phase effects had been discussed many years prior by
\citet{Kuo:1989qe} but it was only recently that
\citet{Fogli:2003PhRvD..68c3005F}, re-examining the profiles of
\citet{Schirato:2002tg}, discovered the phase effects in supernova
neutrinos.  Further examples showing their presence were presented by
\citet{Kneller:2005hf} and then phase effects were discussed
extensively by \citet{Dasgupta:2007PhRvD..75i3002D}.

But the failure of the best one-dimensional (1D), first-principles
simulations indicates that supernova must be a multi-dimensional
phenomena requiring rotation, convection, magnetic fields and/or other
multi-dimensional physics.  If that is the case then one would expect
any explosion to be aspherical with a density profile that varies with
the line of sight.  Observational evidence that supernova are
aspherical has abounded for years: asphericity has been observed or is
implied from observations of Supernova 1987A
\citep{Dotani:1987Natur.330..230D,Sunyaev:1987Natur.330..227S,Matz:1988Natur.331..416M,Erickson:1988ApJ...330L..39E,Li:1993ApJ...419..824L},
from the Vela SNR \citep{Aschenbach:1995Natur.373..587A}, from
Cassiopeia A
\citep{Hughes:2000ApJ...528L.109H,Hwang:2004ApJ...615L.117H} and from
many other supernova remnants
\citep[e.g.][]{Wang:2001ApJ...550.1030W,Leonard:2006Natur.440..505L,Katsuda:2008PASJ...60S.107K}.
And of course pulsars have been observed with astounding velocities
with some approaching $\sim 1500\,\mathrm{km}\,\mathrm{s}^{-1}$
\citep{Taylor:1993ApJS...88..529T,Cordes:1993Natur.362..133C,Lyne:1994Natur.369..127L,Chatterjee:2005ApJ...630L..61C,Winkler:2007ApJ...670..635W}.
As with the 1D simulations, all two-dimensional (2D) simulations
studied so far with the state-of-the-art neutrino transport
\citep{Mezzacappa:1998ApJ...495..911M,Buras:2003PhRvL..90x1101B,Buras:2006A&A...447.1049B}
have yet to explode. In their place, 2D gray codes have
been used to study how asphericity could be generated. Initially the
focus was upon neutrino driven convection but the anisotropy was found
to be small and the pulsar velocities were only of a few hundred
$\mathrm{km}\,\mathrm{s}^{-1}$
\citep{Janka:1994A&A...290..496J,Burrows:1995ApJ...450..830B,Janka:1996A&A...306..167J}.
While not sufficient to explain the observations these studies did
note that their simulations contained other departures from
asphericity: for example, distortions of the stalled shock were
observed by \citet{Burrows:1995ApJ...450..830B} who described the
shock surface in their simulations as botryoidal, i.e.\ resembling a
bunch of grapes. It was further study of these distrotions of the
stalled shock that have led to the discovery by
\citet{Blondin:2003ApJ...584..971B} of a new mechanism for generating
asphericity in supernova. \citet{Blondin:2003ApJ...584..971B} found
that small aspherical perturbations of a stalled, spherical accretion
shock could quickly grow and develop large dipole and, to a lesser
extent quadrapole, modes.  This standing accretion shock instability
(SASI) will play an important part of future supernova simulations and
may be a key ingredient in achieving explosions based upon first
principles.  From the multi-dimensional simulations we find that each
radial slice through the supernova still possess the common features
of forward/reverse shocks and a global hot bubble region identified
from 1D but superimposed will be a mixture of features typical of
asphericity such as additional/internal shocks, local bubbles,
turbulence, sound waves, etc.  Of these additional features in the
profile the one that has generated most interest is the turbulence.
The effects of turbulence in supernova profiles was first considered
by \citet{Sawyer:1990PhRvD..42.3908S,Sawyer:1994PhRvD..50.1167S} and \citet{Loreti:1995ae} then later expanded upon by
\citet{Fogli:2006JCAP...06..012F}.  The correlation function of the
density fluctuations is an important determinant and a typical first
assumption was that the fluctuations were $\delta$-correlated. However,
it was pointed out by \citet{Benatti:2005PhRvD..71a3003B} that such a
correlation is rather idealised and so recently authors such as
\citet{Friedland:2006ta} and \citet{Choubey:2007PhRvD..76g3013C} have
switched to considering Kolmogorov turbulence instead. Whatever the
exact spectrum used the general result of all these studies is that
turbulence of sufficient strength can lead to flavour depolarisation.

In summary, over the period of just a few years the density
profiles used to study supernova neutrino oscillations have progressed
from simple, static parameterisations to profiles taken from
aspherical supernova simulations that contain numerous features that
affect the neutrinos. Later in this review we will examine in greater
detail the dynamics of the pure MSW effect and how the use of improved
density profiles has altered our expectations of the information in
the neutrino signal.

%%%%%%%%%%%%%%%%%%%%%%%%%%%%%%%%%%%%%%%%%%%%%%%%%%%%%%%%%%%%%%%%%%%%%%%%%%%%%%%%%%%

\subsection{Neutrino self-interaction and collective neutrino flavour
  transformation} 

\begin{figure}[t]
\begin{indented}
\item[]% 
\includegraphics*[width=0.5\textwidth,keepaspectratio]%
  {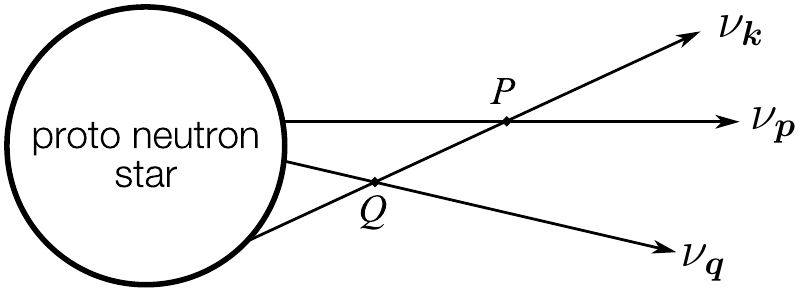}
\end{indented}
%\hspace{2pc}%
%\begin{minipage}[t]{18pc}
\caption{\label{fig:entanglement} With neutrino self-interaction
  flavour evolution histories of neutrinos propagating along different
  but intersecting trajectories, e.g.\ $\nu_\bi{k}$ and $\nu_\bi{q}$,
  are coupled because their flavour evolution histories beyond the
  intersection point, $Q$, will depend on the flavour states of both
  neutrinos at $Q$.  In addition, the flavour evolution histories of
  non-intersecting neutrino beams such as $\nu_\bi{p}$ and
  $\nu_\bi{q}$ can also be coupled because the flavour evolution
  history of $\nu_\bi{p}$ beyond $P$ indirectly depends on the flavour
  state of $\nu_\bi{q}$ at $Q$ through $\nu_\bi{k}$.}
%\end{minipage}
\end{figure}

The contribution from neutrino-neutrino forward scattering, or
neutrino self-interaction, was noted by some of the first studies of
neutrino flavour transformation in stellar collapse
\citep{Fuller:1987aa,Fuller:1992aa} and in the early Universe \citep{Notzold:1988kx}.
Shortly after \citet{Pantaleone:1992xh} pointed out the existence of
off-diagonal elements of the neutrino self-interaction potential
in the flavour basis, and
this observation was then incorporated into the general formalism for
neutrino flavour evolution laid out by \citet{Sigl:1992fn}. From
these early studies it quickly became obvious that the inclusion of
neutrino self-interactions poses a formidable challenge
\citep{Qian:1994wh}. The reason for the complexity is illustrated in
figure \ref{fig:entanglement}.  The flavour evolution histories of any
two neutrino beams becomes coupled due to the self-interaction and, as a
result, in order to study the neutrino oscillations with neutrino
self-interactions one needs to simultaneously follow the flavour
evolution of all the neutrinos (with different energies, initial
flavour states and propagating directions) emitted around the same
time from all points on the neutrino sphere.  We must emphasise that
coupling of neutrino flavour evolution histories discussed in this
review is, however, not quantum entanglement. The possible effects of
the latter were discussed by,
e.g.\ \citet{Bell:2003mg,Friedland:2003dv,Friedland:2003eh}.

Even for an idealised supernova model with a perfect spherical
symmetry one must distinguish between the different neutrino
trajectories along which neutrinos will have travelled 
different distances for the same radius interval.  A simplifying
approximation (but not necessarily a self-consistent treatment) was
proposed by \citet{Qian:1994wh} who considered all the neutrino
trajectories to be equivalent. In this ``single-angle approximation''
only the flavour evolution of neutrinos along one representative
trajectory (e.g.\ the radially directed one) are computed.

In the two-flavour mixing scheme the flavour state of a neutrino can
be represented as a classical spin or ``magnetic dipole'' in the
three-dimensional flavour space \citep{Mikheyev:1986tj,Kim:1987ss}.
Using this spin analogy \citet{Sigl:1992fn} showed that neutrino
self-interaction is represented as coupling between these ``flavour
spins''.  (We stress that a flavour spin is a fictitious spin in
flavour space and that the spin analogy for neutrino oscillations
discussed here is not related to flavour conversion of neutrinos
because of coupling between the neutrino electromagnetic moment and a
physical magnetic field considered by, e.g.\ \citet{Voloshin:1986aa}.)
While the early studies of neutrinos in supernova
\citep{Fuller:1987aa,Qian:1993dg,Qian:1994wh,Qian:1995ua} treated the
self-interaction as an additional matter effect in the MSW mechanism,
it is not inconceivable that the ensemble of flavour spins can exhibit
collective behaviors in a similar fashion as spin gases in condensed
matter physics. In fact, collective neutrino oscillation phenomena
have been discussed in the context of the early Universe
\citep[e.g.][]{Kostelecky:1994dt}.  One collective oscillation
phenomenon which can appear in dense, homogeneous, isotropic neutrino
gases is that, independent of their energy, all the neutrinos
oscillate in the same way as a representative neutrino in vacuum
\citep{Samuel:1993uw}. This phenomenon is known as the synchronised
neutrino oscillation because it can be explained in analogy to ``a
system of magnetic dipoles which are coupled by their
self-interactions to form one large magnetic dipole which then
precesses coherently in a weak external magnetic field''
\citep{Pastor:2002we}.  Another collective oscillation phenomenon is
known to occur in dense gases with approximately equal number of
neutrinos and antineutrinos. Both neutrinos and antineutrinos can
experience ``substantial flavour oscillation even for extremely small
mixing angles'' if the neutrino mass hierarchy is inverted
\citep{Kostelecky:1993dm}.  This behaviour is usually known as the
bipolar neutrino oscillation because it can explained by using an
analogy to a bipolar spin system which consists of two coupled and
approximately oppositely-oriented groups of spins \citep{Duan:2005cp}.

Armed with this insight from neutrino behaviour in the early Universe
and using the single-angle approximation \citet{Pastor:2002we} and
\citet{Balantekin:2004ug} showed that, in a normal neutrino mass
hierarchy scenario, the combination of synchronisation and the MSW
mechanism may indeed cause flavour conversion of both neutrinos and
antineutrinos if the supernova neutrino luminosities are large
enough. This phenomenon requires either a large mass-squared
difference ($\delta m^2\simeq 10\,\mathrm{eV}^2$) or a dilute
supernova envelope so that the synchronised neutrinos encounter a
MSW resonance near the neutrino sphere. Alternatively,
\citet{Fuller:2005ae} envisioned a scenario where the off-diagonal
part of the neutrino self-interaction potential dominates and both neutrinos and
antineutrinos simultaneously achieve maximal mixing even for small
mass-squared differences. This scenario is essentially a
special case of synchronisation. Although \citet{Fuller:2005ae}
identified necessary conditions for this scenario to occur, 
it is not clear if a realistic supernova event could evolve into
this scenario even if these necessary
conditions are satisfied.  Then \citet{Duan:2005cp} showed that a
general bipolar neutrino system (e.g.\ with unequal numbers of
neutrinos and antineutrinos) can exhibit collective flavour
oscillations of either the synchronised or bipolar type if the total
neutrino flux is larger or smaller than some critical value. In the
same study \citet{Duan:2005cp} also showed that ordinary matter has
little effect on collective neutrino oscillations 
 and that bipolar neutrino oscillations can occur very
close to the neutrino sphere even if the matter density is
large. Based on their analyses, \citet{Duan:2005cp} mapped out the
regimes in supernovae where synchronised, bipolar and conventional MSW
neutrino oscillations may occur.

The first multi-angle (i.e.\ treating different neutrino trajectories
as well as different neutrino energies) calculations of neutrino
flavour transformation in supernovae were carried out by
\citet{Duan:2006an,Duan:2006jv}. The results of their calculations
show that, contrary to expectations based on the conventional MSW
mechanism or synchronisation, both neutrinos and antineutrinos can
simultaneously experience significant flavour oscillations near the
neutrino sphere in some \emph{inverted} neutrino mass hierarchy
scenarios. Furthermore, when collective neutrino oscillations end,
neutrinos of different flavours have their energy spectra swapped
above a critical energy $\Es$, a phenomenon now known as ``stepwise
spectral swap'' or ``spectral split''. With density profiles and
neutrino mixing parameters similar to those adopted by
\citet{Balantekin:2004ug}, \citet{Duan:2006an,Duan:2006jv} observed
similar spectral swap/split phenomenon with the normal mass
hierarchy. They also carried out the single-angle calculations with
setups similar to the multi-angle ones, and the results of
single-angle and multi-angle calculations were found to bear many of
the same qualitative features. This suggests that the single-angle
approximation is a valid approach to study, at least qualitatively,
neutrino oscillations in supernovae. Using the single-angle
approximation \citet{Duan:2006an} also showed that the spectral
swap/split phenomenon can be a result of collective neutrino
oscillation mode in which all flavour spins precess collectively.  The
discoveries made by \citet{Duan:2006an,Duan:2006jv} have triggered 
 intense and fruitful studies of collective neutrino
oscillations in the supernova environment. Similar results were found
in calculations using different neutrino mixing parameters and matter
density profiles \citep{Duan:2007bt,Fogli:2007bk} and even for some
full three-flavour mixing scenarios
\citep{Duan:2007sh,Dasgupta:2007ws}.  Signatures of these oscillation
features in supernova neutrino signals have been investigated
\citep{Dasgupta:2008my,Chakraborty:2008zp,Fogli:2008fj,Gava:2009pj,Lunardini:2007vn}
and on the theoretical side collective neutrino oscillations in a
single-angle approximated supernova model are now well understood
\citep{Hannestad:2006nj,Duan:2007mv,Raffelt:2007cb,Raffelt:2007xt,Duan:2007fw,Dasgupta:2007ws,Duan:2008za,Dasgupta:2008cd}.
Collective neutrino oscillations in anisotropic environments
\citep{Hannestad:2006nj,Raffelt:2007yz,EstebanPretel:2007ec,Sawyer:2008zs,Duan:2008fd}
and in non-spherical geometry \citep{Dasgupta:2008cu} have also been
examined. Effects of $CP$-violation
\citep{Balantekin:2007es,Gava:2008rp} and the case of very large
matter density \citep{EstebanPretel:2007yq,EstebanPretel:2008ni} have
also been analysed.  The recent progress in the understanding of
neutrino self-interactions in supernovae has been impressive and, like
the effect of the dynamic density profiles, has completely changed
our expectations of the features in the neutrino signal we shall receive from
the next Galactic supernova.

%%%%%%%%%%%%%%%%%%%%%%%%%%%%%%%%%%%%%%%%%%%%%%%%%%%%%%%%%%%%%%%%%%%%%%%%%%%%%%%%%%%

\subsection{Organisation of the paper}

Due to their importance in the explosion environment the neutrinos and
their flavour transformation touch upon many different aspects of
supernova phenomenology. For lack of space we cannot contemplate a
comprehensive review of the entire field so we shall focus upon the
specific topic of flavour transformation of active neutrinos as they
propagate from the PNS to the supernova surface.  In doing so we have
left out many other interesting subjects such as neutrino nucleus
interactions \citep[e.g.][]{2003JPhG...29.2513B}, neutrino inelastic
scattering \citep[e.g.][]{2008JPhG...35a4057M} and the possibility and
consequences of sterile neutrinos
\citep[e.g.][]{Fuller:2009zz,Hidaka:2007se,Beun:2006ka,Fetter:2002xx}.
We will also not be able to address how one can begin to unpick the
various signatures we now expect in an actual neutrino signal instead
referring the reader to the literature
\citep[e.g.][]{Jachowicz:2006xx,2008PhRvC..77e5501J}.  We begin with
some general formalism of flavour transformation of supernova
neutrinos in section \ref{sec:general}.  Historically the effects of
dynamic matter density profiles and neutrino self-interaction have
been studied in parallel, i.e.\ without inclusion of the effects
caused by each other.  So we will independently discuss the MSW
flavour transformation with dynamic density profiles and collective
flavour transformation in sections \ref{sec:MSW-regime} and
\ref{sec:coll-regime}, respectively.  Finally in section \ref{sec:summary} we
summarise and provide some outlook for where future advances lie.

%%%%%%%%%%%%%%%%%%%%%%%%%%%%%%%%%%%%%%%%%%%%%%%%%%%%%%%%%%%%%%%%%%%%%%%%%%%%%%%%%%%
%%%%%%%%%%%%%%%%%%%%%%%%%%%%%%%%%%%%%%%%%%%%%%%%%%%%%%%%%%%%%%%%%%%%%%%%%%%%%%%%%%%
%%%%%%%%%%%%%%%%%%%%%%%%%%%%%%%%%%%%%%%%%%%%%%%%%%%%%%%%%%%%%%%%%%%%%%%%%%%%%%%%%%%

\section{Neutrino flavour transformation in supernovae: 
general discussions \label{sec:general}}

\subsection{Neutrino flavour transformation without neutrino self-interaction}

Because neutrinos initially propagate coherently outside the neutrino
sphere, the flavour state $|\psi_\bip(\bir)\rangle$ of a neutrino with
momentum $\bip$ at point $\bir$ on its world line can be solved from a
Schr\"odinger-like equation \citep[e.g.][]{Halprin:1986pn,Cardall:2007dy}
%\numparts
\begin{equation}
\fl
\rmi\frac{\rmd}{\rmd \lambda}|\psi_\bip(\bir)\rangle 
=\hat{H}(\bir)
|\psi_\bip(\bir)\rangle
=[\hat{H}_\vac+\hat{H}_\matt(\bir)+\hat{H}_{\nu\nu}(\bir)]
|\psi_\bip(\bir)\rangle,\label{eq:eom}
\end{equation}
%\endnumparts% 
where $\lambda$ is the propagation distance of the
neutrino along its world line, $\hat{H}_\vac$ is the Hamiltonian for
vacuum oscillations, $\hat{H}_\matt(\bir)$ is from the forward
scattering of the neutrino with the ordinary matter, and
$\hat{H}_{\nu\nu}(\bir)$ is the contribution from neutrino
self-interaction. In equation \eref{eq:eom} 
we have adopted the steady-state approximation.
%we have left out the time
%dependence of the neutrino flavour state and the Hamiltonian. 
This is
because neutrinos can traverse regions of interest within very short
time, and the physical conditions in supernovae can barely change
during this period. The flavour evolution of a neutrino along its
world line can, therefore, be solved from a static configuration as
outlined in equation \eref{eq:eom}. We first consider the scenarios
that neutrino self-interaction can be ignored, i.e.\ 
$\hat{H}_{\nu\nu}(\bir)=0$.

There are two fundamental bases for the neutrinos: the flavour states
$|\nu_\alpha\rangle$ ($\alpha=e,\mu,\tau$) which are the states
seen in detectors, and the vacuum mass states $|\nu_i\rangle$
($i=1,2,3$) which are the eigenstates of $\hat{H}_\vac$. In the
latter basis $\hat{H}_\vac$ is --- up to a term proportional to the
identity matrix --- equal to
\begin{equation}
H^\bsv_\vac = \frac{1}{2E}\, \diag[m_1^2, m_2^2, m_3^2],
% \left[\begin{array}{ccc} 
% m_{1}^{2} & 0 & 0 \\ 
% 0 & m_{2}^{2} & 0 \\ 
% 0 & 0 & m_{3}^{2} 
% \end{array}\right],
\end{equation}
where the elements of $H^\bsv_\vac$ are defined as
$(H^\bsv_\vac)_{ij}\equiv\langle\nu_i|\hat{H}_\vac|\nu_j\rangle$,
superscript ``$\bsv$'' stands for the vacuum mass basis, $E=|\bip|$ is
the energy of the neutrino, and $m_i$ is the mass of $\nu_i$.  The two
bases are related by a unitary transformation
\begin{equation}
|\nu_\alpha\rangle=\sum_i U_{\alpha i}^* |\nu_i\rangle.
\label{eq:U-trans}
\end{equation}
%% A unitary matrix can generally be expressed in terms of six phase
%% factors and the trigonometric functions of three angles. We note that
%% the phase factor of a flavour state $|\nu_\alpha\rangle$ is not
%% measurable and can be of any value. We can fix the phase factors of
%% the flavour states by setting three of the six phases in $U$ to
%% zero. With this elimination of three of the phases the unitary matrix
%% $U$ in equation \eref{eq:U-trans} can be 
The unitary matrix $U$ in equation \eref{eq:U-trans} is conventionally
parametrised with three
vacuum mixing angles $\theta_{12}$, $\theta_{23}$ and $\theta_{13}$,
and the three remaining phases $\delta$, $\alpha_1$ and $\alpha_2$
\citep{Amsler:2008zzb} as:
\begin{eqnarray}
%\fl
U&=&\left[\begin{array}{ccc} c_{12} c_{13} & s_{12} c_{13} & s_{13}
    \rme^{-\rmi\delta} \\ -s_{12} c_{23} - c_{12} s_{23} s_{13}
    \rme^{\rmi\delta} & c_{12} c_{23} - s_{12} s_{23} s_{13}
    \rme^{\rmi\delta} & s_{23} c_{13} \\ s_{12} s_{23} - c_{12} c_{23}
    s_{13} \rme^{\rmi\delta} & -c_{12} s_{23} - s_{12} c_{23} s_{13}
    \rme^{\rmi\delta} & c_{23} c_{13}
\end{array}\right]\nonumber\\
&\quad&\times\diag[\rme^{\rmi\alpha_1/2},\rme^{\rmi\alpha_2/2},1],
\label{eq:U}
\end{eqnarray}
where $c_{ij}=\cos\theta_{ij}$ and $s_{ij}=\sin\theta_{ij}$. We will
ignore the two Majorana phases $\alpha_1$ and $\alpha_2$ from this
point because they have no effect for flavour transformation of
ultra-relativistic neutrinos in matter \citep[e.g.][]{Strumia:2006db}. Various
neutrino oscillation experiments have measured or put constraints on
most of the neutrino mixing parameters: $\delta
m_{12}^2=m_2^2-m_1^2\simeq\delta m_\odot^2\simeq 8\times
10^{-5}\,\mathrm{eV}^2$, $|\delta m_{23}^2|=|m_3^2-m_2^2|\simeq\delta
m_\mathrm{atm}^2\simeq 2.4\times 10^{-3}\,\mathrm{eV}^2$,
$\sin^22\theta_{12}\simeq0.86$, $\sin^22\theta_{23}\simeq1$ and
$\sin^22\theta_{13}\lesssim0.19$
\citep{Fogli:2006PrPNP..57..742F}.  The $CP$ phase $\delta$ and
the sign of $\delta m_{23}^2$ are undetermined. It is conventional to
call a mixing scheme a normal mass hierarchy if $\delta
m_{23}^2>0$ and an inverted mass hierarchy if $\delta
m_{23}^2<0$.

The Hamiltonian $\hat{H}_\matt(\bir)$ is a diagonal matrix in the
flavour basis
\begin{equation}
H_\matt^\bsf(\bir)=\diag[V_e(\bir),0,0] =\sqrt{2}\GF\NA
\rho(\bir)\,\diag[Y_e(\bir), 0, 0],
\label{eq:Hmatt}
\end{equation} 
where superscript ``$\bsf$'' denotes the flavour basis,
$\GF\simeq1.166\times10^{-5}\,\mathrm{GeV}^{-2}$ is Fermi's constant,
$\NA\simeq6.022\times10^{23}\,\mathrm{g}^{-1}$ is Avogadro's number,
$\rho(\bir)$ is the matter density, and $Y_e(\bir)$ is the electron
fraction per baryon. We note that the matter temperature outside the
neutrino sphere is too low to have $\mu$ or $\tau$ produced. (However,
see, e.g.\ \citealp{Dighe:1999bi,EstebanPretel:2007yq} for the effects of
nonzero ``effective'' 
$\mu$-abundance in the presence of a very large matter density.)  The
eigenstates $|\nu^\rmm_i(\bir)\rangle$ ($i=1,2,3$) of the
Hamiltonian $\hat{H}(\bir)=\hat{H}_\vac+\hat{H}_\matt(\bir)$
constitute the ``matter basis''.
%%  and are related to the flavour states by
%% another unitary transformation
%% \begin{equation}
%% |\nu_\alpha\rangle=\sum_i [U^\rmm_{\alpha i}(\bir)]^*
%% |\nu^\rmm_i(\bir)\rangle.
%% \label{eq:U-trans-matt}
%% \end{equation}
We define matter eigenstates $|\nu^\rmm_i(\bir)\rangle$ to be in the
same order as that of vacuum mass eigenstates $|\nu_i\rangle$, i.e.\ 
the eigenvalues $k_i(\bir)$ associated with $|\nu^\rmm_i(\bir)\rangle$
satisfy $k_1(\bir)<k_2(\bir)<k_3(\bir)$ for the normal mass hierarchy and
$k_3(\bir)<k_1(\bir)<k_2(\bir)$ for the inverted mass hierarchy. 
%% It is worthy noting that, in 
%% constrast to the vacuum mixing matrix $U$ in equation \eref{eq:U}, the
%% unitary matrix $U^\rmm(\bir)$ must be, in general, parameterised with
%% all six phases \citep{Kneller2009}. 
Unlike the flavour or vacuum mass
states, the eigenstates $|\nu^\rmm_i(\bir)\rangle$ as well as the
corresponding eigenvalues $k_i(\bir)$ vary with location as the
electron number density $n_e(\bir)=\NA\rho(\bir) Y_e(\bir)$
changes.  All the above discussions also apply to antineutrinos except
that there is a sign difference between $\hat{H}_\matt(\bir)$ for
neutrinos and for antineutrinos, i.e.\ one should take
$V_e(\bir)\rightarrow-V_e(\bir)$ in equation \eref{eq:Hmatt} for
antineutrinos.

As an alternative to solving equation \eref{eq:eom} the evolution of
a neutrino can also be found by solving for the evolution operator
$\hat{S}_\bip(\bir,\bir_0)$. This operator obeys a similar equation,
\begin{equation}
\rmi\frac{\rmd}{\rmd\lambda}\hat{S}_\bip(\bir,\bir_0)
=\hat{H}(\bir)\hat{S}_\bip(\bir,\bir_0).
\label{eq:S}
\end{equation}
and has the additional property that it obeys the product rule:
\begin{equation}
\hat{S}_\bip(\bir,\bir_0)=\hat{S}_\bip(\bir,\bir^\prime)
\hat{S}_\bip(\bir^\prime,\bir_0). \label{eq:product_rule}
\end{equation}
%% The probability that a neutrino with initial state
%% $|\nu_j(\bir_0)\rangle$ is later detected as state
%% $|\nu_i(\bir)\rangle$ is
%% %%
%% \begin{equation}
%% P(|\nu^\rmm_j\rangle \rightarrow |\nu^\rmm_i\rangle) =
%% |\hat{S}_\bip(\bir,\bir_0)_{ij}|^2.
%% \end{equation}
Equation \eref{eq:S} has a formal solution
\begin{equation}
\hat{S}_\bip(\bir,\bir_0)=\mathbb{T}\exp\left[-\rmi
  \int_{\lambda_0}^\lambda\hat{H}(\bir^\prime)\rmd\lambda^\prime\right],
\label{eq:S-exp}
\end{equation}
where $\mathbb{T}$ is the space/time-ordering operator, and
$\lambda_0$, $\lambda$ and $\lambda^\prime$ are the distances along
the world line of the neutrino that correspond to $\bir_0$, $\bir$ and
$\bir^\prime$, respectively.  In practice, $\hat{S}_\bip(\bir,\bir_0)$
is approximated by a sum of truncated series
\citep{Kneller:2005hf,Kneller2009}.  

\subsection{The two-flavour approximation}
\begin{figure}[t]
\begin{indented}
\item[]% 
$\begin{array}{@{}c@{\hspace{0.02 \textwidth}}c@{}}
  \includegraphics*[scale=0.48, keepaspectratio]{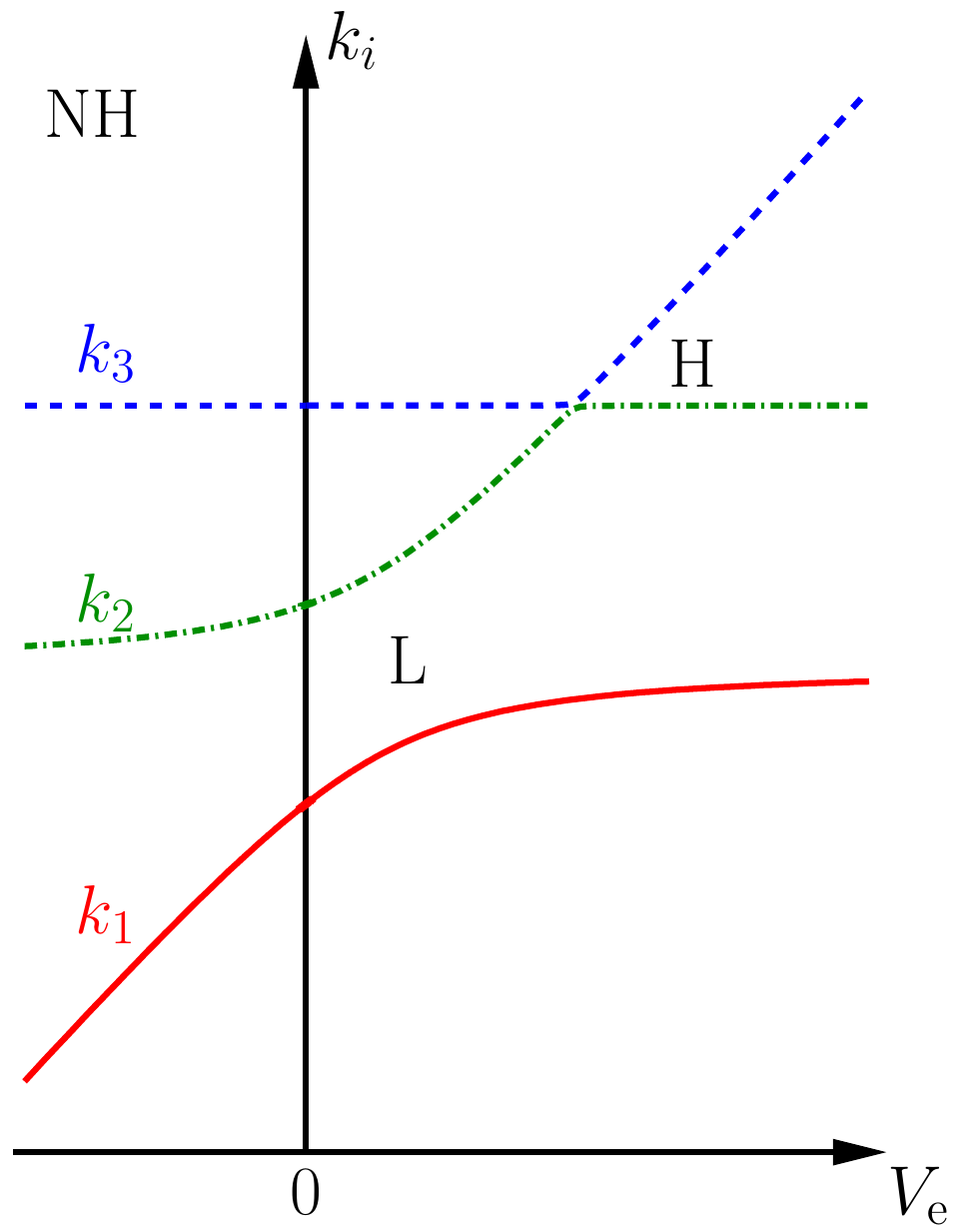} &
  \includegraphics*[scale=0.48, keepaspectratio]{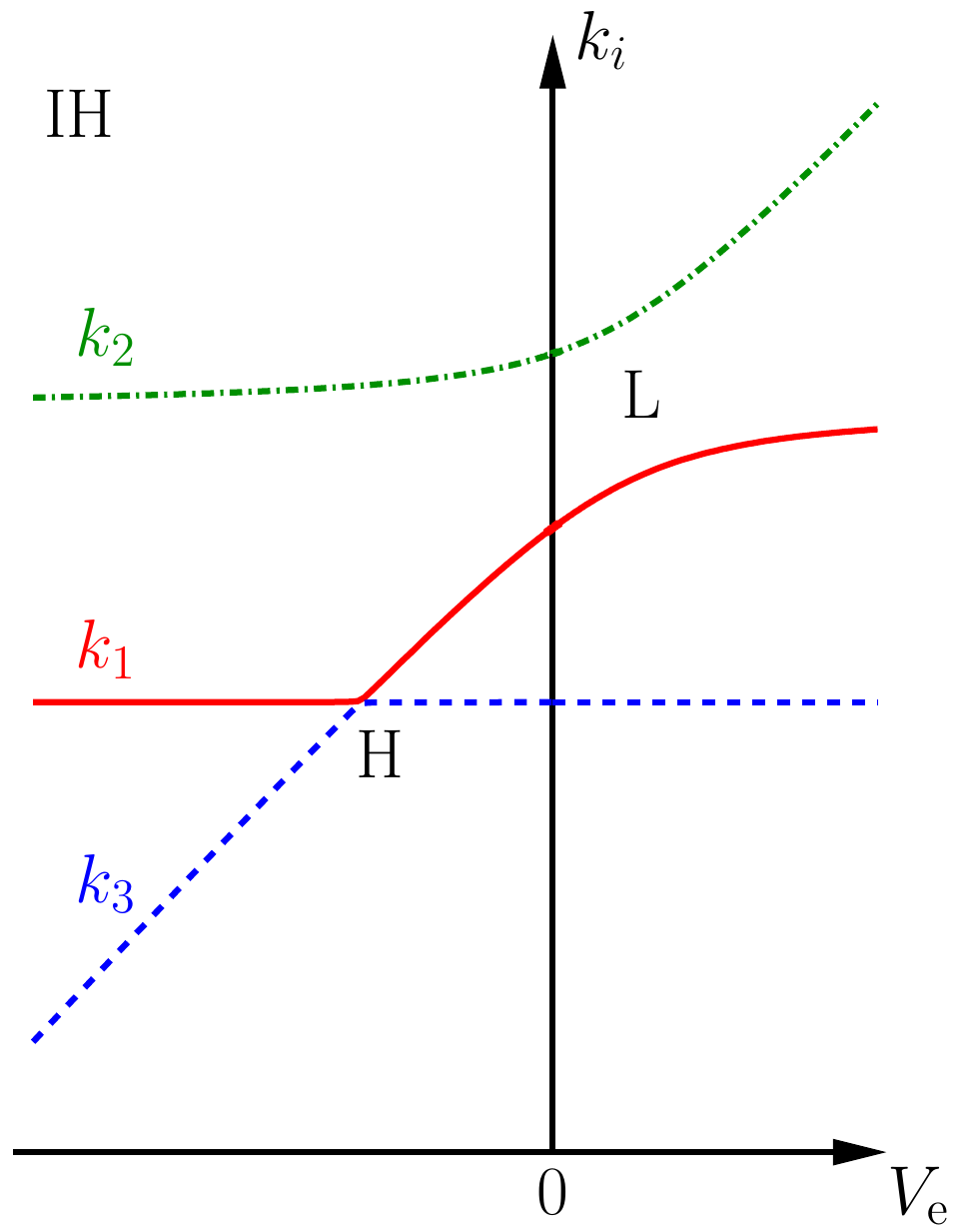}
\end{array}$
\end{indented}
\caption{\label{fig:kV} Schematic plots of $k_i$, the eigenvalues of
  Hamiltonian $\hat{H}$, as functions of $V_e$ for the normal 
  neutrino mass hierarchy (NH, the left panel) and
  the inverted neutrino mass hierarchy (IH, the right panel),
  respectively. The ``H'' and ``L'' MSW resonances are labelled
  correspondingly. The resonances occur at $V_e>0$ and $V_e<0$
  are for neutrinos and antineutrinos, respectively. The eigenvalues
  of the two distinct matter states that are involved in the H
  resonance are close to each other (but do not equal) at the resonance.}
\end{figure}

We have, so far, described everything in terms of three flavours but a
quick scan through the literature by the reader will reveal that many
studies have used only two. The reduction in number of flavours comes
from observation of how neutrinos evolve in the matter basis.  
The evolution operator $\hat{S}_\bip$ in this basis is a diagonal
matrix and the probability for the neutrino to be in each matter
eigenstate is constant
except in the vicinity of the resonant regions where two of
the three eigenvalues $k_i$ of $\hat{H}$ approach one another.  The
evolution of the eigenvalues as a function of the potential is shown
in figure \ref{fig:kV} and from it we see that there are two
resonances: the ``H'' resonance at a larger matter density associated
with the $\delta m_\mathrm{atm}^2$-scale, and the resonance at the $\delta
m_\odot^2$-scale known as the ``L'' resonance which occurs at a
smaller matter density.  As a result of the large difference between
the two measured mass-squared differences, $\delta m_\odot^2$ and
$\delta m_\mathrm{atm}^2$, neutrino mixing at any resonant location
occurs principally between the two matter states whose eigenvalues are
the closest, and the third matter state remains decoupled
\citep{Kuo:1989qe,Dighe:1999bi}.  
Therefore, the full three-flavour mixing problem can be approximated as
two successive two-flavour mixing scenarios that occur at the H and L
resonances, respectively.
Here the L resonance always describes the
mixing between $|\nu^\rmm_1(\bir)\rangle$ and $|\nu^\rmm_2(\bir)\rangle$ but for
the H resonance the mixing states differ being $|\nu^\rmm_2(\bir)\rangle$
and $|\nu^\rmm_3(\bir)\rangle$ for normal mass hierarchy and the states
$|\bar\nu^\rmm_1(\bir)\rangle$ and $|\bar\nu^\rmm_3(\bir)\rangle$ for
inverted mass hierarchy. 

Either the H or L resonance is said to be ``adiabatic'' or
``non-adiabatic'' depending upon whether the crossing probability $P_\rmC$ is
close to zero or closer to unity and indeed these are the two natural
values. If the electron number density $n_e=\rho Y_e$ has a power-law
dependence on the distance along the world line of the neutrino,
i.e.\ $n_e(\lambda)\propto 1/(\lambda-\lambda_0)^n$, 
the crossing probability $P_\rmC$ in the two-flavour approximation has an
analytical solution
\begin{equation}
P_\rmC = \exp\left( -\frac{\pi\gamma_{\star}\,F(n,\thetav)}{2}
\right) \label{eq:PC.KP}.
\end{equation}
In equation \eref{eq:PC.KP}
\begin{equation}
\gamma_{\star}=\sin^{2}2\thetav\,\frac{\delta m^{2}}{E^{2}}\,
\left|\frac{\rmd V_e(\bir)}{\rmd
  \lambda}\right|^{-1}_{\bir=\bir_{\star}} \label{eq:gammastar}
\end{equation}
is the adiabaticity parameter evaluated at the resonant position
$\bir_\star$ where
\begin{equation}
\frac{\delta
  m^2}{2E}\cos2\thetav-V_e(\bir_\star)=0. \label{eq:2nuresonancecondition}
\end{equation}
Here $\delta m^2$ and $\thetav$ are the mass-squared difference and
effective vacuum mixing angle in the two-flavour approximation.
The function $F(n,\thetav)$ in equation \eref{eq:PC.KP} depends solely on
$n$ and $\thetav$ and can be found in \citet{Kuo:PhysRevD.39.1930} and
\citet{Kachelriess:2001PhRvD..64g3002K}. It is common to find authors
using the Landau-St\"{u}ckelberg-Zener
\citep{Landau:1977oj,Stuckelberg1932,Zener:1932RSPSA.137..696Z}
formula for a linear density profile for which $F=1$.  

When equation \eref{eq:PC.KP} is applied to the L resonance with
$\delta m^2\simeq\delta m_\odot^2$ and $\thetav\simeq\theta_\odot$ it
turns out that the crossing probability is always very small --- as in
the sun --- and the neutrino transformation is
adiabatic \citep{Dighe:1999bi}. As a result, one always has
\begin{equation}
%\fl
S_\bip \sim \left[\begin{array}{ccc}
1 & 0 & 0 \\
0 & \alpha_\rmH & \beta_\rmH \\
0 & -\beta_\rmH^* & \alpha_\rmH^*
\end{array}\right]
\quad\mathrm{and}\quad
\bar{S}_\bip \sim \left[\begin{array}{ccc}
\bar\alpha_\rmH & 0 & \bar\beta_\rmH \\
0 & 1 & 0 \\
-\bar\beta_\rmH^* & 0 &  \bar\alpha_\rmH^*
\end{array}\right],
\label{eq:S3}
\end{equation}
where $S_\bip$ ($\bar{S}_\bip$) is the matrix of the evolution
operator in the matter basis that evolves the neutrino from the
neutrino sphere to the surface of the supernova, and ``$\sim$'' means
that the two matrices are different only by some diagonal matrices
which do not change the probability for the neutrino or antineutrino
to be in each matter eigenstate \citep{Kneller:2008PhRvD..77d5023K}.
In equation \eref{eq:S3} $\alpha_\rmH$ and $\beta_\rmH$
($\bar\alpha_\rmH$ and $\bar\beta_\rmH$) are the Cayley-Klein
parameters for $S_\rmH$ ($\bar{S}_\rmH$), the evolution operator in
the matter basis that describes the two-flavour mixing scenario for the
H resonance of the neutrino (antineutrino).  Up to a common phase
factor which is irrelevant for neutrino oscillations we can write
\begin{equation}
S_\rmH=
\left[\begin{array}{cc} 
\alpha_\rmH & \beta_\rmH \\  
-\beta_\rmH^* & \alpha_\rmH^*
\end{array}\right]
\quad\mathrm{and}\quad
\bar{S}_\rmH=
\left[\begin{array}{cc} 
\bar\alpha_\rmH & \bar\beta_\rmH \\  
-\bar\beta_\rmH^* & \bar\alpha_\rmH^*
\end{array}\right],
 \label{eq:SH}
\end{equation}
where the dependence of Cayley-Klein parameters on neutrino momentum
is implicit. The Cayley-Klein parameters satisfy the unitary condition
\begin{equation}
|\alpha_\rmH|^2+|\beta_\rmH|^2=1
\quad\mathrm{and}\quad
|\bar\alpha_\rmH|^2+|\bar\beta_\rmH|^2=1.
\label{eq:alpha-beta}
\end{equation}
From equations \eref{eq:SH} and \eref{eq:alpha-beta} we see that
the crossing probability at the H resonance is
\begin{equation}
P_\rmH= 1-|\alpha_\rmH|^2=|\beta_\rmH|^2
\end{equation}
for neutrinos and
\begin{equation}
\bar{P}_\rmH= 1-|\bar\alpha_\rmH|^2=|\bar\beta_\rmH|^2
\end{equation}
for antineutrinos. 

For the H resonance the appropriate substitutions into equation
\eref{eq:2nuresonancecondition} are $\thetav\simeq\theta_{13}$ and
$\delta m^2\simeq\pm\delta m_\mathrm{atm}^2$, where the plus and minus
signs are for normal mass hierarchy and inverted mass hierarchy,
respectively. But, as figure \ref{fig:kV} illustrates, we cannot have
a situation where an H resonance exists for both neutrinos \emph{and}
antineutrinos. For this reason we must have $\bar\beta_\rmH=0$ for a
normal hierarchy and $\beta_\rmH=0$ for an inverted hierarchy .

%%%%%%%%%%%%%%%%%%%%%%%%%%%%%%%%%%%%%%%%%%%%%%%%%%%%%%%%%%%%%%%%%%%%%%%%%%%%%%%%%%%

\subsection{Neutrino self-interaction potential\label{sec:self-interaction}}

Assuming that the effects of the neutrino medium on the flavour
evolution of a test neutrino depend on the average flavour content of
background neutrinos
\citep{Friedland:2003dv,Friedland:2003eh,Balantekin:2006tg}, 
Hamiltonian $\hat{H}_{\nu\nu}$ for a neutrino with momentum $\bip$ can
be written in the flavour basis as
\begin{equation}
H^\bsf_{\nu\nu,\bip}(\bir)=
\sqrt{2}\GF\int\!\frac{\rmd^3\bip^\prime}{(2\pi)^3}
     [1-\cos\vartheta_{\bip\bip^\prime}(\bir)]
     [\varrho_{\nu,\bip^\prime}(\bir)
       -\varrho_{\bar\nu,\bip^\prime}^*(\bir)],
\label{eq:Hnunu}
\end{equation}
where $\bip$ and $\bip^\prime$ are the momenta of the test and
background neutrinos, respectively, $\vartheta_{\bip\bip^\prime}$ is
angle between the propagation directions of these neutrinos, and
$\varrho_{\nu(\bar\nu),\bip^\prime}$ are the matrices of density
describing the flavour states of background neutrinos (antineutrinos).
(In \cite{Sigl:1992fn,Strack:2005ux} the density of matrix for an
  antineutrino is actually defined as
  $[\bar\varrho_{\bip}(\bir)]_{\alpha\beta}=[\varrho_{\bar\nu,\bip}]_{\beta\alpha}
  =[\varrho_{\bar\nu,\bip}^*]_{\alpha\beta}$. This definition of the
  density of matrix with a reversed order in subscripts is helpful if
  one uses matrices of densities instead of wavefunctions to describe
  the flavour states of neutrinos. See equation \eref{eq:rho-s} and
  discussions in section \ref{sec:nfis}.)
Note that we use $\vartheta$ and $\theta$ (with or without subscripts
or superscripts) to denote angles that are in coordinate space and
flavour space, respectively.
Similarly, the Hamiltonian for
an antineutrino with momentum $\bip$ is
\begin{equation}
\bar{H}^\bsf_{\nu\nu,\bip}(\bir)=
\sqrt{2}\GF\int\!\frac{\rmd^3\bip^\prime}{(2\pi)^3}
     [1-\cos\vartheta_{\bip\bip^\prime}(\bir)]
     [\varrho_{\bar\nu,\bip^\prime}(\bir)
       -\varrho_{\nu,\bip^\prime}^*(\bir)].
\label{eq:Hanuanu}
\end{equation}

%In equation \eref{eq:Hnunu}
%$\varrho_{\bip^\prime}(\bir)$ and
%$\bar\varrho_{\bip^\prime}(\bir)$ are the ``matrices of densities''
%for neutrinos and antineutrinos, respectively
%\cite{Sigl:1992fn,Strack:2005ux}. 
Assuming that the flavour state of each neutrino can be fully described
by a wavefunction (i.e.\ without any quantum decoherence), the
elements of the matrices of densities for neutrinos and antineutrinos
can be written as 
%\numparts
\begin{eqnarray}
\left[\varrho_{\nu,\bip}(\bir)\right]_{\alpha\beta} & =
\sum_{\nu^\prime_\bip} F_{\nu^\prime_\bip}(\bir)
\langle\nu_\alpha|\psi_{\nu^\prime_\bip}(\bir)\rangle
\langle\psi_{\nu^\prime_\bip}(\bir)|\nu_\beta\rangle,
\\ \left[\varrho_{\bar\nu,\bip}(\bir)\right]_{\alpha\beta} & =
\sum_{\bar\nu^\prime_\bip} F_{\bar\nu^\prime_\bip}(\bir)
\langle\bar\nu_\alpha|\psi_{\bar\nu^\prime_\bip}(\bir)\rangle
\langle\psi_{\bar\nu^\prime_\bip}(\bir)|\bar\nu_\beta\rangle,
\end{eqnarray}
%\endnumparts 
where $\nu^\prime_\bip$ ($\bar\nu^\prime_\bip$)
represents a neutrino (antineutrino) beam with momentum $\bip$,
$F_{\nu^\prime_\bip(\bar\nu^\prime_\bip)}(\bir)$ is its number flux at
$\bir$, and $|\psi_{\nu^\prime_\bip(\bar\nu^\prime_\bip)}\rangle$ is
the flavour state of the neutrino (antineutrino).  The diagonal
elements of a matrix of density give the number densities of neutrinos
in the corresponding flavours.  For example,
\begin{equation}
n_{\nu_e}^\tot(\bir)=
\int\!\frac{\rmd^3\bip^\prime}{(2\pi)^3}[\varrho_{\nu,\bip^\prime}(\bir)]_{ee}
\end{equation}
is the total number density of $\nu_e$ at $\bir$. The off-diagonal
elements of a matrix of density contain information about neutrino
flavour mixing.

Like in pure MSW flavour transformation, the problem of three-flavour
neutrino oscillation with neutrino self-interaction is also believed
to be factorisable into two successive two-flavour mixing scenarios under most
circumstances \citep{Dasgupta:2008cd,Dasgupta:2007ws,Duan:2008za}. In
this review will focus on the two-flavour mixing scenario at
the $\delta m^2_\mathrm{atm}$-scale which is the mostly likely neutrino mixing
scenario to affect, if at all, supernova dynamics and/or nucleosynthesis.

%%%%%%%%%%%%%%%%%%%%%%%%%%%%%%%%%%%%%%%%%%%%%%%%%%%%%%%%%%%%%%%%%%%%%%%%%%%%%%%%%%%
%%%%%%%%%%%%%%%%%%%%%%%%%%%%%%%%%%%%%%%%%%%%%%%%%%%%%%%%%%%%%%%%%%%%%%%%%%%%%%%%%%%
%%%%%%%%%%%%%%%%%%%%%%%%%%%%%%%%%%%%%%%%%%%%%%%%%%%%%%%%%%%%%%%%%%%%%%%%%%%%%%%%%%%

\section{Neutrino flavour transformation in supernovae: the electron dominated regime\label{sec:MSW-regime}}

%% Once the neutrino density is sufficiently small --- beyond $r \sim
%% 1000\;{\rm km}$ --- we enter the electron dominated --- pure MSW --- regime.
%% Now the neutrino flavour transformation is driven solely by the
%% density and its derivative so in order to appreciate how the flavour
%% oscillations will change with time we must simultaneously consider the
%% evolution of the profile through which they pass.

In this section we review the effects of dynamical density profiles on
flavour transformation of supernova neutrinos. In sections
\ref{sec:forward-shock} and \ref{sec:reverse-shock} we will discuss
the effects of the forward and reverse shocks and hot bubbles. In
doing so we will 
utilise density profiles that are derived from 1D supernova
simulations and we will focus on neutrinos that are emitted radially
from the PNS. In section \ref{sec:aspherical-profiles} we will discuss
the features in aspherical profiles and flavour transformation of
neutrinos that propagate along non-radial trajectories.

%%%%%%%%%%%%%%%%%%%%%%%%%%%%%%%%%%%%%%%%%%%%%%%%%%%%%%%%%%%%%%%%%%%%%%%%%%%%%%%%%%%

\subsection{The forward shock \label{sec:forward-shock}}

Until recently any study of supernova neutrino oscillations assumed
the density profile observed by the outrushing neutrinos to be that of
the progenitor star and to be static.  Roughly this profile obeys an
inverse power law i.e.\ $\rho \propto 1/r^{n}$ with common values of
$n$ being $2- 3$
\citep{Brown:1982NuPhA.375..481B,Notzold:1987PhLB..196..315N,Kuo:1988PhRvD..37..298K,Dighe:1999bi}
though upon closer inspection one finds that profiles of progenitors
from simulations show deviations from this form
\citep{Notzold:1987PhLB..196..315N,Takahashi:2003APh....20..189T,Kneller:2008PhRvD..77d5023K}.
Whatever progenitor profile used, we must first establish as a
baseline the effect of the progenitor upon the neutrinos because, a)
for the first second or so of the neutrino signal this is the profile
at the H resonance, and b), it is changes from this baseline that
reveal the evolution of the density profile.

An insightful study of the effects of the progenitor was made by
\citet{Lunardini:2003eh}.  Using equation \eref{eq:PC.KP} with the
appropriate mixing parameters for the H resonance and adopting an
inverse power law profile they distinguished three regimes in the
parameter space of $\theta_{13}$:
\begin{itemize}
\item The adiabatic regime where
\begin{equation} 
\sin^{2} \theta_{13}\gtrsim10^{-4}\left(\frac{E}{10\;{\rm
    MeV}}\right)^{2/3}. \label{eq:adthat13}
\end{equation}
In this regime $\gamma_{\star} \gg 1$ and $P_\rmH\sim 0$ for the 
normal hierarchy or $\bar{P}_\rmH\sim 0$ for the inverted hierarchy.
\item The non-adiabatic regime where
\begin{equation}
\sin^{2} \theta_{13}\lesssim10^{-6}\left(\frac{E}{10\;{\rm
    MeV}}\right)^{2/3}.
\end{equation}
In this regime $\gamma_{\star} \ll 1$ and $P_\rmH\sim 1$ (or
$\bar{P}_\rmH\sim 1$).
\item The transition regime where
\begin{equation}
\sin^{2} \theta_{13}\sim(10^{-6}-10^{-4})\left(\frac{E}{10\;{\rm
    MeV}}\right)^{2/3}.
\end{equation}
In this regime $\gamma_{\star} \sim 1$ and $P_\rmH$ (or $\bar{P}_\rmH$) takes on
intermediate values.
\end{itemize}

The most significant change in the study of the pure-MSW in supernovae
over recent years has been the realisation that the profile seen by
the neutrinos at later times is not the progenitor profile. It is now
apparent that prominent features appear in the profile of the
supernova that are not present in the progenitor and these features
will affect the neutrinos.  The first, generic feature of any
supernova density profile is the forward shock.  Snapshots of density
profiles with solely this feature are shown in figure \ref{fig:shock
  only profile} which is taken from
\citet{Kneller:2008PhRvD..77d5023K}. 
The origin of time $t$ is taken to be the formation of the shock, an event
often called the ``bounce''. 
Observationally the bounce would be
indicated by the neutronisation burst which occurs a millisecond or so
afterwards. 
\citet{Kneller:2008PhRvD..77d5023K} mapped into a
hydrodynamical code a spherically symmetric (1D) density profile
resembling the state of the supernova approximately at the moment when the
forward shock has stalled at $r \sim 200\;\mathrm{km}$.  The material
between a $100\;\mathrm{km}$ gain radius and the stalled shock was
heated in a fashion resembling neutrino energy deposition in order to
revive the outward motion of the shock. By adjusting the energy
deposition rate \citet{Kneller:2008PhRvD..77d5023K} could control the
features of the explosion.  We should mention that density profiles
that \emph{only} contain a forward shock are not typical of iron
core-collapse supernova, nevertheless, such a simple profile is a
useful place to begin to understand the manner in which the explosion
affects the neutrinos.  In the figure the forward shock is the step in
density and we can see it racing out through the supernova and through
the H resonances.  For normal shocks the density jump,
$\Delta\rho_{\rm sh}/\rho_{\rm sh}$, is related to the Mach number, $M$,
of the shock and the ratio of specific heats, $C_P/C_V$,
of the medium via
\begin{equation}
\frac{\Delta \rho_{\rm sh}}{\rho_{\rm sh}} =
\frac{2\,\left(M^{2}-1\right)}{\left(C_P/C_V-1\right)\,M^{2}+2}. \label{eq:deltarhooverrho}
\end{equation} 
For strong shocks the density jump becomes independent of the Mach
number.

\begin{figure}
\begin{indented}
\item[]% 
\includegraphics[width=4in,clip]{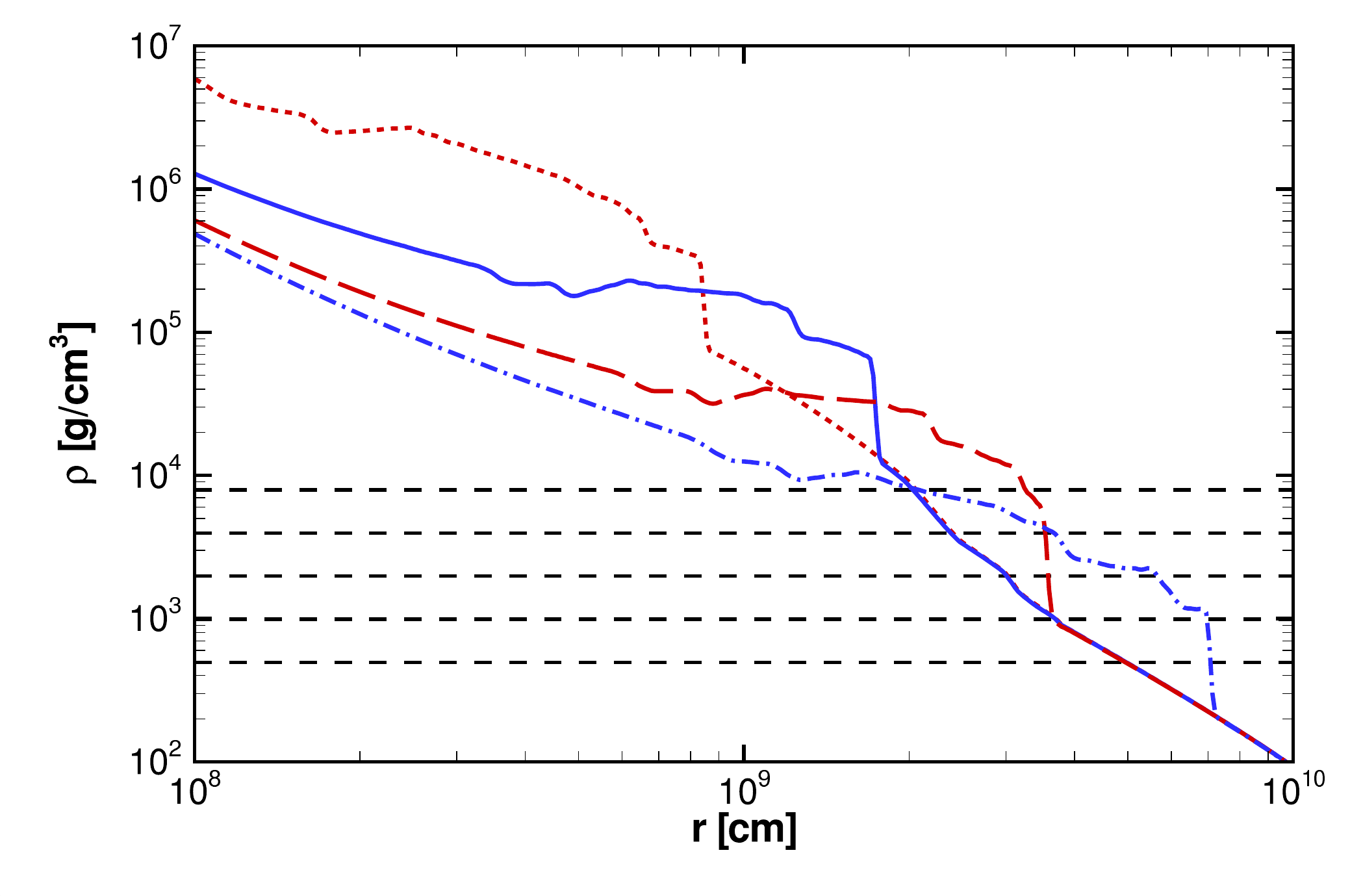}
\end{indented}
\caption{The density profile of a weak explosion in a spherically
  symmetric simulation taken from
  \citet{Kneller:2008PhRvD..77d5023K}. Displayed are the radial
  profile at a series of snapshot times which are: $t=0.9\;{\rm s}$
  (dotted), $1.8\;{\rm s}$ (solid), $3.6\;{\rm s}$ (dashed) and
  $7.2\;{\rm s}$ (dash dot). The horizontal dashed lines are the H
  resonance densities of, from top to bottom, neutrinos with energies
  of $5,\;10,\;20,\;40$ and $80\;{\rm MeV}$.
\label{fig:shock only profile}}
\end{figure}

As the shock reaches the H resonance for any given neutrino energy the
density derivative $\rmd\rho/\rmd r$ becomes abruptly steeper.  This will
change the adiabaticity of the resonance according to equation
\eref{eq:gammastar}.  If the mixing parameters are such that neutrino
propagation through the progenitor is adiabatic, i.e.\ $\theta_{13}$
lies in the adiabatic regime so that $P_\rmH\sim 0$, then the arrival
of the shock at any given resonance means that neutrinos of the same
energy now propagate through the profile non-adiabatically,
i.e.\ $P_\rmH\sim 1$.  It is the possibility, first considered by
\citet{Schirato:2002tg}, that $P_\rmH$ can evolve with time that
allows us to peer inside the supernova as it explodes.  Of course if
$\theta_{13}$ is very small so that it lies in the non-adiabatic
regime then no change will be observed as the shock passes the H
resonance region because the H resonance is already non-adiabatic for
the undisturbed/progenitor profile, i.e.\ $\gamma_{\star} \ll 1,\;
P_\rmH\sim 1$.  When the shock arrives all that happens is that
$\gamma_{\star}$ will become even smaller and but this has no effect upon
$P_\rmH$.  Such a situation would be disappointing but at least we
would be able to set an upper limit on $\theta_{13}$ that is
competitive with any possible terrestrial experiment.  So for the rest
of this discussion of the dynamic MSW effect we shall consider only
the case where $\theta_{13}$ lies in the adiabatic regime,
i.e.\ $\theta_{13}$ is such that equation \eref{eq:adthat13} is
satisfied, and more specifically adopt a value
$\sin^{2}2\theta_{13}=4\times 10^{-4}$.

With our understanding of how $P_\rmH$ can change as the star explodes
we can begin to map out our expectations for the evolution of $P_\rmH$
as a function of time and energy. Let us consider the particular
example shown in figure \ref{fig:shock only profile}.  For the first
couple of seconds the profile at the H resonances of neutrinos with
energy $5-80\;{\rm MeV}$ is still the undisturbed progenitor and only
after this initial delay does the shock appear in the H resonance
region.  Then at $t=2.0\;{\rm s}$ the shock will change the
adiabaticity of the $5\;{\rm MeV}$ neutrinos but will not begin to affect
the $40\;{\rm MeV}$ neutrinos until later at $3.6\;{\rm s}$ because it takes
longer for the shock to reach their H resonance density.  Also around
$3.5\;{\rm s}$ the shock ceases to intersect the $5\;{\rm MeV}$
resonance density so for this energy the adiabaticity will return to
$P_\rmH\sim 0$.  Only later at $t=7.2\;{\rm s}$ does the shock cease
affecting the $40\;{\rm MeV}$ neutrinos.

\begin{figure}
\begin{indented}
\item[]% 
\includegraphics[width=4in,clip]{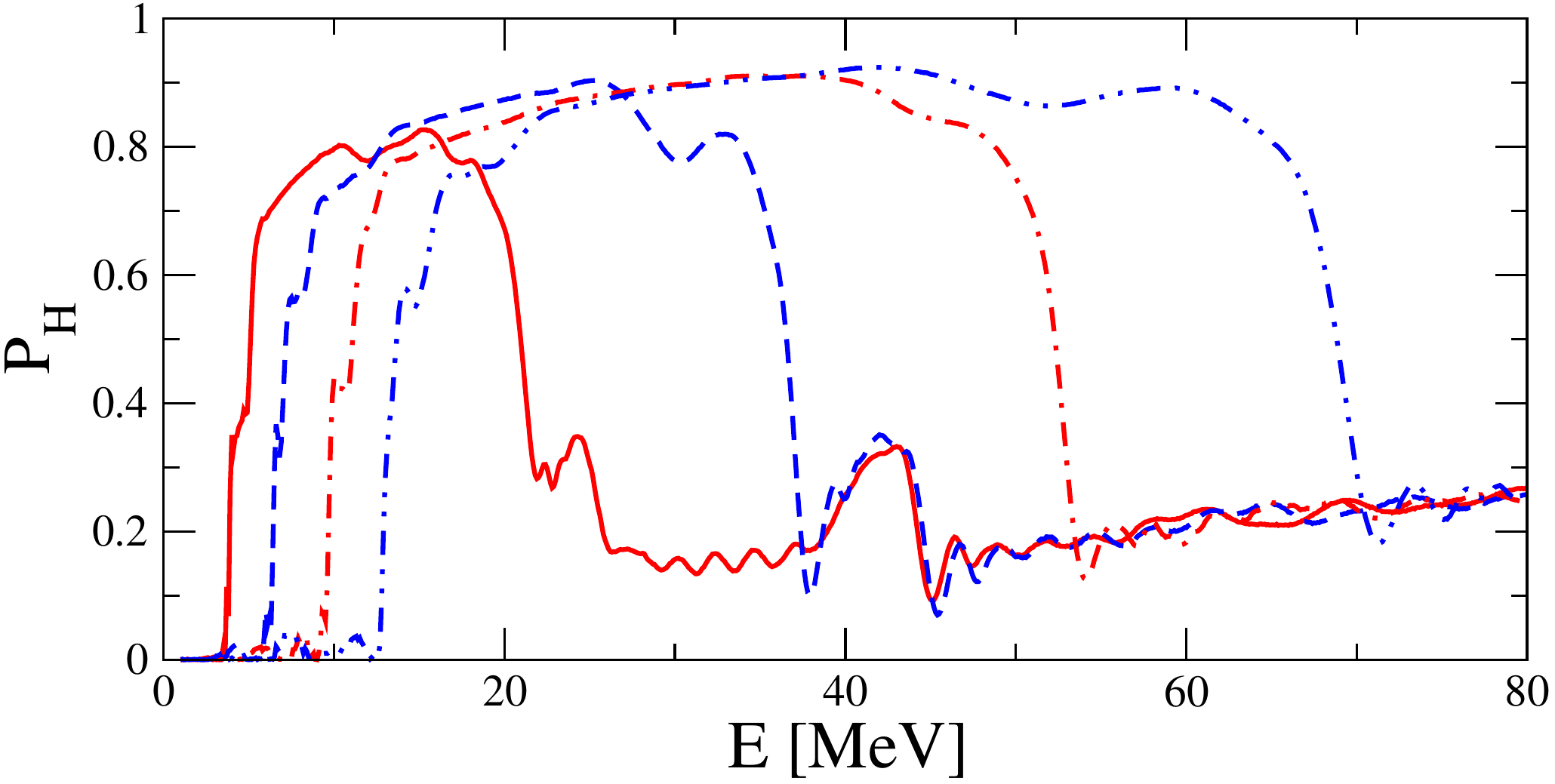}
\end{indented}
\caption{The crossing probability as a function of neutrino energy for
  a supernova simulation containing only a forward shock. Density
  profiles from this simulation are shown in figure \ref{fig:shock
    only profile} and the snapshot times are: $t=3.0\;{\rm s}$
  (solid), $3.5\;{\rm s}$ (dashed), $4.0\;{\rm s}$ (dash-dot) and
  $4.5\;{\rm s}$ (dash-double dot). The figure is taken from 
  \citet{Kneller:2008PhRvD..77d5023K}.
\label{fig:PH versus E.shock only}}
\end{figure}
\begin{figure}
\begin{indented}
\item[]% 
\includegraphics[width=4in,clip]{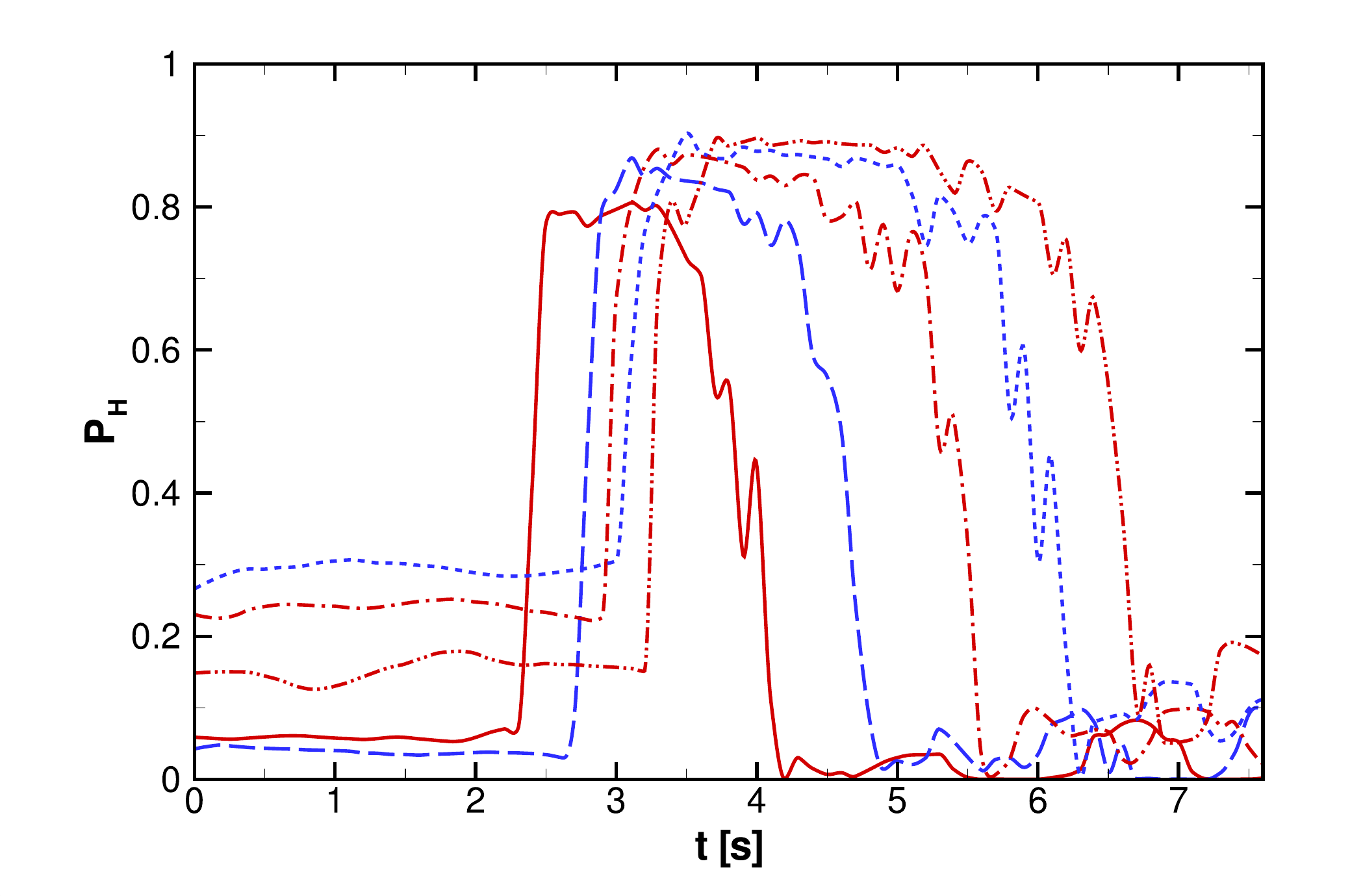}
\end{indented}
\caption{The crossing probability as a function of time for selected
  neutrino energies as they pass through the density profiles taken
  from the weak explosion shown in figure \ref{fig:shock only
    profile}. The energies are: $E=10\;{\rm MeV}$ (solid), $15\;{\rm
    MeV}$ (dashed), $20\;{\rm MeV}$ (dash dot), $25\;{\rm MeV}$
  (dotted) and $30\;{\rm MeV}$ (dash double dot). The figure is
  taken from \citet{Kneller:2008PhRvD..77d5023K}
\label{fig:PH versus t.shock only}}
\end{figure}

In figures \ref{fig:PH versus E.shock only} and \ref{fig:PH versus
  t.shock only} we show the actual numerical results for the crossing
probability $P_\rmH$ as a function of energy at various snapshot
times, and as a function of time for fixed neutrino energies
respectively as neutrinos pass through density profiles taken from the
weak supernova shown in figure \ref{fig:shock only profile}.  We
clearly see the arrival of the shock at early times as the sudden jump
in $P_\rmH$ at low energies which then sweeps up through the spectrum.
Note how the width of the shock feature increases with time.  We also
see that higher energy neutrinos are affected after the lower energies
and that the effect is of longer duration as expected.  
From figures \ref{fig:PH versus E.shock only} and \ref{fig:PH versus
  t.shock only} we can identify three basic quantities that describe
the evolution of $P_\rmH$:
\begin{itemize}
\item $E_{\rm sh}(t)$, the highest energy affected by the shock at any
  given moment $t$,
\item $\Delta E_{\rm sh}(t)$, the width of the shock feature in the
  spectrum at any given $t$, and
\item $\Delta t_{\rm sh}(E)$, the time period over which the shock
  affects neutrinos of energy $E$.
\end{itemize}
From figures \ref{fig:PH versus E.shock only} and
\ref{fig:PH versus t.shock only} we see that both $E_{\rm sh}(t)$ and
$\Delta E_{\rm sh}(t)$ increase with $t$ and that $\Delta t_{\rm sh}(E)$
increases with $E$.  This behaviour is easily understood.

The highest energy affected by the shock, $E_{\rm sh}(t)$, at time $t$
corresponds to the moment when the bottom of the shock reaches the
resonance density for that particular energy as shown in figure
\ref{fig:shock only profile}. $E_{\rm sh}(t)$ is a monotically increasing
function of $t$ which is a generic prediction and occurs simply because the
resonance density is inversely proportional to the neutrino energy,
the density profile is a monotonically decreasing function of distance
and because the shock wave is generated at the core of the star where
the density is highest. 

The width of the shock feature, $\Delta E_{\rm sh}(t)$, also
monotonically increase with $t$.  This can be understood as follows.
The resonance condition \eref{eq:2nuresonancecondition} allows us to relate
$E_{\rm sh}(t)$ to the value of the neutrino potential at the bottom of
the shock, $V_{\rm sh}(t)$, by
\begin{equation}
V_{\rm sh}=\frac{\delta m^{2}\,\cos2\thetav}%
{2\,E_{\rm sh}}.\label{eq:Vs}
\end{equation}
The lowest energy affected by the shock, $E_{\rm sh}-\Delta E_{\rm sh}$ are those neutrinos whose
resonance density is the top of the shock, i.e.\ $V_{\rm sh}+\Delta
V_{\rm sh}$ where $\Delta V_{\rm sh}$ is the jump in the neutrino
potential across the shock. Using equation \eref{eq:Vs} and
exploiting the fact that 
$\Delta V_{\rm sh}/V_{\rm sh}= \Delta \rho_{\rm sh}/\rho_{\rm sh}$ --- where $\rho_{\rm sh}(t)$ is the density at the
bottom of the shock and $\Delta \rho_{\rm sh}(t)$ the density jump --- we
can derive the simple relationship that
\begin{equation}
\frac{\Delta E_{\rm sh}}{E_{\rm sh}-\Delta E_{\rm sh}}=\frac{\Delta
  \rho_{\rm sh}}{\rho_{\rm sh}}.
\end{equation}
Note that this is independent of the mixing parameters. If the ratio
of specific heats of the material does not change rapidly with
distance then $\Delta \rho_{\rm sh}/\rho_{\rm sh}$ is constant for a
strong shock --- equation \eref{eq:deltarhooverrho} --- so the ratio
$\Delta E_{\rm sh} / (E_{\rm sh}-\Delta E_{\rm sh})$ is also
invariant. This relationship shows us that $\Delta E_{\rm sh} \propto
E_{\rm sh}$. Thus, if $E_{\rm sh}$ monotonically increases with $t$ then so
must $\Delta E_{\rm sh}$.

Finally, that $\Delta t_{\rm sh}(E)$ increases with $E$ is also just a
consequence of $E_{\rm sh}$ increasing with $t$.  The density
$\rho_{\rm sh}(t)$ is the point where the shock attaches to the
progenitor profile so if we assume a form for the density profile of
the progenitor then we can relate $\rho_{\rm sh}(t)$ to the position of
the shock, $r_{\rm sh}(t)$.  For illustrative purposes we shall adopt
for the profile the power law $\rho=C_{\star}\,(r_{\star}/r)^{n}$ with
$C_{\star}$ the constant of proportionality and $r_{\star}$ some scale
but it is possible to do better if the progenitor star can be
identified. 
Using the adopted profile we find that $E_{\rm sh}$ is related to the
shock radius by
\begin{equation}
r_{\rm sh}=r_{\star}\left(\frac{\sqrt{8}\,\GF\,Y_e\,C_{\star}\,
E_{\rm sh}}{m_{N}\,\delta
  m^{2}\,\cos2\thetav}\right)^{1/n}. \label{eq:rs(Es)}
\end{equation}
At a time $t'$ the shock will have moved outwards through the star and
will be located at a position $r_{\rm sh}'=r_{\rm sh}(t')$.  As it does
so the shock feature will have swept up through the neutrino spectrum
with $E_{\rm sh}'=E_{\rm sh}(t')$ now the highest affected energy and
the spectral width will now be $\Delta E_{\rm sh}'=\Delta E_{\rm
  sh}(t')$.  From equation \eref{eq:rs(Es)} we see that $r_{\rm sh}/r_{\rm sh}' = (Y_e\,E_{\rm sh}/ Y_e'\,E_{\rm sh}')^{1/n}$ where
$Y_e'$ is the electron fraction at $r_{\rm sh}'$.  If we select $t'$
to be the moment when neutrinos of energy $E_{\rm sh}$ cease to be
affected by the shock then we must have $E_{\rm sh} = E_{\rm sh}' -
\Delta E_{\rm sh}'$.  The time difference $t'-t= \Delta t(E_{\rm sh})$. If $\bar{v}_{\rm sh}$ is the average shock velocity over this
time period then we can derive that
\begin{equation}
\Delta t(E_{\rm sh}) = \frac{r_{\rm sh}' - r_{\rm sh}}{\bar{v}_{\rm sh}} =
\frac{r_{\rm sh}'}{\bar{v}_{\rm sh}}\,\left[\left(\frac{E_{\rm sh}'}{E_{\rm sh}'-\Delta E_{\rm sh}'}\right)^{1/n}\,\left(\frac{Y_e'}{Y_e}\right)^{1/n}
  -1\right]. \label{eq:Delta t}
\end{equation}
The term in square brackets is a constant for a constant $\Delta
\rho_{\rm sh}/\rho_{\rm sh}$ and a medium with constant $Y_e$ thus
$\Delta t(E_{\rm sh})$ is proportional to $r_{\rm sh}'$. With this discovery 
we immediately see that if $r_{\rm sh}$ increases with $E_{\rm sh}$, so must $\Delta t(E_{\rm sh})$
increase with $E_{\rm sh}$. 

While equation \eref{eq:Delta t} shows us that 
$\Delta t(E_{\rm sh})$ must increase with $E_{\rm sh}$ perhaps what is more interesting
is that we can turn the equation around and use the observations of 
$E_{\rm sh}(t)$, $\Delta E_{\rm sh}(t)$, and $\Delta t_{\rm sh}(E)$
to determine the average velocity of the shock as it travelled from
$r_{\rm sh}$ to $r_{\rm sh}'$; that is
\begin{equation}
\bar{v}_{\rm sh}= \frac{r_{\rm sh}(E_{\rm sh})}{\Delta t_{\rm sh}(E_{\rm sh})}\left[\left(\frac{E_{\rm sh}'}{E_{\rm sh}'-\Delta E_{\rm sh}'}\right)^{1/n}\,\left(\frac{Y_e'}{Y_e}\right)^{1/n}
  -1\right]\label{eq:vbar}.
\end{equation} 
As an alternative, we could simply exploit equation \eref{eq:rs(Es)} to relate the
observed $E_{\rm sh}(t)$ to construct $r_{\rm sh}(t)$ which can then be
fit with a parametric from.  The shock velocity would then be found as
the derivative of $r_{\rm sh}(t)$ but there is additional information
in the fit.  The shock is formed at the PNS and current
models of supernova indicate it stalls at $r\sim 200\;{\rm km}$.
Somehow the outward motion of the shock is revived and thereafter it
races through the mantel of the supernova.  The time spent by the
shock in the stalled position is not known but suggestions typically
lie in the range of $\sim 0.5\;{\rm s}$.  The velocity of the shock
through the star is also not known but a reasonable initial guess is
that the shock velocity is approximately constant because the profile is close to a
$1/r^{3}$ form \citep{1959sdmm.book.....S}.  Thus our expectation
might be that the position of the shock after its revival is a simple
linear model given by
\begin{equation}
r_{\rm sh}(t) = r_{\rm stall}+v_{\rm sh}\,(t-t_{\rm
  stall}) \label{eq:rs(t)}
\end{equation} 
where $r_{\rm stall}$ is the radius at which the shock stalls and
$t_{\rm stall}$ is the time at which the shock was revived.  The shock
velocity $v_{\rm sh}$ is of course the gradient of equation \eref{eq:rs(t)} and the 
intercept is equal to $r_{\rm stall}-v_{\rm sh}\,t_{\rm stall}$.
So if we set $r_{\rm stall}$ to zero then we derive that 
$t_{\rm stall}=r_{\rm sh}/v_{\rm sh}$ and obtain the earliest time that the
shock can have been revived.  Thus, if the neutrino signal from the
next Galactic supernova is of sufficient quality we might be able to
determine a lower bound for $t_{\rm stall}$ and in doing so we would
be able to confirm or refute a fundamental component of the
core-collapse supernova paradigm: that the shock stalls and is
revived. 

%%%%%%%%%%%%%%%%%%%%%%%%%%%%%%%%%%%%%%%%%%%%%%%%%%%%%%%%%%%%%%%%%%%%%%%%%%%%%%%%%%%

\subsection{Hot bubbles and reverse shocks \label{sec:reverse-shock}}

\begin{figure}
\begin{indented}
\item[]% 
\includegraphics[width=4in,clip]{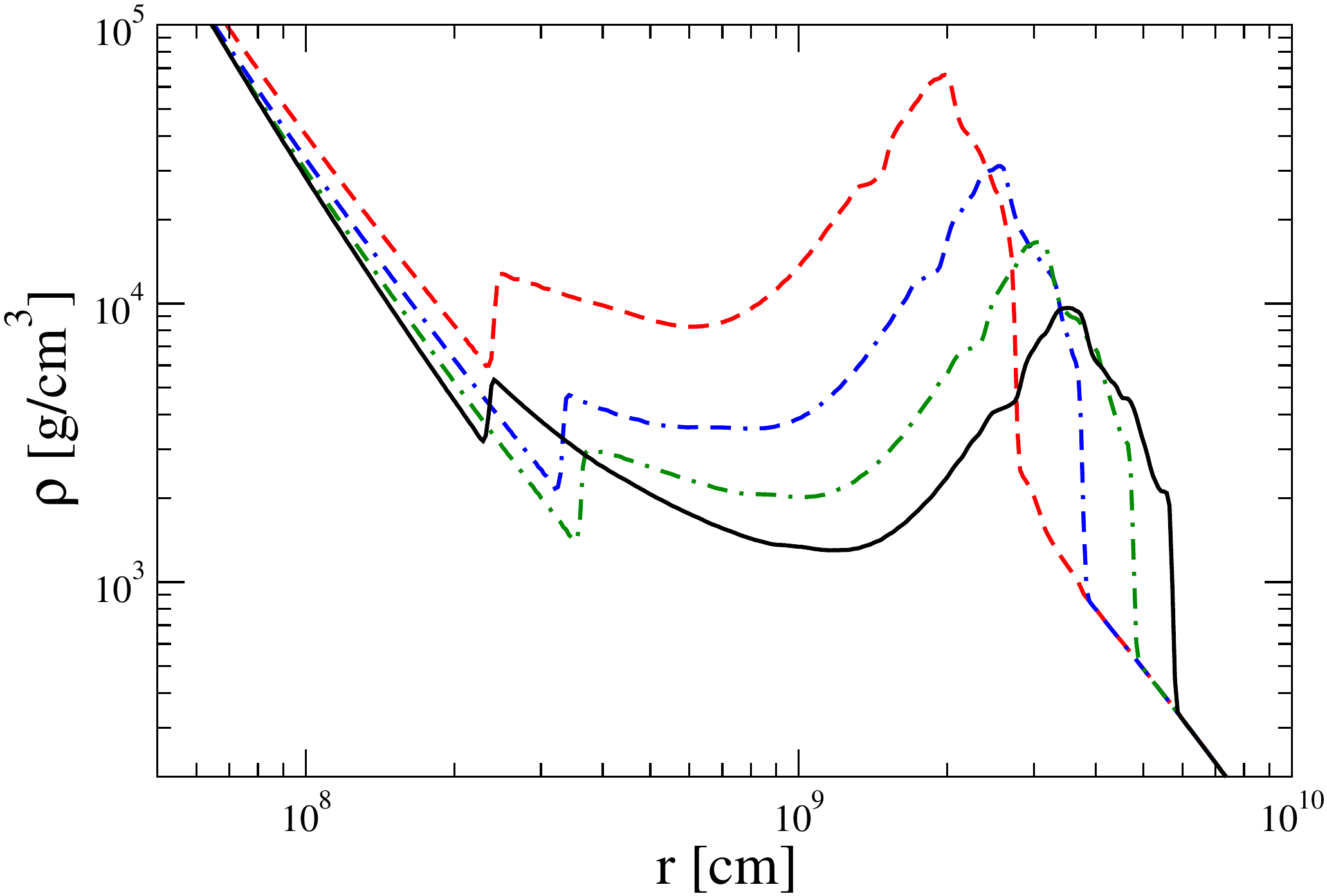}
\end{indented}
\caption{Density profiles containing a reverse shock. This figure is
  adapted from a figure in \citet{Kneller:2008PhRvD..77d5023K}. The
  snapshot times are: $t=1.5\;{\rm s}$ (dashed), $2.0\;{\rm s}$
  (dot double-dashed), $2.5\;{\rm s}$ (dot dashed) and $3.0\;{\rm s}$ (solid).
\label{fig:reverse shock profiles}}
\end{figure} 

The density profiles of the supernova simulation used in the previous
section were from a very weak supernova simulation and contained only
a forward shock.  The shock was revived by heating the material above
the PNS but the energy deposition does not not switch
off once the forward shock has been revived but rather will continue
for some time after.  The continued heating creates a wind that pushes
outwards to create a hot bubble behind the forward shock.  It was
actually profiles of this type that were studied by 
\citet{Schirato:2002tg}.
If we up the rate of energy deposition then strength of the wind can
grow to the point where its velocity becomes greater than the sound
speed.  This situation leads to the formation of a second shock which
faces the PNS rather than the exterior
\citep{Janka:1995ApJ...448L.109J,Burrows:1995ApJ...450..830B}, i.e.\ it
is reversed compared to the forward shock.  Profiles with reverse
shocks are shown in figure \ref{fig:reverse shock profiles}.  The
behaviour of the reverse shock has been studied by
\citet{Tomas:2004JCAP...09..015T}, \citet{Arcones:2007A&A...467.1227A}
and \citet{Kneller:2008PhRvD..77d5023K} in 1D and 2D using hydro codes
of varying degrees of sophistication.  It has been found that its
behaviour is much richer than the forward shock due to its greater
sensitivity to the mechanism by which it was formed, i.e.\ the energy
deposition.  If the heating is sustained for a long period then the
reverse shock trails the forwards shock through the star, but if the
energy deposition is over a briefer period then the reverse shock has
been seen to diminish in size, stall and then to turn around and head
back to the core.  This can be seen in figure \ref{fig:reverse shock profiles}. 
Also we see from the figure how the densities either side of the reverse shock
will change in relation to the forward shock as a function of time.

The hot bubble and the reverse shock also affect the neutrinos.  If a
profile contains either feature then we see that, at any given moment,
there are some neutrino energies that will experience three (or more)
resonances.  In order to compute the effects of multiple resonance one
must be careful less certain effects are ignored. 
\begin{figure}
\begin{indented}
\item[]% 
\includegraphics[width=4in,clip]{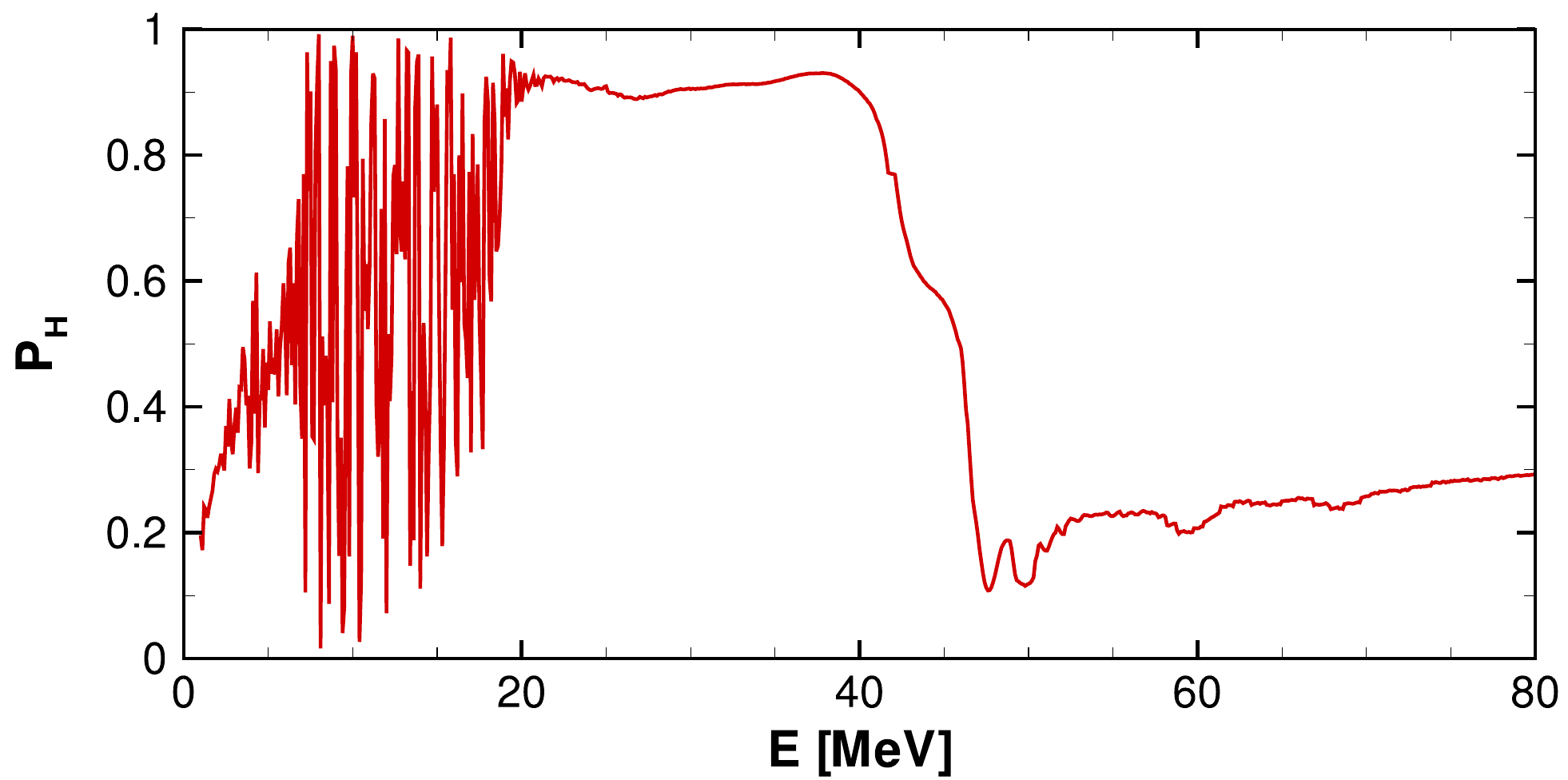}
\end{indented}
\caption{The crossing probability as a function of neutrino energy for
  a supernova simulation containing a reverse shock.  Density profiles
  for this simulation are shown in figure \ref{fig:reverse shock profiles}.
  The snapshot time is $t=2.0\;{\rm s}$ and the energy resolution is
  $100\;{\rm keV}$.  The figure is adapted from
  \citet{Kneller:2008PhRvD..77d5023K} and corresponds to the
  1D model with a total energy deposition $Q=3.36 \times
  10^{51}\;\mathrm{erg}$.
\label{fig:earlyphaseeffects1}}
\end{figure} 
\begin{figure}
\begin{indented}
\item[]% 
\includegraphics[width=4in,clip]{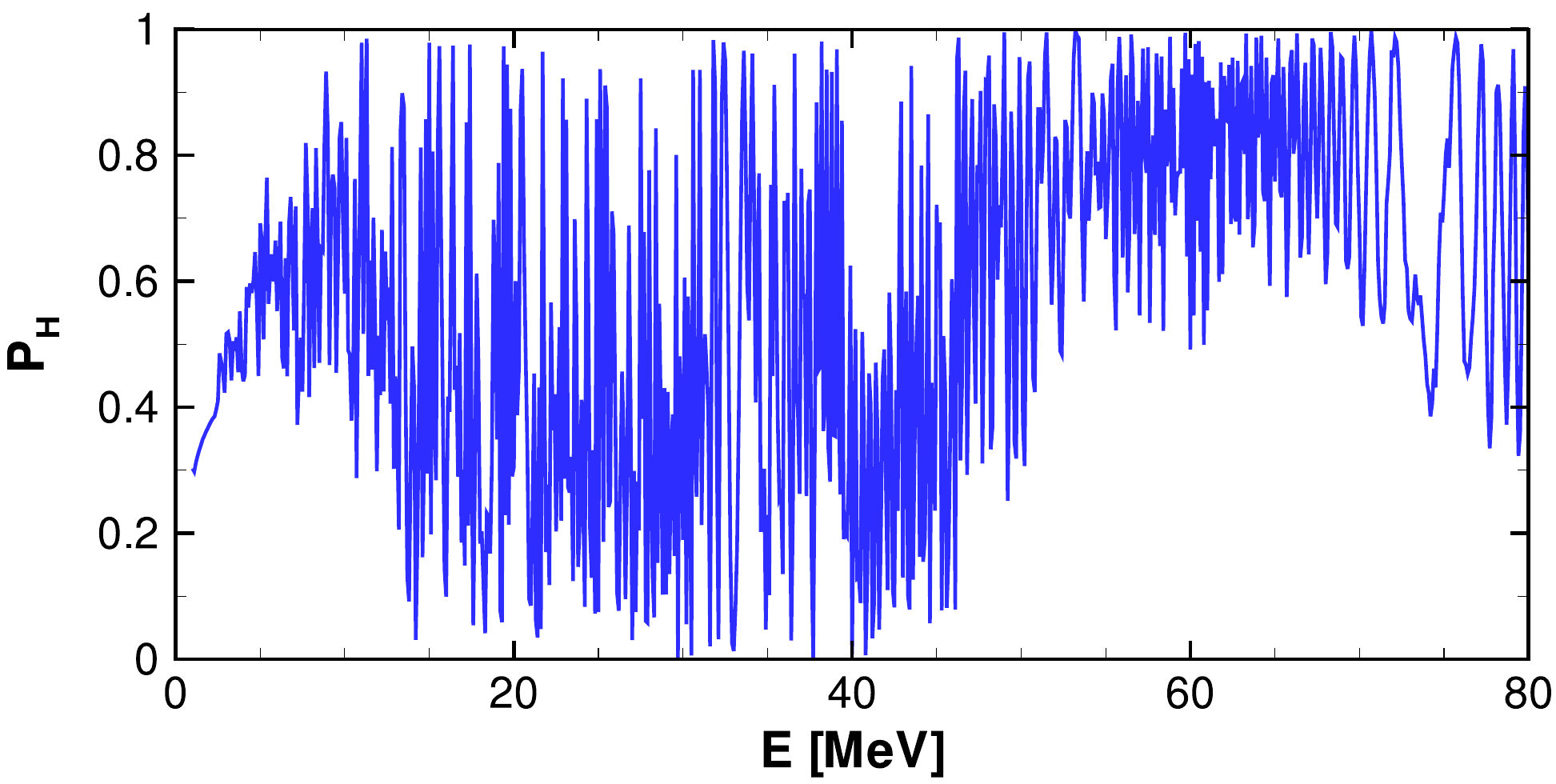}
\end{indented}
\caption{The crossing probability as a function of neutrino energy for
  a supernova simulation containing a reverse shock.  Density profiles
  from this simulation are shown in \citet{Kneller:2008PhRvD..77d5023K} and
  correspond to the model where the total energy deposition is $Q=4.51
  \times 10^{51}\;\mathrm{erg}$.  The snapshot time is $t=2.0\;{\rm
    s}$ and the energy resolution is $100\;{\rm keV}$.
\label{fig:earlyphaseeffects2}}
\end{figure} 
Examples of calculations of $P_\rmH$ as a function of neutrino energy 
through supernova profiles containing reverse shocks are shown 
in figures \ref{fig:earlyphaseeffects1} and \ref{fig:earlyphaseeffects2}.
While some parts of figure \ref{fig:earlyphaseeffects1} in particular resemble 
figure \ref{fig:PH versus E.shock only} we see also that something has noticeably changed. 
The crossing probability is seen to oscillate rapidly for some ranges of 
neutrino energies and the reason is due to ``phase effects''.
That phase effects are present in $P_\rmH$ for supernova profiles was
first noticed by \citet{Fogli:2003PhRvD..68c3005F}.  They were found
again by \citet{Kneller:2005hf} who also presented some basic
arguments for why the phase effects appear.  The phase effects and
their detectability were then the focus of \citet{Dasgupta:2007PhRvD..75i3002D}.

While, perhaps, initially surprising the origin of the phase effects is quite 
simple. The neutrino wavefunction after passing through multiple H resonances is
$\psi(r_\star) = S_\rmH(r_\star,R_\nu)\psi(R_\nu)$ where $r_\star$ is
some point between the H resonances and the L resonance, and $R_\nu$
is the radius of the neutrino sphere. We can
factorise $S_\rmH$ by dividing up the profile into regions
within which there is just one H resonance, i.e.\ just one location in
each region where equation \eref{eq:2nuresonancecondition} is
satisfied. Using the product rule, equation \eref{eq:product_rule},
$S_\rmH$ is the product of evolution operators over the sub-domains:
$S_\rmH=...\,S_{\rmH 2}\,S_{\rmH 1}$ where
$S_{\rmH i}$ is the $S$ matrix for passage through the
$i$'th H resonance encountered by the neutrino. 
All these $S_{\rmH i}$ matrices, up to a phase which is
irrelevant for neutrino oscillations, have the same structure in the two flavour
approximation,
\begin{equation}
S_{\rmH i}= \left[ \begin{array}{cc} 
\alpha_{\rmH i} & \beta_{\rmH i} \\
-\beta_{\rmH i}^* & \alpha_{\rmH i}^* 
\end{array} \right]
\end{equation}
where $\alpha_{\rmH i}$ and $\beta_{\rmH i}$ are, again, Cayley-Klein
parameters. The crossing probability for the $i$'th resonance is
$P_{\rmH i}=|\beta_{\rmH i}|^{2} = 1-|\alpha_{\rmH i}|^{2}$. Even for the case of three 
H resonances we obtain a long and complicated expression for the net effect 
of the multiple resonances so to keep things simple we shall consider the example of just two
resonances. In this case the net H resonance crossing probability, $P_\rmH$, is just 
%%
%\numparts
\begin{eqnarray}
\fl P_\rmH&=P_{\rmH 2}\left(1-P_{\rmH 1}\right)+\left(1-P_{\rmH
  2}\right)\,P_{\rmH 1}+
2\Re\left(\alpha_{\rmH 1}\alpha_{\rmH 2}
\beta_{\rmH 1}\beta_{\rmH 2}^*\right) \\ \fl
&=P_{2}\left(1-P_{1}\right)+\left(1-P_{2}\right)P_{1}
+2\sqrt{P_{1}P_{2}\left(1-P_{1}\right)\left(1-P_{2}\right)}
\cos\phi_{\rmH}
\label{eq:PH phase effects}
\end{eqnarray} 
%\endnumparts
%%
where $\phi_{\rmH}$ is a phase formed from the phases of the
$\alpha$'s and $\beta$'s. The first two terms are
what we would expect if there were no correlations; the last term
represents interference and depends upon the relative phases of the
$\alpha$'s and $\beta$'s, hence the name ``phase
effects''.  Note the amplitude of the oscillatory term is always
smaller than the constant and maximal when $P_{\rmH 1}=P_{\rmH 2}=1/2$. 
From this result we can begin to understand the results of
\citet{Tomas:2004JCAP...09..015T} for profiles containing both reverse
and forward shocks. The net effect upon the neutrinos of a given energy 
depends upon the relationship between the shocks and the resonance density. 
If the two shocks intersect completely different resonance densities then 
we have a situation where, at most, either $|\beta_{1}|=1$ or
$|\beta_{2}|=1$ but not both. But if we have a case where the two shocks affect 
the same resonance densities then we can have a situation where $|\beta_{1}|=|\beta_{2}|=1$ and the two
non-adiabatic transitions cancel. As we see in figure \ref{fig:reverse shock profiles}, 
the two shocks do not always overlap completely and so, due to the relative proportions of
the forward and reverse shocks, the cancellation may be over just a limited range 
of neutrino energies while energies slightly larger or smaller may experience just one 
shock. This can create what \citet{Tomas:2004JCAP...09..015T} referred to as the ``double dip'' in
the crossing probability $P_\rmH$.

But in the more general case with many H resonances some of which are neither 
exactly adiabatic nor non-adiabatic we must consider the effect of the oscillatory terms such 
as the one in equation \eref{eq:PH phase effects}. 
\citet{Dasgupta:2007PhRvD..75i3002D} show that the phase $\phi_\rmH$ appearing in equation \eref{eq:PH phase effects} is approximately
\begin{equation}
\fl \phi_\rmH(E) \approx  
\int_{r_{\rmH 1}(E)}^{r_{\rmH 2}(E)}\,\sqrt{\left(\frac{\delta
    m^{2}\,\cos 2\thetav}{2\,E}-2V_e(r)\right)^{2} +
  \left(\frac{\delta m^{2}\,\sin 2\thetav}{2\,E}\right)^{2}
}\rmd r. \label{eq:Phi12}
\end{equation}
where the $r_{\rmH i}(E)$ are the (energy dependent) positions of the
resonances i.e.\ those locations that satisfy equation
\eref{eq:2nuresonancecondition}.  Equation \eref{eq:Phi12} is just the
integral of the difference between the eigenvalues over the distance
between the two resonances.  In order to compute it properly one needs
the density profile $V_e(r)$ but 
for our purposes we approximate the result by neglecting the first term under the square
root since we are in the vicinity of the resonances where this term
vanishes.
Thus $\phi_\rmH$ is approximately
\begin{equation}
\phi_\rmH \approx \frac{|\delta m^{2}|\,\sin 2\thetav\,L_\rmH}{2\,E}
\end{equation}
where $L_\rmH = r_{\rmH 2} - r_{\rmH 1}$ is the distance between the
resonances.  Typically this distance is much greater than the
wavelength of the oscillations ($\sim E / \delta m^{2}\sin
2\thetav$), so that $\phi_\rmH$ is a large number ($\phi_\rmH \gg 1$).
Now let us consider a change in $E$ and/or $L_\rmH$.  The resulting
phase change $\delta \phi_\rmH$ is
\begin{equation}
\delta \phi_\rmH = \phi_\rmH\,\left[ \frac{\delta L_\rmH}{L_\rmH} -
  \frac{\delta E}{E}\right].
\end{equation}
We require just a change $\delta \phi_\rmH = 2\pi$ in order for the
oscillatory term in equation \eref{eq:PH phase effects} to cycle
through one period.  If $\phi_\rmH$ is large, say $\phi_\rmH \sim
10^{3}$, then a fractional change in the energy of just $\delta E/ E
\sim 10^{-3}$ is enough to give us $\delta \phi_\rmH \sim 2\pi$. For neutrinos with 
energy $E \sim 10\;{\rm MeV}$ we see that $P_\rmH$ will cycle over one period
with a change of energy of just $\delta E \sim 10\;{\rm keV}$. 
%%We can even get some idea of how fast $P_\rmH$ at a given energy will vary:
%%\begin{equation}
%%\frac{\rmd \phi_\rmH}{\rmd t} \approx \frac{\phi_\rmH}{L_\rmH}\,\frac{\rmd L_\rmH}{\rmd t} = \frac{|\delta m^{2}|\,\sin 2\theta}{2\,E}\,\frac{\rmd L_\rmH}{\rmd t}.
%%\end{equation}
%%The distance $L_\rmH$ will vary as something like the shock speed $v_{\rm sh}$. 
%%Thus $P_\rmH$ will cycle over one period in the amount of time it takes for the distance between the resonances to change by one oscillation length. 
%%For $v_{\rm sh} \sim 10^{4}\;{\rm km/s}$, $\theta_{13} \sim 10^{-2}$, $E = 10\;{\rm MeV}$ and $\delta m_{2} \sim 10^{-3}\;{\rm eV^2}$ this time may be as small as $1/100\;{\rm s}$.

From our understanding of equation \eref{eq:PH phase effects} we can
begin to digest the results for $P_\rmH$ as a function of neutrino
energy shown in figure \ref{fig:earlyphaseeffects1} and
\ref{fig:earlyphaseeffects2}. In the first figure we again observe
the presence of the forward shock in the spectrum because the crossing
probability makes the transition from adiabatic to non-adiabatic
propagation at $E\sim 50\;\mathrm{MeV}$ but in this same figure we
also see that below $E \sim 20\;{\rm MeV}$ the crossing probability
starts to oscillate as a function of $E$: this occurs because those
energies now passes through multiple H resonances and so phase effects
appear. In figure \ref{fig:earlyphaseeffects2} phase effects dominate
the entire spectrum.  In both cases the oscillations are very rapid which is simply
an indication that the resonances are spaced by a large number of oscillation wavelengths.

%%%%%%%%%%%%%%%%%%%%%%%%%%%%%%%%%%%%%%%%%%%%%%%%%%%%%%%%%%%%%%%%%%%%%%%%%%%%%%%%%%%

\subsection{Aspherical and turbulent profiles \label{sec:aspherical-profiles}}

\begin{figure}
\begin{indented}
\item[]% 
\includegraphics[width=4in,clip]{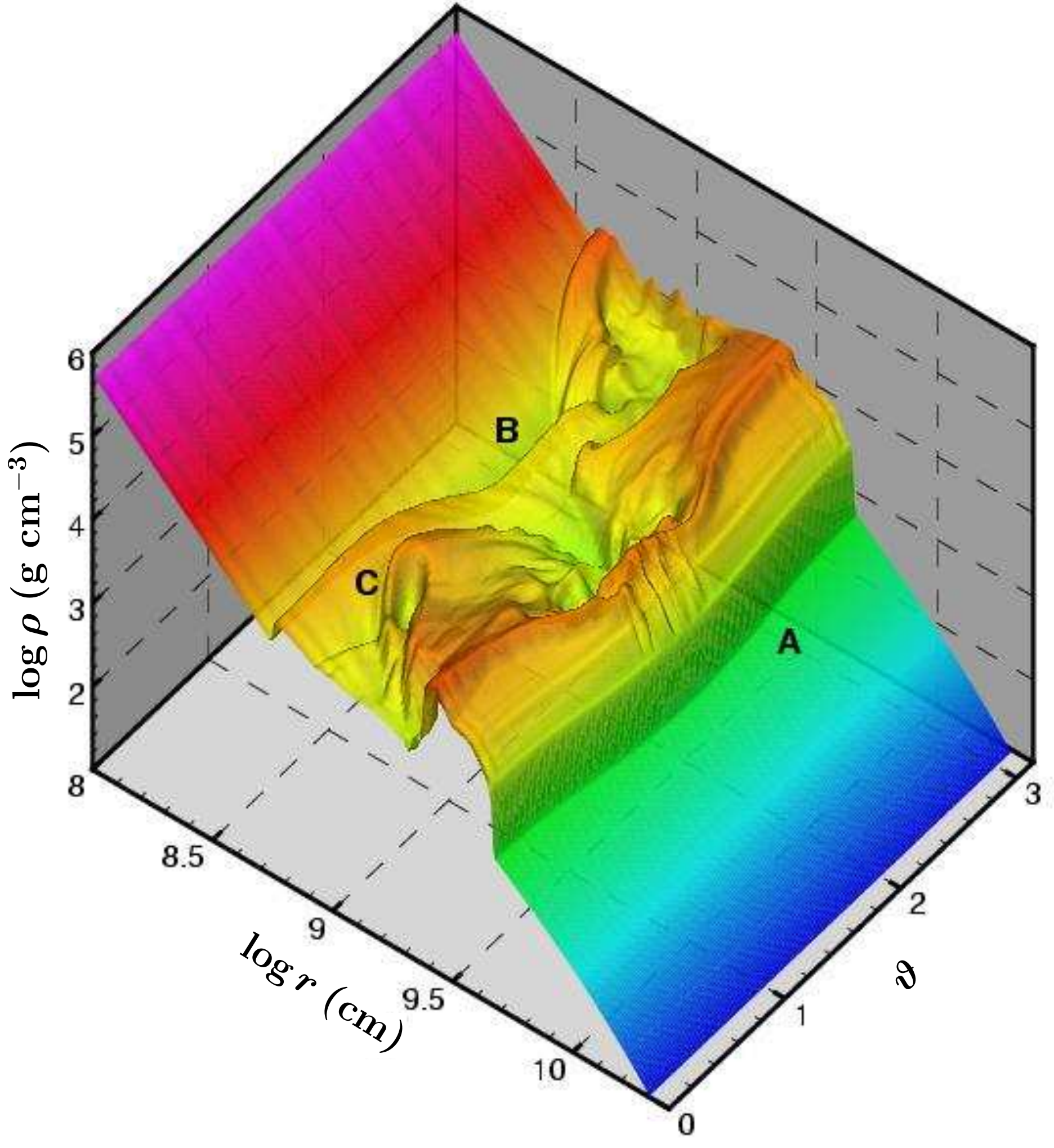}
\end{indented}
\caption{The density as a function of the radius and angle in a 2D
  supernova model at $t=2.5\;{\rm s}$ taken from
  \citet{Kneller:2008PhRvD..77d5023K}. The forward shock is located to
  the left of ``A'', the reverse shock is the step-up in density found
  to the right of ``B'', and one of the many localised features in the
  profile is to the right of ``C''.\label{fig:2D}}
\end{figure}
\begin{figure}
\begin{indented}
\item[]% 
\includegraphics[width=4in,clip]{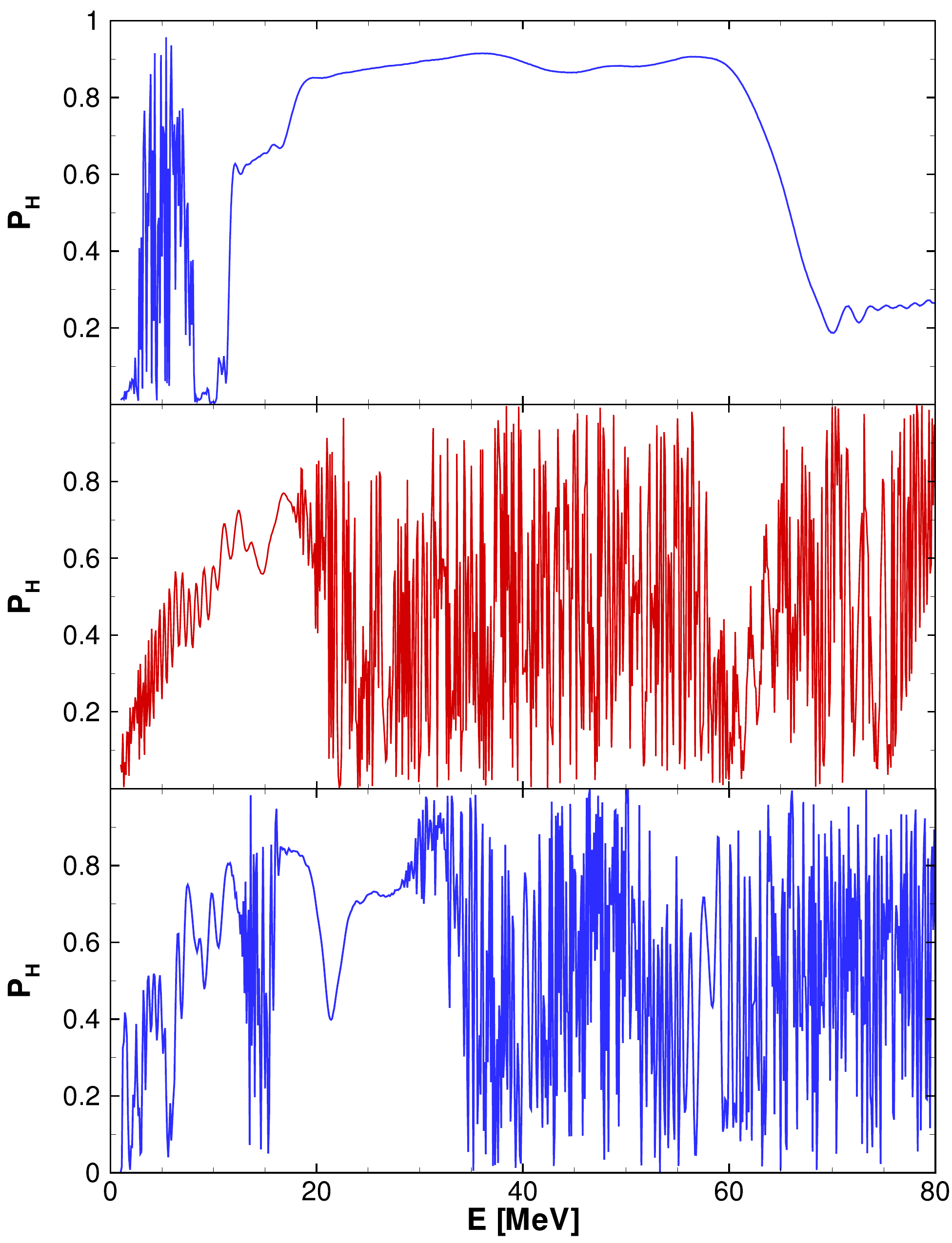}
\end{indented}
\caption{The H resonance crossing probability $P_\rmH$ as a function
  of neutrino energy for a radial slice at $\vartheta=25^{\circ}$ through
  the 2D supernova model.  From top to bottom the snapshot times are
  $t=2.4\;{\rm s}$, $5.4\;{\rm s}$ and $6.4\;{\rm s}$.  The figure
  is taken from \citet{Kneller:2008PhRvD..77d5023K}.\label{fig:early
    2D}}
\end{figure}

As we indicated back in section \ref{sec:introduction} there is now
ample evidence that supernovae are aspherical and there has been
significant recent progress in identifying what processes may be the
root of the asphericity.  The standing accretion shock instability was
found by \citet{Blondin:2003ApJ...584..971B} to generate large dipolar
and quadrapolar modes from small perturbations of a stalled, spherical
accretion shock.  When the simulations were repeated in
three-dimensional (3D) differential, post-shock flow was found indicating
the ``spinning up'' the PNS \citep{2005JPhCS..16..370B}.  More recent
work
\citep{Blondin:2006ApJ...642..401B,Blondin:2007Natur.445...58B,Blondin:2007ApJ...656..366B,Scheck:2006A&A...457..963S,2008JPhCS.112d2018O,2008ApJ...678.1207I}
have confirmed the result and furthered the understanding of the SASI.

As an example of an aspherical density profile we show in figure
\ref{fig:2D} a snapshot, taken from
\citet{Kneller:2008PhRvD..77d5023K}, of an aspherical explosion. For
this calculation the same ``initial'' density profile used in
1D was mapped into a 2D hydrodynamical code
and heated to revive the motion of the forward shock.  The total
energy deposition was $Q \approx 3 \times 10^{51}\;\mathrm{erg}$, the
canonical explosion energy for the matter portion of the supernova. 
To generate the asphericity the
heating was inhomogeneous with preferential energy deposition along
the equator.  From the figure we see that the density profile along
each line of sight can be very different but within the density
profile along any given radial slice we can still identify the forward
and reverse shocks and a hot bubble region between them.  The radial
positions of the two shocks now depends upon the polar angle $\vartheta$. Upon
closer inspection we find even in 2D that the reverse shock also
abates and moves back to the core.  Superimposed upon the generic
template are ``fluctuations'' generated by more shocks, localised
bubbles, sound waves, etc.

Snapshots of $P_\rmH$ as a function of neutrino energy are shown in figure \ref{fig:early 2D}. 
The transition from the progenitor profile to the forward shock
appears once again as the transition to $P_\rmH \sim 1$ and we also
notice how $P_\rmH$ drops suddenly around $E\sim 10\;{\rm MeV}$ for
$t=2.4\;{\rm s}$ when the profile becomes adiabatic again. Below
$E\sim 10\;{\rm MeV}$ phase effects appear as the profile develops
multiple resonances for these neutrino energies. 
Given the complexity of these profiles and our understanding of how
multiple resonances affect neutrinos it comes as little surprise then
that phase effects are prominent. By $t=5.4\;{\rm s}$
they dominate the entire spectrum above $20\;{\rm MeV}$. After
$t=6.4\;{\rm s}$ the forward shock, hot bubble, etc.\ have largely swept
through the spectrum and we are able to notice a narrow range of
energies surrounding $E\sim 15\;{\rm MeV}$ where phases effects again
appear moving down through the spectrum. This is the signature of the
reverse shock returning to the core.

\begin{figure}
\begin{indented}
\item[]% 
\includegraphics[width=4in]{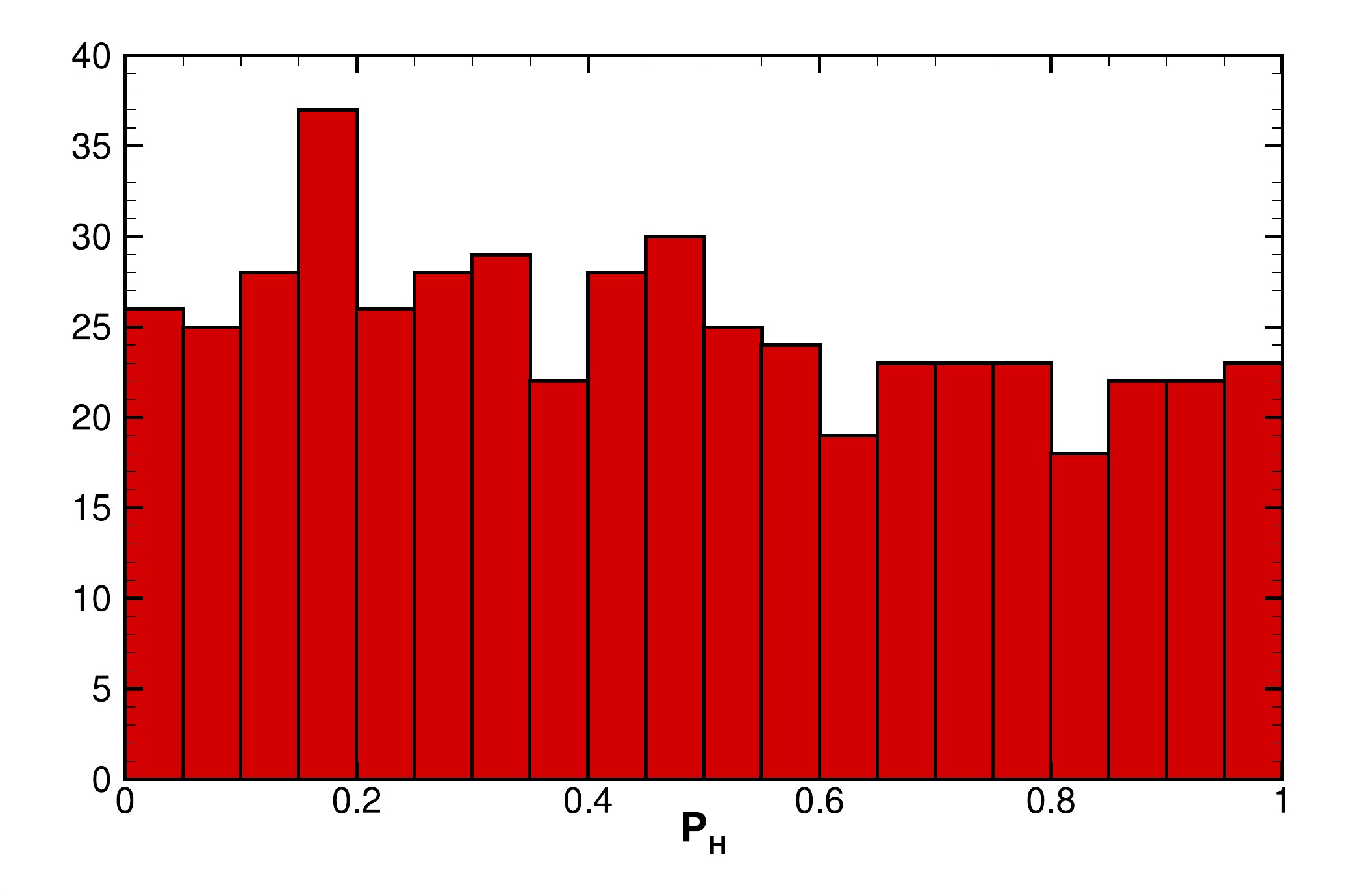}
\end{indented}
\caption{A histogram of $P_{\rmH}$ for neutrino energies between $30$ and $80\;{\rm MeV}$ 
for neutrinos passing through the $t=5.4$ density profile shown in figure \ref{fig:early 2D}.
The histogram is formed from 501 points and there are 20 bins so a uniform distribution would 
give an average count of 25 in each bin.\label{fig:histogram}}
\end{figure}

The crossing probabilities shown in figure \ref{fig:early 2D} are
very similar to those shown earlier in figures
\ref{fig:earlyphaseeffects1} and \ref{fig:earlyphaseeffects2}. After
comparing figure \ref{fig:early 2D} with those figures one has a sense that
the phase effects are somewhat stronger in 2D.  In fact the phase
effects are so strong that the crossing probability over wide swathes
of the neutrino energy range appears to be essentially random with,
more-or-less, uniform distribution from zero to unity. 
This impression of the uniformity of the distribution of $P_\rmH$ 
is confirmed by figure \ref{fig:histogram} where we plot the histogram of 
$P_{\rmH}$ from the middle panel of figure \ref{fig:early 2D} for neutrino 
energies between $30\;{\rm MeV}$ and $80\;{\rm MeV}$. While not proof, figure \ref{fig:histogram} 
hints that the density profiles from 2D supernova simulations may cross over into the regime of turbulence.
The effects upon neutrinos of a turbulent, noisy, density profile
either in general or upon the solar density profile date back to
\citet{Schafer:1987PhLB..185..417S},
\citet{Krastev:1989PhLB..226..341K},
\citet{Sawyer:1990PhRvD..42.3908S}, \citet{Loreti:PhysRevD.50.4762},
\citet{Balantekin:1996PhRvD..54.3941B} and many others thereafter.
Turbulence in supernova was first considered by \citet{Loreti:1995ae}
and then later by \citet{Fogli:2006JCAP...06..012F} in light of the
fact that the evolving density profile was found to leave its imprint
on the neutrino signal. In both cases $\delta$-correlated density
fluctuations were assumed. More recently \citet{Friedland:2006ta} and
\citet{Choubey:2007PhRvD..76g3013C} have used Kolmogorov fluctuation
spectra.  Whatever the fluctuation spectra used the same result
emerges: if the turbulence is strong then $P_\rmH$ becomes a random
variate uniformly distributed over the range $[0,1]$. While this prediction
seems to be in general agreement with figure \ref{fig:histogram} we remind the reader 
that what is plotted is the histogram of $P_{\rmH}$ for different neutrino energies
at one snapshot time which is something different. 
For \citet{Fogli:2006JCAP...06..012F} fluctuations were restricted to a
scale of less than $10\;{\rm km}$ while the spectra used by
\citet{Friedland:2006ta} permitted fluctuations on much larger
scales. \citet{Friedland:2006ta} showed that it is the density
fluctuations on the scale of the oscillation wavelength at the
resonances that contribute most and that the amplitude of the density
fluctuations need only be greater than $\delta\rho/\rho \gtrsim
0.1(\theta_{13})^{1/3}$ in order to enter the strong turbulence
limit. For the value of $\theta_{13}$ used when computing the results in 
figure \ref{fig:early 2D} we require fluctuations of just $\sim 1\%$ to 
reach this limit so indeed turbulence should be expected. 
Detecting turbulence in a neutrino signature would be a clear
signature that the supernova exploded aspherically but one must not
forget that we shall receive this signal along just one line of sight
which makes demonstrating that the signal is the result of turbulence
quite difficult.

%%%%%%%%%%%%%%%%%%%%%%%%%%%%%%%%%%%%%%%%%%%%%%%%%%%%%%%%%%%%%%%%%%%%%%%%%%%%%%%%%%%
%%%%%%%%%%%%%%%%%%%%%%%%%%%%%%%%%%%%%%%%%%%%%%%%%%%%%%%%%%%%%%%%%%%%%%%%%%%%%%%%%%%
%%%%%%%%%%%%%%%%%%%%%%%%%%%%%%%%%%%%%%%%%%%%%%%%%%%%%%%%%%%%%%%%%%%%%%%%%%%%%%%%%%%

\section{Neutrino flavour transformation in supernovae: effects of neutrino
  self-interaction\label{sec:coll-regime}}

In this section we review neutrino oscillations with neutrino
self-interaction in supernovae. For lack of space, we will focus on
the key results of the two-flavour numerical simulations of supernova
neutrino oscillations by \citet{Duan:2006an,Duan:2006jv}
and the analytical explanations of the results obtained under the
single-angle approximation. 
In section \ref{sec:numerical-results} we describe in detail the two
major approximations, i.e.\ the single-angle and multi-angle
approximations, that are now employed to treat neutrino oscillations
with neutrino self-interaction in supernovae. We also highlight the
key results in the numerical simulations carried out by
\citet{Duan:2006an,Duan:2006jv}. 
The Schr\"{o}dinger-like equation that governs neutrino flavour
transformation, although well suited for studying flavour evolution of
a single neutrino, is inconvenient when the neutrino potential
$\hat{H}_{\nu\nu}$ is not negligible.
In section \ref{sec:nfis} we introduce the ``language'', the notation of
neutrino flavour isospin (NFIS), that we will use to analyse and visualise 
collective neutrino oscillations. 
In section \ref{sec:MSW-like} we describe
an adiabatic MSW-like flavour evolution of dense neutrino gases in the
presence of 
ordinary matter. This type of neutrino oscillation echos the early
explorations \citep[e.g.][]{Qian:1994wh}
where neutrino self-interaction was included as an additional matter
effect in the conventional MSW mechanism.  
In section \ref{sec:pendulum} we analyse the stability of bipolar
neutrino systems such as supernova neutrinos by utilising the pendulum
analogy as well as the simple concept of energy conservation.
In section \ref{sec:bipolar-regime} we consider the matter effects and
try to gain some insights into the
qualitative behaviour of neutrino systems as neutrino number densities
decrease. We explain why supernova neutrinos
do not follow the MSW-like flavour evolution all the way through. 
In section \ref{sec:prec-sol} we discuss a collective
neutrino oscillation mode which is believed to cause the spectral
swap/split, a novel phenomenon revealed in numerical calculations. 
In section \ref{sec:progress} we briefly report current 
understandings of collective neutrino oscillations with full three
flavours and/or in realistic
supernova environments which are highly inhomogeneous and anisotropic.

\subsection{Approximations and numerical results%
\label{sec:numerical-results}}

It is clear that neutrino oscillations in the collective regime must
be treated in a way different from that in the pure MSW regime. In the
pure MSW regime, the flavour evolution of any single neutrino can be
calculated without knowing the flavour states of other neutrinos, as
long as the matter densities along the world line (i.e.\ the matter
profile) of this neutrino is given. For a complicated
3D matter profile, the flavour evolution histories of
neutrinos propagating along different trajectories can be different
even if they are initially in the same flavour state and have the same
energy. In the pure MSW regime, this means that the same algorithm for
calculating flavour evolution of a single neutrino needs to be run
repeatedly for each distinct neutrino beam.  In the collective regime,
however, because the flavour evolution histories of neutrinos
propagating along different trajectories are all coupled, the flavour
states of all distinct neutrino beams must be followed simultaneously.
While the former problem is ready to be solved given enough computing
time, the latter poses such a great numerical challenge that it has
not been tackled yet.

Partly because of the complexity of the problem, the supernova models
adopted so far to treat neutrino flavour transformation with neutrino
self-interaction are all simple and ideal. 
In these models the supernova environment is spherically symmetric and
neutrinos and antineutrinos are emitted isotropically
from an infinitely thin neutrino sphere with radius $R_\nu$. 
Neutrinos
and antineutrinos are in pure flavour states at the neutrino sphere and
encounter only forward-scattering with other particles outside the
neutrino sphere. The relativistic effects
such as redshift of neutrino energies and gravitational bending of
neutrino trajectories are ignored. 

In such supernova models nonequivalent neutrino trajectories at radius $r$ can
be distinguished by $\vartheta$, the angle between the radial direction
and the propagation direction of the neutrino.
Equation \eref{eq:Hnunu}, therefore, becomes
%\newpage
\begin{eqnarray}
\fl [H^\bsf_{\nu\nu,\vartheta}(r)]_{\alpha\beta} 
& =
\frac{\sqrt{2}\GF n_\nu^\tot(r)}%
{1-\cos\vartheta_{\max}(r)}
\int_{\cos\vartheta_{\max}(r)}^1%
(1-\cos\vartheta\cos\vartheta^\prime)
\rmd(\cos\vartheta^\prime)\nonumber\\
\fl &\quad\times
\Big[\sum_{\alpha^\prime}\xi_{\nu_{\alpha^\prime}}
\int_0^\infty\rmd E^\prime 
f_{\nu_{\alpha^\prime}}(E^\prime)
\langle\nu_\alpha|\psi_{\nu_{\alpha^\prime}}(r)\rangle
\langle\psi_{\nu_{\alpha^\prime}}(r)|\nu_\beta\rangle
\nonumber\\
\fl &\quad 
%-\sqrt{2}\GF n_{\nu_e}^\eff(r)
-\sum_{\alpha^\prime}
\xi_{\bar\nu_{\alpha^\prime}}
\int_0^\infty\rmd E^\prime 
f_{\bar\nu_{\alpha^\prime}}(E^\prime)
\langle\bar\nu_\beta|\psi_{\bar\nu_{\alpha^\prime}}(r)\rangle
\langle\psi_{\bar\nu_{\alpha^\prime}}(r)|\bar\nu_\alpha\rangle\Big],
\label{eq:Hnunu-multi-angle}
\end{eqnarray}
where 
$\vartheta_{\max}(r)=\arcsin(R_\nu/r)$ is the maximum possible value for
$\vartheta$ at $r$,
%\begin{equation}
%n_\nu^\tot(r)=\frac{1}{2\pi R_\nu^2} 
%[1-\cos\vartheta_{\max}(r)]
%%\left[1-\sqrt{1-\left(\frac{R_\nu}{r}\right)^2}\right]
%\left(\sum_\alpha\frac{L_{\nu_\alpha}}{\langle E_{\nu_\alpha}\rangle}
%+\sum_\alpha\frac{L_{\bar\nu_\alpha}}{\langle E_{\bar\nu_\alpha}\rangle}\right)
%\end{equation}
$n_\nu^\tot(r)$ is the total number density of neutrinos and
antineutrinos at $r$, 
$\xi_{\nu_\alpha(\bar\nu_{\alpha^\prime})}$ and
$f_{\nu_\alpha(\bar\nu_{\alpha^\prime})}(E)$ are the fractions of total number flux and
normalised energy distributions of neutrinos (antineutrinos) with
flavour $\alpha^\prime$ at the neutrino sphere, and
$|\psi_{\nu_{\alpha^\prime}(\bar\nu_{\alpha^\prime})}(r)\rangle$
is the flavour state of a neutrino (antineutrino) initially in
flavour $\alpha^\prime$ and (implicitly) with 
energy $E^\prime$ and trajectory angle $\vartheta^\prime$.

Even with great simplifications, the ``multi-angle approximation'' or
treatment discussed 
above is hard to analyse and can be computationally intensive.
For these reasons, the early works on the subject 
adopt what is called the ``single-angle approximation''. 
Under the single-angle approximation neutrino trajectories with
different $\vartheta$ are assumed to be equivalent, and the flavour
evolution of neutrinos propagating along a representative
trajectory (e.g.\ the radial trajectory) are computed.
The single-angle approximation further simplifies equation
\eref{eq:Hnunu-multi-angle} and one obtains
\begin{eqnarray}
\fl [H^\bsf_{\nu\nu}(r)]_{\alpha\beta}
&=
\frac{\mu(r)}{2}
%\sqrt{2}\GF n_{\nu}^\tot(r) D(r/R_\nu)
%\nonumber\\
%\fl &\quad\times
\Big[\sum_{\alpha^\prime}\xi_{\nu_{\alpha^\prime}}
\int_0^\infty\rmd E^\prime 
f_{\nu_{\alpha^\prime}}(E^\prime)
\langle\nu_\alpha|\psi_{\nu_{\alpha^\prime}}(r)\rangle
\langle\psi_{\nu_{\alpha^\prime}}(r)|\nu_\beta\rangle,
\nonumber\\
\fl &\quad 
%-\sqrt{2}\GF n_{\nu_e}^\eff(r)
-\sum_{\alpha^\prime}\xi_{\bar\nu_{\alpha^\prime}}
\int_0^\infty\rmd E^\prime 
f_{\bar\nu_{\alpha^\prime}}(E^\prime)
\langle\bar\nu_\beta|\psi_{\bar\nu_{\alpha^\prime}}(r)\rangle
\langle\psi_{\bar\nu_{\alpha^\prime}}(r)|\bar\nu_\alpha\rangle\Big],
\end{eqnarray}
\label{eq:Hnunu-single-angle}%
where 
\begin{equation}
\mu(r)=2\sqrt{2}\GF n_{\nu}^\tot(r) C(r)
\label{eq:mu}
\end{equation}
is the ``effective strength'' of neutrino self-interaction at $r$. 
In equation \eref{eq:mu} $C(r)$ is a geometric
factor that takes partly into account the angle effects and is 
\begin{equation}
\fl
C(r)
=\frac{\int_{\cos\vartheta_{\max}(r)}^1%
(1-\cos\vartheta^\prime)\rmd(\cos\vartheta^\prime)}%
{\int_{\cos\vartheta_{\max}(r)}^1%
\rmd(\cos\vartheta^\prime)}
=\frac{1}{2}\left[1-\sqrt{1-\left(\frac{R_\nu}{r}\right)^2}\right]
\end{equation}
if the representative neutrino trajectory is radially oriented.
(See \citealp{Dasgupta:2008cu} for a discussion of the single-angle
approximation in non-spherical geometry.)

\begin{figure}
%\begin{center}
$\begin{array}{@{}c@{\hspace{0.02 \textwidth}}c@{}}
\includegraphics*[width=0.49 \textwidth, keepaspectratio]{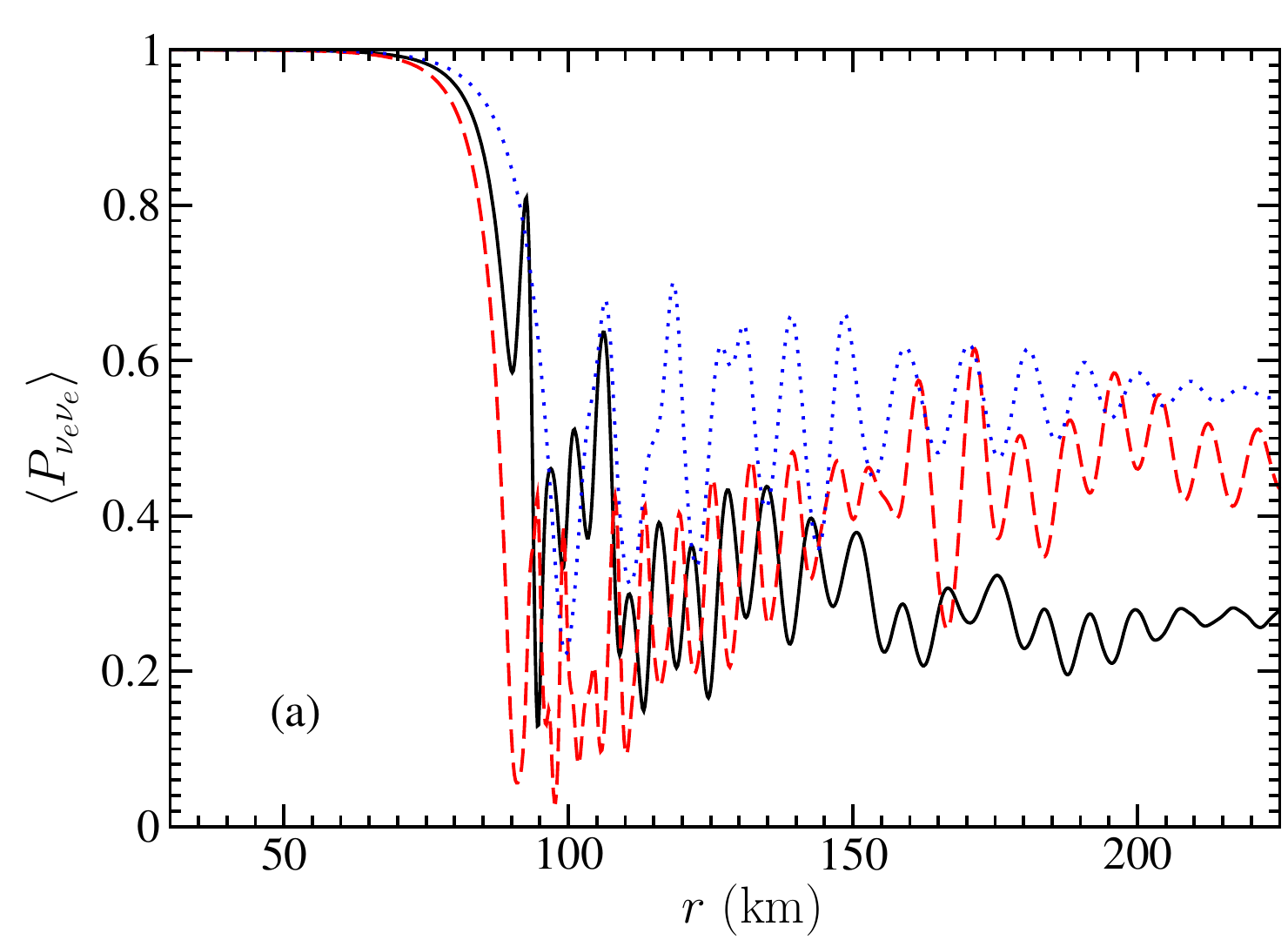} &
\includegraphics*[width=0.49 \textwidth, keepaspectratio]{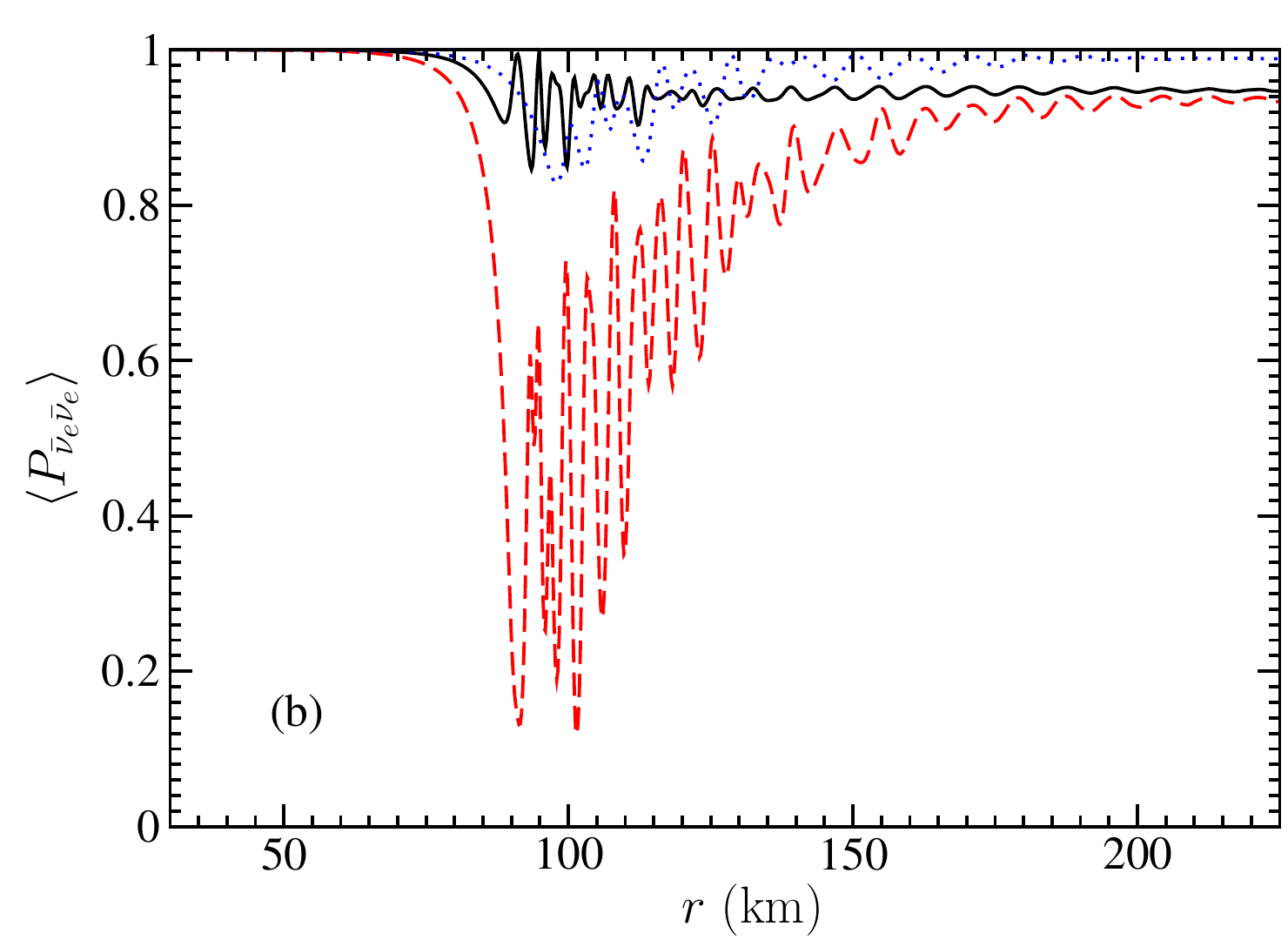} \\
\includegraphics*[width=0.49 \textwidth, keepaspectratio]{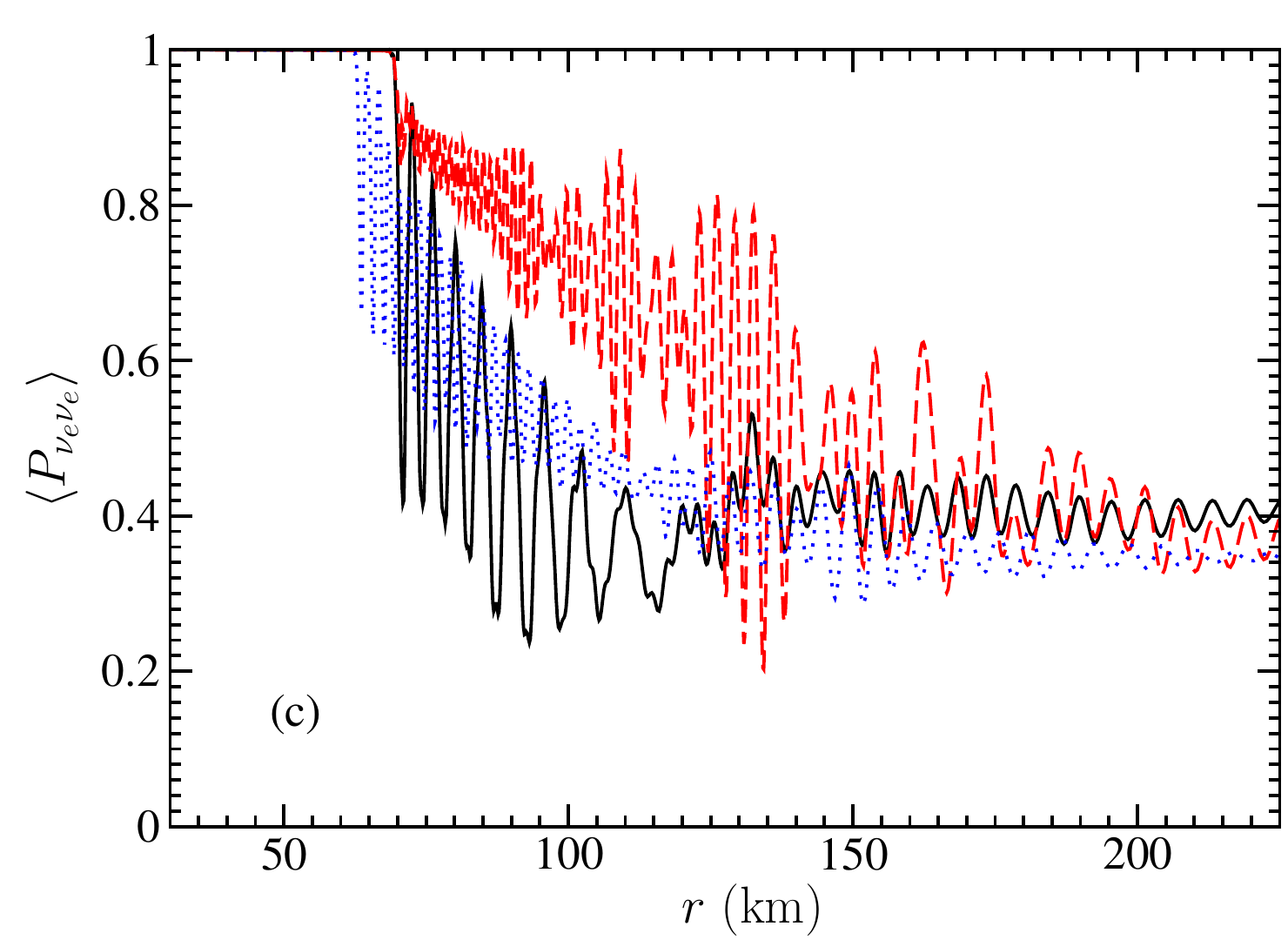} &
\includegraphics*[width=0.49 \textwidth, keepaspectratio]{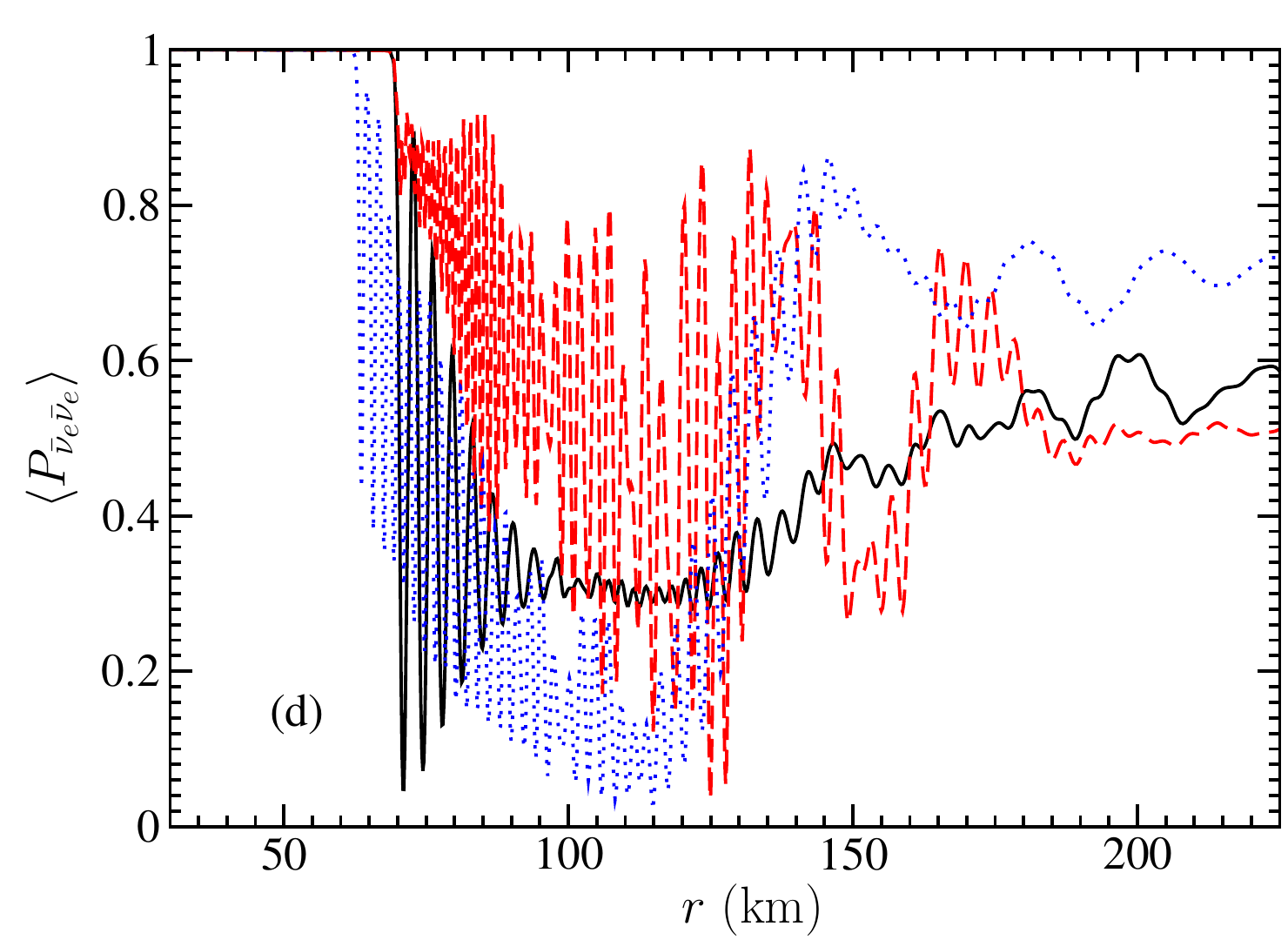}
\end{array}$
%\end{center}
\caption{\label{fig:P-r}
Energy-averaged survival probabilities $\langle P_{\nu\nu}\rangle$ 
for $\nu_e$ (left panels) 
and $\bar\nu_e$ (right panels) as functions of radius $r$
for the normal (upper panels) and
inverted (lower panels) neutrino mass hierarchies, respectively, when
neutrino self-interaction is taken into account.
The solid and dashed lines give average survival probabilities
along the radial and tangential trajectories,
respectively, as computed in the multi-angle simulations. The
dotted lines give the average survival probabilities
computed in the single-angle simulations. Figure adapted from
\citet{Duan:2006jv}.} 
\end{figure}

\begin{figure}
%\begin{center}
$\begin{array}{@{}c@{\hspace{0.01 \textwidth}}l@{}}
\includegraphics*[scale=0.41, keepaspectratio]{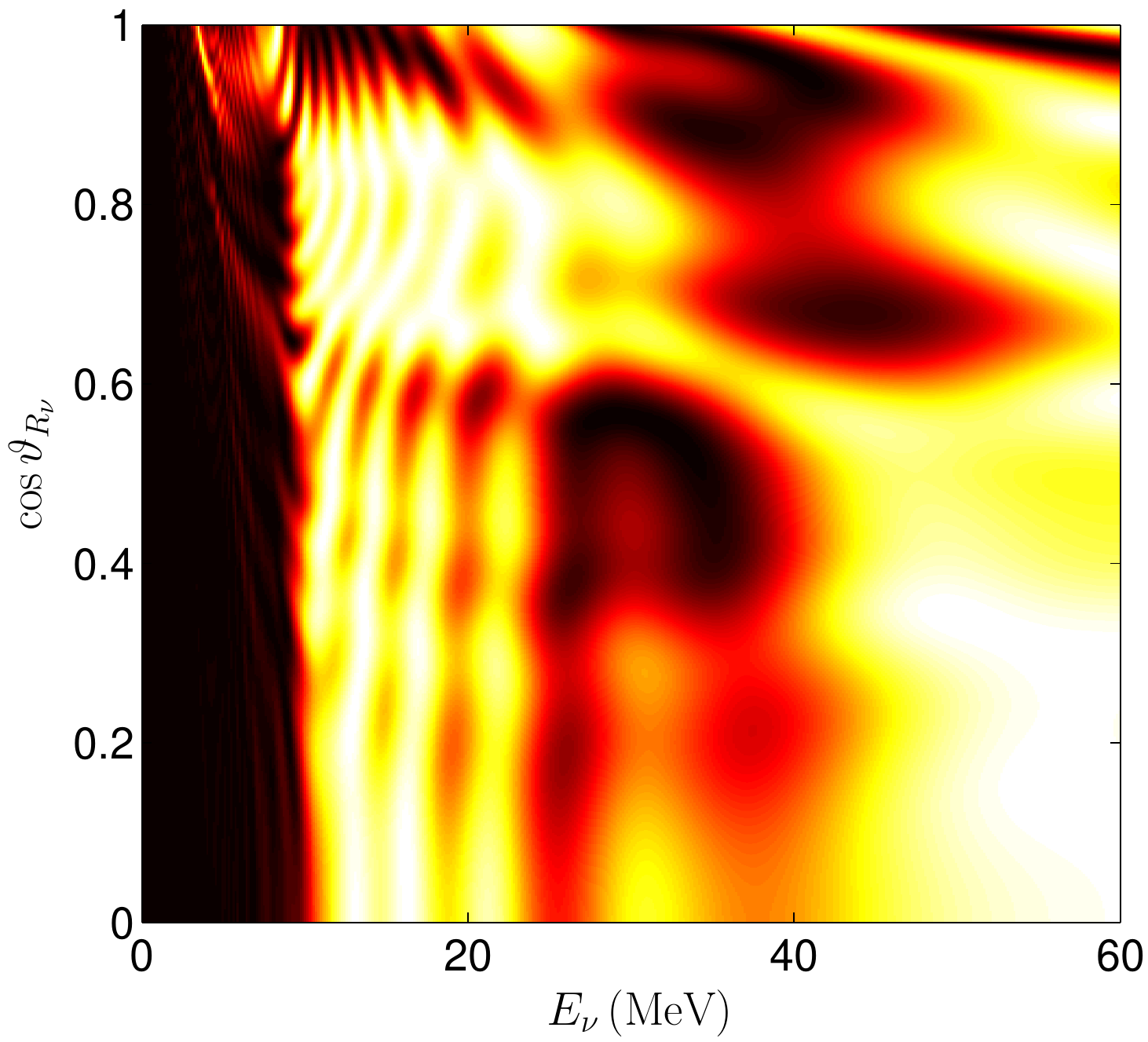} &
\includegraphics*[scale=0.41, keepaspectratio]{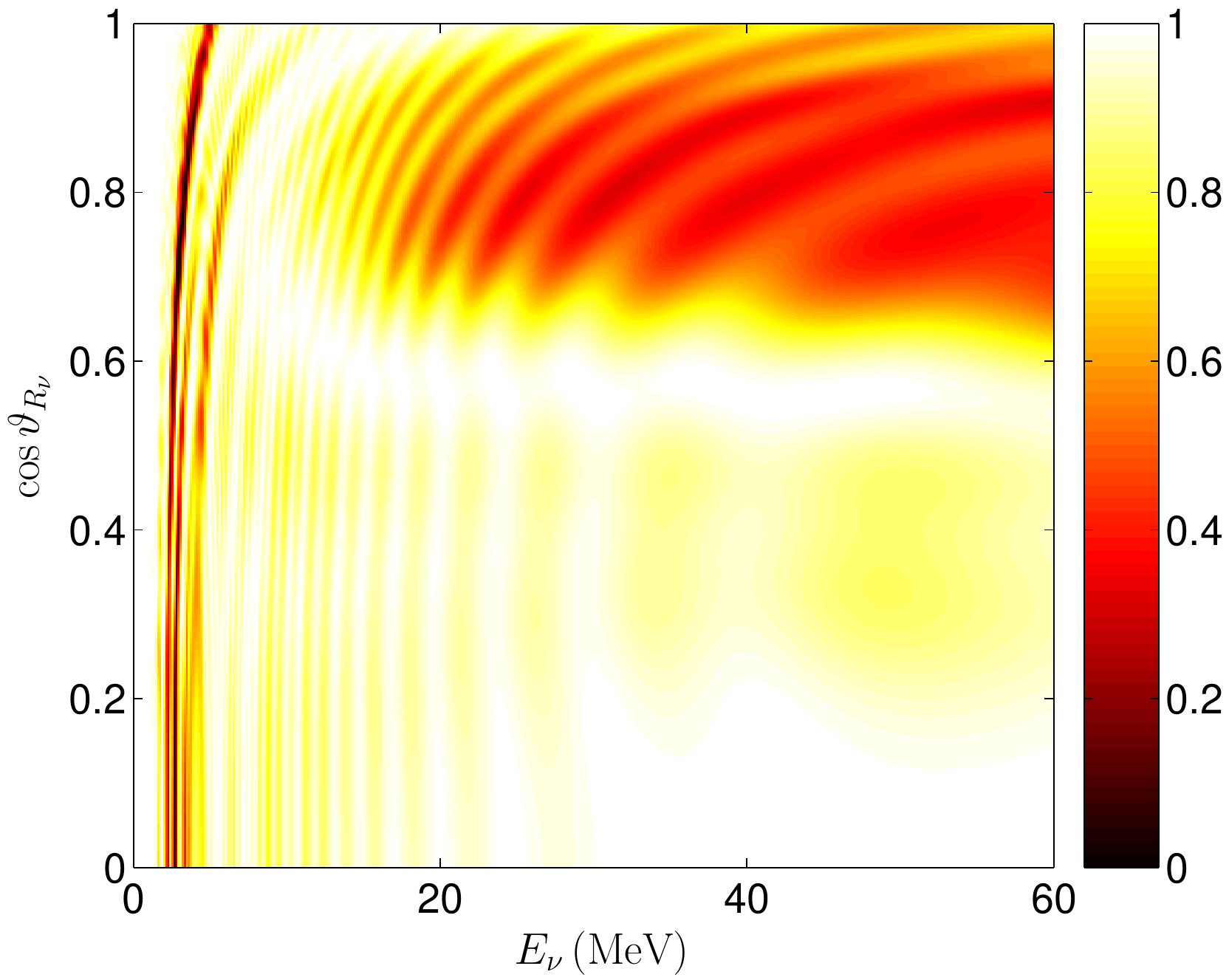} \\
\includegraphics*[scale=0.41, keepaspectratio]{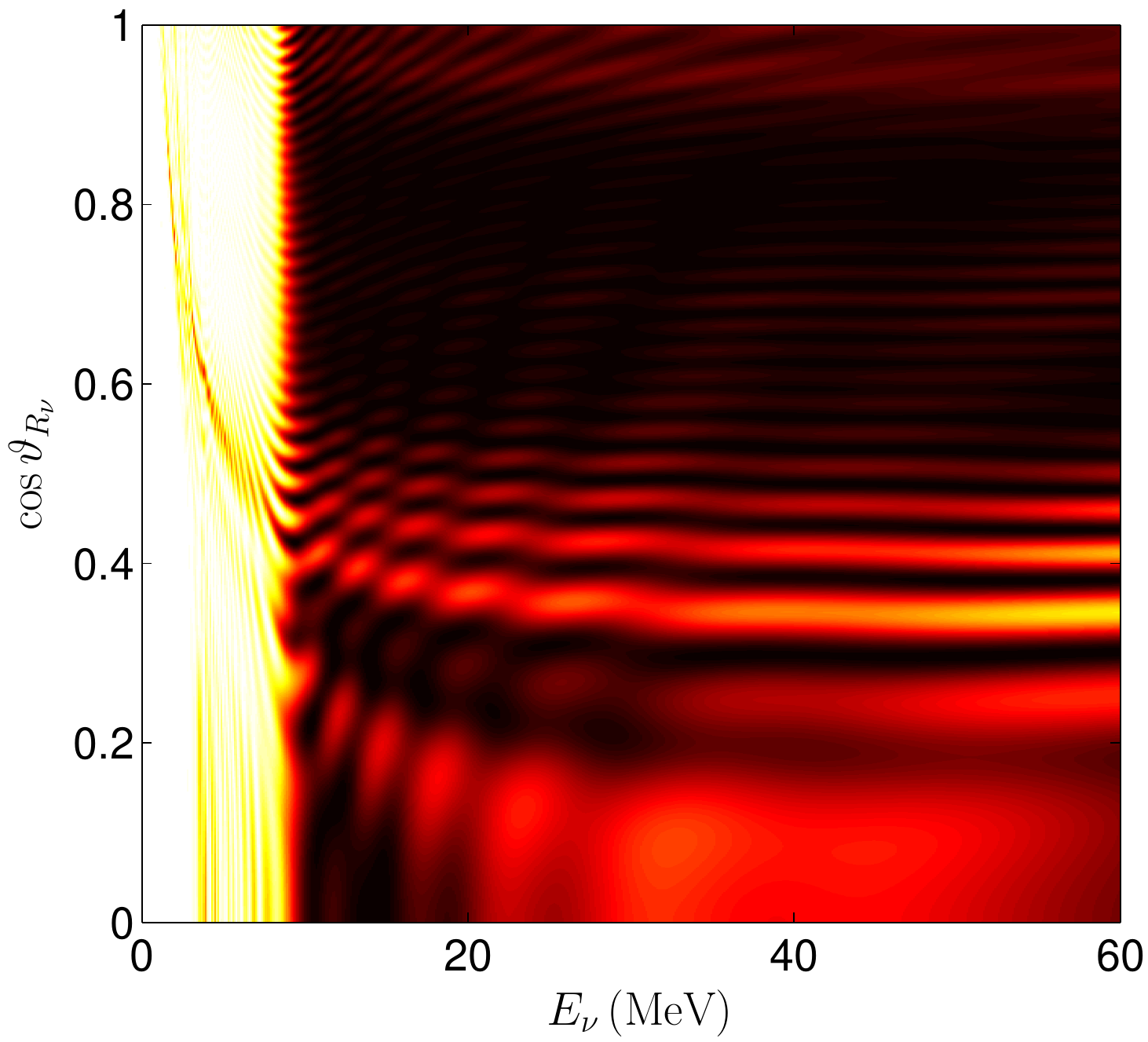} &
\includegraphics*[scale=0.41, keepaspectratio]{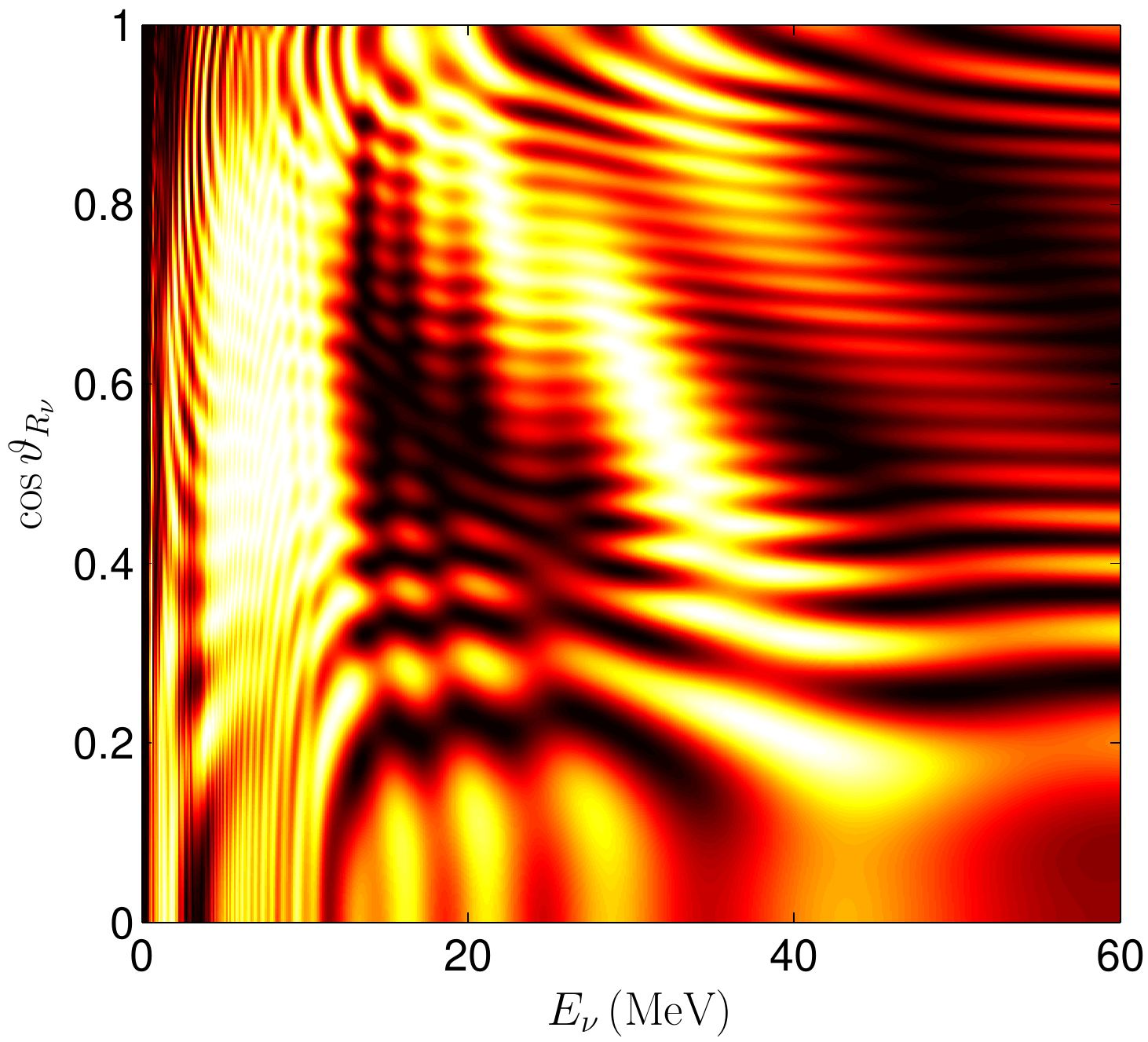}
\end{array}$
%\end{center}
\caption{\label{fig:P-E-c}
Results from the same multi-angle simulations as shown in figure \ref{fig:P-r}.
This figure
shows survival probabilities $P_{\nu\nu}$ for neutrinos (left panels) and
antineutrinos (right panels) 
at radius $r=225$ km
as functions of both neutrino energy $E_\nu$
and emission angle $\vartheta_{R_\nu}$, the angle between the
propagation direction of the neutrino and the normal to the neutrino sphere.
The upper panels employ a normal neutrino mass hierarchy,
and the lower panels employ an inverted neutrino mass hierarchy.
The colour scale in the plot denotes the neutrino survival probability
with $P_{\nu\nu}=1$ being the lightest. 
Figure adapted from \citet{Duan:2006jv}.}
\end{figure}

Figure \ref{fig:P-r} shows the results of the first numerical
simulations by \citet{Duan:2006an,Duan:2006jv}  of flavour
transformation of supernova neutrinos employing the multi-angle
treatment. (Also see
\citealp{Duan:2008eb} for the movies for the entire simulations.)
For comparison the results of the corresponding
single-angle calculations are also shown in the same figure. 
These calculations adopted two-flavour mixing schemes
with mass-squared differences $\delta m^2\simeq\pm\delta m_\mathrm{atm}^2$ and a
small mixing angle ($\thetav=0.1$) representing the unknown value of
$\theta_{13}$. The mixing occurs between the electronic flavour and 
the $\tau^\prime$ flavour, a
linear combination of the muon and tau flavours.
The matter profile and the neutrino fluxes/spectra used
by these calculations correspond to a late-time epoch when r-process
nucleosynthesis is supposed to occur.

Figure \ref{fig:P-r} shows that, in both the normal and inverted neutrino
mass hiearchy cases, neutrinos and antineutrinos with various energies
propagating along different trajectories can simultaneously experience
significant flavour transformation near the PNS.
This is clearly different from a standard MSW-flavour-transformation
phenomenon. Interestingly, although flavour evolution histories of
neutrinos propagating along various trajectories are different in
multi-angle calculations, the single-angle calculations seem to share
some key qualitative features with their multi-angle counterparts.

Figure \ref{fig:P-E-c} reveals an interesting result of collective
oscillations of supernova neutrinos which is obtained using the
multi-angle treatment. It shows that, when neutrino fluxes have
dropped off and collective oscillations end, survival probability
$P_{\nu\nu}$ for the neutrino becomes approximately a step function of
neutrino energy $E_\nu$.  In addition, the directions of these step
functions are opposite for the normal and inverted neutrino mass
hierarchy cases.  The corresponding single-angle calculations yield
results that are qualitatively consistent with those employing the
multi-angle treatment. The phenomenon seen in figure \ref{fig:P-E-c}
is called ``stepwise spectral swapping'' because $\nu_e$'s and
$\nu_{\tau^\prime}$'s apparently swap energy spectra at energies below
(above) a critical energy $\Es$ in a normal (inverted) neutrino mass
hierarchy case \citep{Duan:2006an}.  This phenomenon is also known as
the ``spectral split'' since $\Es$ ``splits the transformed spectrum
sharply into parts of almost pure but different flavours''
\citep{Raffelt:2007cb}.

Neutrino flavour transformation in supernova with neutrino
self-interaction has since been studied using both the single-angle
and multi-angle treatments. All these calculations confirm that
single-angle and multi-angle calculations do share the key common
features shown here. Further study by \citet{Duan:2007bt} showed that the
swap/split energy $\Es$ decreases significantly with $\thetav$ in the
normal mass hierarchy case and is not sensitive to $\thetav$ in the
inverted mass hierarchy case. 
Adopting a different matter profile with a thick
supernova envelope, \citet{Fogli:2007bk} found no
significant flavour transformation of supernova neutrinos in the
normal mass hierarchy case and a similar spectral-swap/split
phenomenon in the inverted mass hierarchy case. 

The single-angle and multi-angle treatments are clearly not equivalent.
In fact,  the single-angle
approximation is not even a self-consistent treatment of the problem.
For example, neutrinos propagating along trajectories with different
values of $\vartheta$ travel  different distances 
for any given radius interval.
Therefore, if one uses the results from a
single-angle computation and calculates flavour
evolution for neutrinos propagating along a trajectory that is
different from the 
representative one, one expects to find results that will contradict the
single-angle assumption. 

Meanwhile, it is also clear that the single-angle treatment is a
much simpler model than the multi-angle one.  Although the multi-angle
treatment approaches the inhomogeneous and
anisotropic nature of the supernova environment in a self-consistent
way, it is very difficult 
to study analytically. Under the single-angle assumption, however,
supernova neutrinos are 
essentially treated as a homogeneous, isotropic neutrino gas 
with effective neutrino number density 
\begin{equation}
n_\nu^\eff(r) = n_\nu^\tot(r) C(r)
\end{equation}
that expands with ``time'' $r$.
Because the results of the single-angle and multi-angle numerical calculations
available so far share many qualitative features,
\citet{Duan:2006an} conjectured that collective neutrino oscillations in
supernovae can be understood, at least qualitatively, by studying
similar phenomena in isotropic, homogeneous neutrino gases. 

For this reason we shall discuss collective flavour transformation of
supernova neutrinos under the single-angle
treatment or, equivalently, a homogeneous, isotropic neutrino gas which
expands with ``time'' $r$. In the following discussions neutrino number
densities must be 
understood as the effective ones with the geometric factor $C(r)$
included. [For truly homogeneous, isotropic neutrino gases one has
  $C(r)=1$ and, therefore, $n_\nu^\eff(r)=n_\nu^\tot(r)$.]
These analyses shall offer valuable insights into
the full-fledged problem of collective neutrino oscillations in realistic
supernova environments. 
In section \ref{sec:progress} we will briefly summarise
the current understandings of collective neutrino flavour
transformation in inhomogeneous, anisotropic environments.

\subsection{Neutrino flavour isospin\label{sec:nfis}}

Because $\hat{H}_{\nu\nu}$ is essentially a sum of matrices of
densities for all background neutrinos and antineutrinos,
it is convenient to use matrices of
densities instead of wavefunctions to describe the flavour states of
neutrinos when neutrino self-interaction is important. For a
two-flavour mixing scheme, a matrix of density 
\begin{equation}
\varrho_{\nu,E}=\frac{1}{2}
       [n_{\nu,E}+\vec{\sigma}\cdot\vP_{\nu,E}]
\label{eq:P}
\end{equation}
is equivalent to polarisation vector $\vP_{\nu,E}$ because the
trace of $\varrho_{\nu,E}$
is not relevant for neutrino oscillations. 
In equation \eref{eq:P} we have used energy $E$ to identify a
neutrino (or antineutrino) mode in a homogeneous, isotropic neutrino
gas, and $n_{\nu,E}$ is the number density of neutrinos with energy $E$.
Note that we use 
``$\vec{X}$'' to indicate a vector in flavour space
(as compared to vector ``$\bi{X}$'' in coordinate space), 
and $\vec{\sigma}$ is 
such a vector consisting of the three Pauli matrices. Viewing
polarisation vectors as ``magnetic spins'' and neutrino
self-interaction as coupling between these spins, 
\citet{Pastor:2001iu} elucidate the physics mechanism behind
synchronisation, a collective neutrino oscillation phenomenon
discovered in earlier numerical simulations 
\citep{Samuel:1993uw}.
To fully exploit this spin analogy, we will adopt the notation of neutrino
flavour isospin (NFIS) defined by \citet{Duan:2005cp}.

For a neutrino (antineutrino) state described by wavefunction
$\psi_{\nu(\bar\nu),E}$, the corresponding NFIS is defined as
\begin{equation}
\vs_{\omega}=\left\{
\begin{array}{ll}
\psi_{\nu,E}^\dagger\frac{\vec{\sigma}}{2}\psi_{\nu,E}
&\mathrm{for\ neutrino},\\
(\sigma_y\psi_{\bar\nu,E})^\dagger\frac{\vec{\sigma}}{2}
(\sigma_y\psi_{\bar\nu,E})
\quad&\mathrm{for\ antineutrino},\\
\end{array}
\right.
\label{eq:s-def}
\end{equation}
where
\begin{equation}
\omega=\pm\frac{\delta m^2}{2E}
\end{equation}
with the plus and minus signs for the neutrino and the antineutrino,
respectively. 
The $\sigma_y$-transformation of $\psi_{\bar\nu,E}$ in equation
\eref{eq:s-def} 
eliminates the superficial distinction between neutrinos and
antineutrinos in the two-flavour mixing scheme. In this notation
flavour states of neutrinos and antineutrinos are represented by spins
in flavour space with different values of $\omega$.
 In the flavour basis  
(i.e.\ $\psi_{\nu,E}=\langle\nu_\alpha|\psi_{\nu,E}\rangle$),
the probability for a neutrino to be in $|\nu_e\rangle$ is given by
\begin{equation}
|\langle\nu_e|\psi_{\nu,E}\rangle|^2=\frac{1}{2}+\vs_\omega\cdot\ve_z^\bsf,
\label{eq:sz-nu}
\end{equation}
where $\ve_z^\bsf$ is the flavour-basis, $z$-direction unit vector in flavour
space. Similarly, the probability for an antineutrino to be 
$|\bar\nu_e\rangle$ is given by 
\begin{equation}
|\langle\bar\nu_e|\psi_{\bar\nu,E}\rangle|^2
=\frac{1}{2}-\vs_\omega\cdot\ve_z^\bsf.
\label{eq:sz-anu}
\end{equation}

NFIS can also be defined from the matrix of density for a neutrino
$\nu$ or antineutrino $\bar\nu$:
\begin{equation}
\fl
\vs_{\omega}=\left\{
\begin{array}{ll}
\frac{1}{2}\nf_\omega^{-1}
\Tr(\varrho_{\nu,E}\vec{\sigma})
&\mathrm{for\ neutrino},\\
\frac{1}{2}\nf_\omega^{-1}
\Tr(\sigma_y\varrho_{\bar\nu,E}\sigma_y^\dagger\vec{\sigma})
=-\frac{1}{2}\nf_\omega^{-1}
\Tr(\varrho_{\bar\nu,E}^*\vec{\sigma})
\ &\mathrm{for\ antineutrino},\\
\end{array}
\right.
\label{eq:rho-s}
\end{equation}
where 
\begin{equation}
\nf_\omega
=\sqrt{\sum_{i=x,y,z}[\Tr(\varrho_{\nu(\bar\nu),E}\sigma_i)]^2}.
\end{equation}
Here NFIS $\vs_\omega$ represents the average flavour of the
neutrino (antineutrino) mode $\omega$, and 
$\nf_\omega$ is the ``net number density'' of the neutrino
(antineutrino) mode $\omega$ that is in the flavour state represented by
$\vs_\omega$. For example, if all neutrinos with energy $E$ are
either in pure $|\nu_e\rangle$ state or in pure $|\nu_{\taup}\rangle$
with number densities $n_{\nu_e,E}$ and $n_{\nu_{\taup},E}$,
respectively, then 
\[\vs_\omega=\sgn(n_{\nu_e,E}-n_{\nu_{\taup},E})\frac{\ve_z^\bsf}{2}
\quad\mathrm{and}\quad
\nf_\omega=|n_{\nu_e,E}-n_{\nu_{\taup},E}|.\]

Using $\nf_\omega$ we can define the distribution function of NFIS to be
\begin{equation}
\ff_\omega=\frac{\nf_\omega(r)}{n_\nu^\eff(r)}.
\label{eq:f-nfis}
\end{equation}
We emphasise that, for flavour transformation of supernova neutrinos
under the single-angle treatment, all quantities that are proportional
to neutrino number densities such as $n_\nu^\eff(r)$ 
and $\nf_\omega(r)$ are computed with the geometric factor
$C(r)$ included.
Because we consider only forward scattering between neutrinos and
background particles, 
$\ff_\omega$ do not change with $r$.
We note that $\nf_\omega$ is usually less than the neutrino
(antineutrino) number density with energy $|\delta m^2/2\omega|$. They
are equal only if all neutrinos or antineutrinos with this energy 
are in the same flavour state. As a result, the NFIS distribution
function $\ff_\omega$ is not normalised to unity but, rather, 
$\int_{-\infty}^\infty\ff_\omega\rmd\omega\leq 1$.

Under the NFIS notation equation \eref{eq:eom} becomes
\begin{equation}
\fl
\frac{\rmd}{\rmd r}\vs_\omega(r)
=\vs_\omega(r)\times\vH_\omega(r)
=\vs_\omega(r)\times[\omega\vH_\vac+\vH_\matt(r)
-\mu(r)\langle\vs(r)\rangle],
\label{eq:eom-s}
\end{equation}
where the misalignment between the ``vacuum field''
\begin{equation}
\vH_\vac=-\ve_x^\bsf\sin2\thetav+\ve_z^\bsf\cos2\thetav
\end{equation}
and $\ve_z^\bsf$ indicates the mismatch between vacuum mass eigenstates
and pure flavour states of neutrinos, the ``matter field''
\begin{equation}
\vH_\matt(r) = -\sqrt{2}\GF \NA \rho(r) Y_e(r)\ve_z^\bsf
\end{equation}
represents the change of refractive indices of neutrinos caused by
ordinary matter, and 
\begin{equation}
\langle\vs(r)\rangle = 
\int_{-\infty}^\infty\rmd\omega
\ff_\omega\vs_\omega(r)
\end{equation}
is the average NFIS. Equation \eref{eq:eom-s} shows that NFIS's are
 ``antiferromagnetically'' coupled with each other with strength
$\mu(r)=2\sqrt{2}\GF n_\nu^\eff(r)$.

In vacuum, NFIS $\vs_\omega$ precesses around $\vH_\vac$ with angular
frequency $|\omega|$. When
projected to the $\ve_z^\bsf$-axis, this precession motion 
is interpreted as the oscillation of neutrino flavours
[see equations \eref{eq:sz-nu} and \eref{eq:sz-anu}]. 
 
Because of the popular usage of the neutrino polarisation
vector, it shall be helpful to discuss the relationship between this
notation and the NFIS notation. From equations \eref{eq:P} and
\eref{eq:rho-s} it is clear that NFIS's and neutrino flavour
polarisation vectors are connected to each other by simple relations
\begin{equation}
\vP_{\nu,E}(r)=2n_\nu^\eff(r)\ff_\omega\vs_\omega(r)
\quad\mathrm{and}\quad
\vP_{\bar\nu,E}(r)=-2n_\nu^\eff(r)\ff_\omega\vs_\omega(r).
\end{equation}
A slightly different definition 
\begin{equation}
\vP_\omega(r)\propto\sgn(\omega\delta m^2)\ff_\omega\vs_\omega(r)
\end{equation} 
is used in some recent literature 
\citep[e.g.][]{Raffelt:2007cb}. With this new definition, $\vP_\omega(r)$ and
$\vs_\omega(r)$ are 
related by some constant factor (i.e.\ independent of $r$)
 which is positive for a neutrino 
($\omega\delta m^2>0$) and negative for an antineutrino ($\omega\delta m^2<0$).
The e.o.m.\ for $\vP_\omega(r)$ is
\begin{equation}
\frac{\rmd}{\rmd r}\vP_\omega(r)=
[\omega\vec{B}+\lambda(r)\vec{L}+\mu^\prime(r)
  \vec{D}(r)]\times\vP_\omega(r),
\label{eq:eom-P}
\end{equation}
where $\vec{B}=-\vH_\vac$, $\lambda(r)\vec{L}=-\vH_\matt(r)$,
$\mu^\prime(r)$ is different from $\mu(r)$ by a constant factor
depending on the normalisation of $\vP_\omega(r)$, and 
\begin{equation}
\vec{D}(r)=\int_{-\infty}^\infty
\sgn(\omega\delta m^2)\vP_\omega(r) \,\rmd\omega 
\propto\langle\vs(r)\rangle.
\end{equation}
We note that the orders of the cross products on the right-hand sides
of equations \eref{eq:eom-s} and \eref{eq:eom-P} are different with the former
closely imitating the e.o.m.\ for magnetic spins in magnetic fields.

The notations of the polarisation vector and the NFIS are fully
equivalent. However, there is a caveat when the ``corotating frame''
technique is applied \citep{Duan:2005cp}. In the absence of ordinary
matter (i.e.\ $\vH_\matt=0$) the e.o.m.\ for the NFIS system is
essentially unchanged when observed in a reference frame that rotates
about $\vH_\vac$ with a constant angular frequency $\omega_0$. The only
difference is that the angular precession frequency $\omega$ of each
NFIS $\vs_\omega(r)$ is
shifted by $-\omega_0$. This is a powerful technique to analyse
collective flavour transformation in neutrino systems. For example, 
consider a neutrino gas that initially consists of pure $\nu_e$ and
$\nu_{\tau^\prime}$ with energies $|\delta m^2/3\omega_0|$ and
$|\delta m^2/\omega_0|$, respectively. Applying the corotating-frame
transformation, one 
can see that the flavour evolution of this system is similar to that
of the system which initially consists of pure, mono-energetic $\nu_e$
and $\bar\nu_e$ with energy $|\delta m^2/\omega_0|$. We note that a
corotating-frame transformation can change the sign of the angular precession
frequency $\omega$ of polarisation vector $\vP_\omega(r)$, and the
direction of $\vP_\omega(r)$ must 
be simultaneously inverted when this occurs. In the NFIS notation,
however, the superficial distinction between particles and
antiparticles
(i.e.\ the positive and negative frequency modes)
 is eliminated, and $\vs_\omega(r)$ is invariant under such a
corotating-frame transformation. In some very recent literature
\citep[e.g.][]{Raffelt:2008hr} polarisation vectors for antineutrinos
are defined in a way similar to NFIS's and have their directions
inverted. The caveat discussed above, therefore, disappears under this
new definition, and $\vP_\omega(r)$ and $\vs_\omega(r)$ are only
different by a positive, constant factor that is proportional to $\ff_\omega$.

\subsection{Adiabatic MSW-like flavour evolution\label{sec:MSW-like}}

Without knowing the results from numerical computations, one may think that the
effects of background neutrinos can be treated simply as another potential 
added to $V_e(r)$ in the MSW mechanism 
\citep[e.g.][]{Qian:1994wh}. In particular, if 
both $\rho(r)$ and $n_\nu^\eff(r)$ vary slowly with $r$, one would
expect that a
neutrino or antineutrino stays in the same matter eigenstate which
is essentially a pure flavour state at the neutrino sphere. Of course, a
matter eigenstate in this case is an eigenstate of the full Hamiltonian which
includes neutrino self-interaction.

Similar to the energy eigenstates of an electron in the presence of a
magnetic field, 
in the NFIS notation the matter eigenstates of neutrinos are
represented by spins that are
aligned or antialigned with the total effective field $\vH_\omega(r)$
in flavour space. If supernova neutrinos experience the adiabatic
MSW-like flavour evolution described above, one should have
\begin{equation}
\vs_\omega(r)=\frac{\epsilon_\omega}{2}\frac{\vH_\omega(r)}{H_\omega(r)},
%\label{eq:alignment-msw}
\end{equation}
or
%\numparts
\begin{eqnarray}
s_{\omega,x}^\bsv(r) &=
\frac{\epsilon_\omega}{2 H_\omega(r)}
[H_{\matt,x}^\bsv(r)-\mu(r)\langle s_x^\bsv(r)\rangle],
\label{eq:alignment-msw-sx}\\
s_{\omega,y}^\bsv(r) &=-
\frac{\epsilon_\omega}{2 H_\omega(r)}
\mu(r)\langle s_y^\bsv(r)\rangle,
\label{eq:alignment-msw-sy}\\
s_{\omega,z}^\bsv(r) &=
\frac{\epsilon_\omega}{2 H_\omega(r)}
[\omega+H_{\matt,z}^\bsv(r)-\mu(r)\langle s_z^\bsv(r)\rangle],
\label{eq:alignment-msw-sz}
\end{eqnarray}
%\label{eq:alignment-msw}
%\endnumparts%
where the alignment factor $\epsilon_\omega$ is constant and equals to
$+1$ ($-1$) if $\vs_\omega(R_\nu)$ is aligned (antialigned) with
$\vH_\omega(R_\nu)$, $H_\omega(r)=|\vH_\omega(r)|$, and
$X_i^\bsv$ ($i=x,y,z$) stand for the components of vector $\vec{X}$ in
the vacuum mass basis:
%\numparts
\begin{eqnarray}
\ve_x^\bsv &=
\ve_x^\bsf\cos2\thetav+\ve_z^\bsf\sin2\thetav,\\
\ve_y^\bsv &=
\ve_y^\bsf,\\
\ve_z^\bsv &= \vH_\vac =
-\ve_x^\bsf\sin2\thetav+\ve_z^\bsf\cos2\thetav.
\end{eqnarray}
%\endnumparts

Averaging equations
\eref{eq:alignment-msw-sx}--\eref{eq:alignment-msw-sz} over all
neutrino modes one obtains 
%\numparts
\begin{eqnarray}
\langle s_x^\bsv(r)\rangle
&= \frac{1}{2}[H_{\matt,x}^\bsv(r)-\mu(r) \langle s_x^\bsv(r)\rangle]
\int_{-\infty}^\infty\rmd\omega \frac{\epsilon_\omega
  \ff_\omega}{H_\omega(r)},
\label{eq:avg-sx}\\
\langle s_y^\bsv(r)\rangle
&= -\frac{1}{2}\mu(r) \langle s_y^\bsv(r)\rangle
\int_{-\infty}^\infty\rmd\omega \frac{\epsilon_\omega
  \ff_\omega}{H_\omega(r)},
\label{eq:avg-sy}\\
\langle s_z^\bsv(r)\rangle
&= \frac{1}{2}
\int_{-\infty}^\infty\rmd\omega \frac{\epsilon_\omega
  \ff_\omega}{H_\omega(r)}
[\omega+H_{\matt,z}^\bsv(r)-\mu(r) \langle s_z^\bsv(r)\rangle].
\label{eq:avg-sz}
\end{eqnarray}
%\endnumparts
Equations \eref{eq:avg-sx} and \eref{eq:avg-sy} imply that,
if $\vH_\matt(r)\neq0$, then
$\langle s_y^\bsv(r)\rangle=0$ and, therefore,
$s_{\omega,y}^\bsv(r)=0$ for all neutrinos [equation \eref{eq:alignment-msw-sy}].

\begin{figure}
%\begin{center}
$\begin{array}{@{}c@{\hspace{0.02 \textwidth}}c@{}}
\includegraphics*[width=0.49 \textwidth, keepaspectratio]{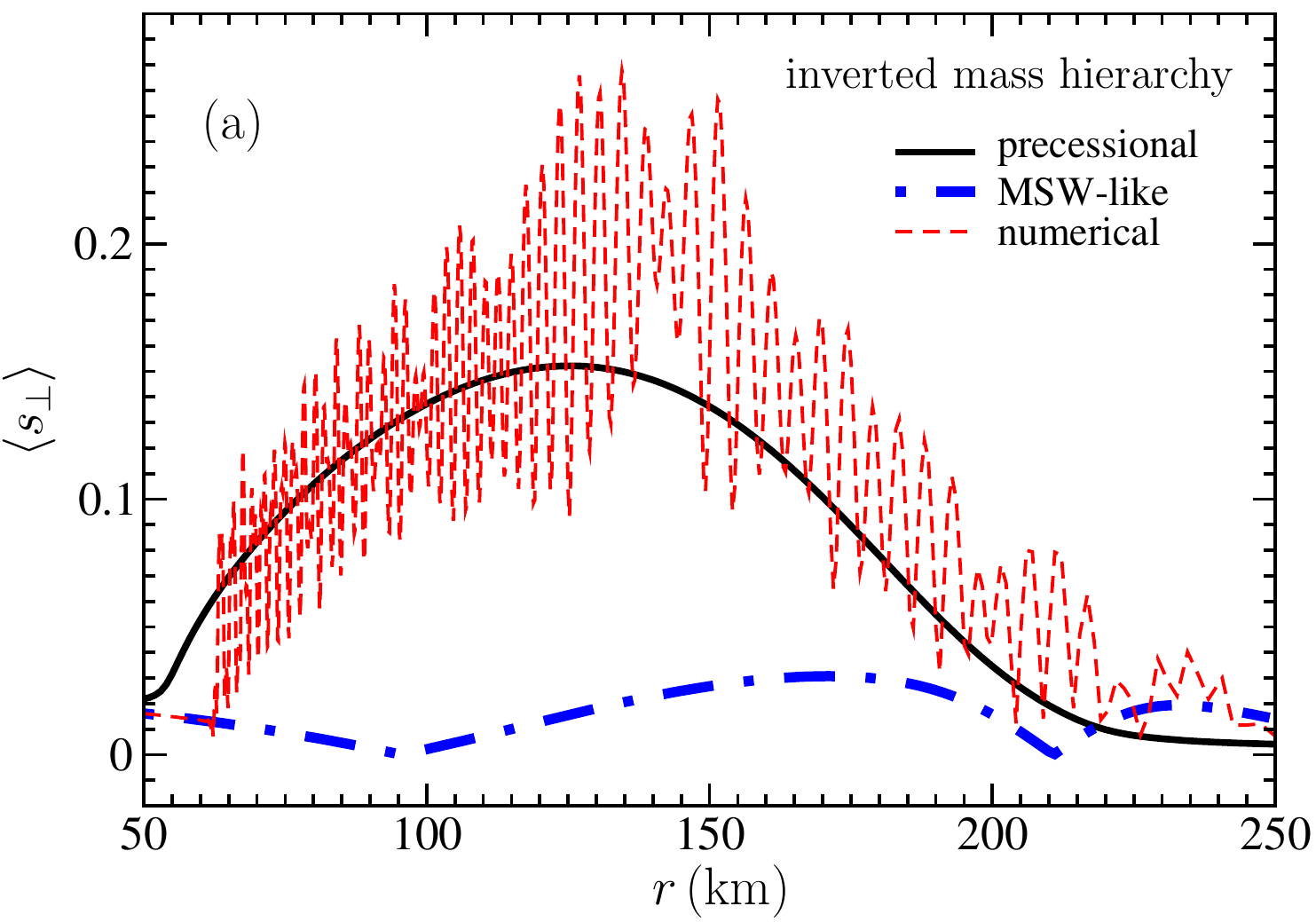} &
\includegraphics*[width=0.49 \textwidth, keepaspectratio]{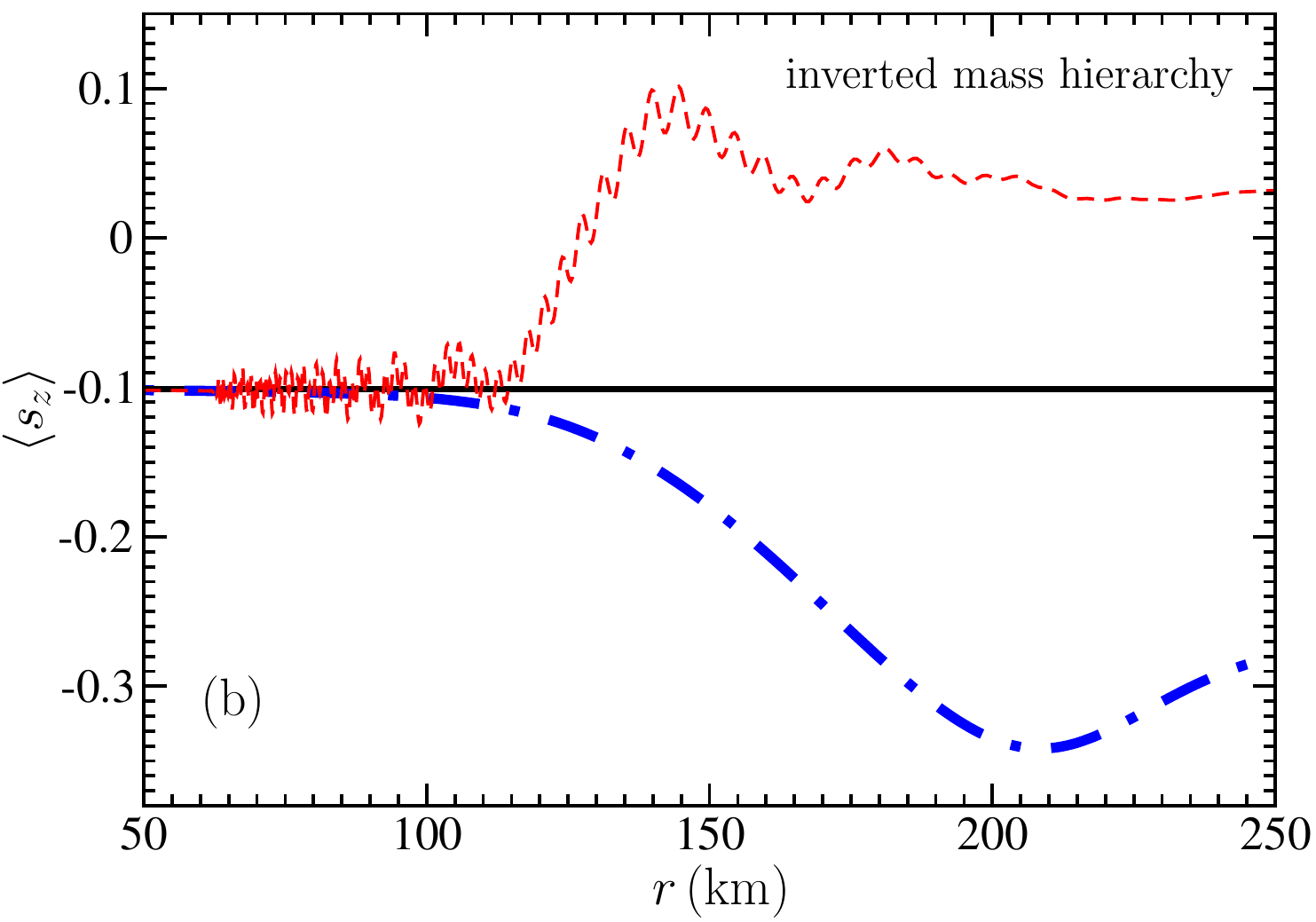} \\
\includegraphics*[width=0.49 \textwidth, keepaspectratio]{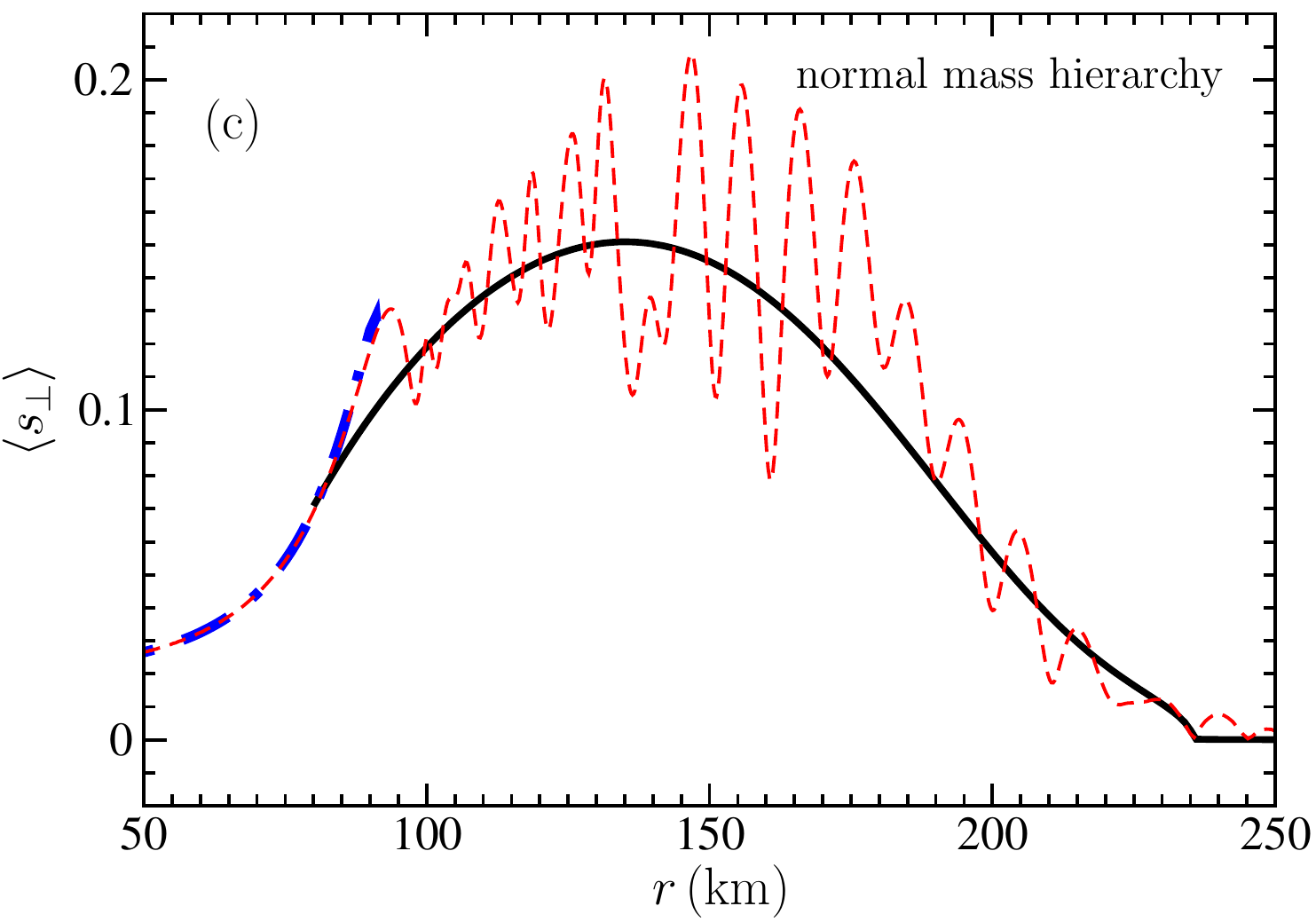} &
\includegraphics*[width=0.49 \textwidth, keepaspectratio]{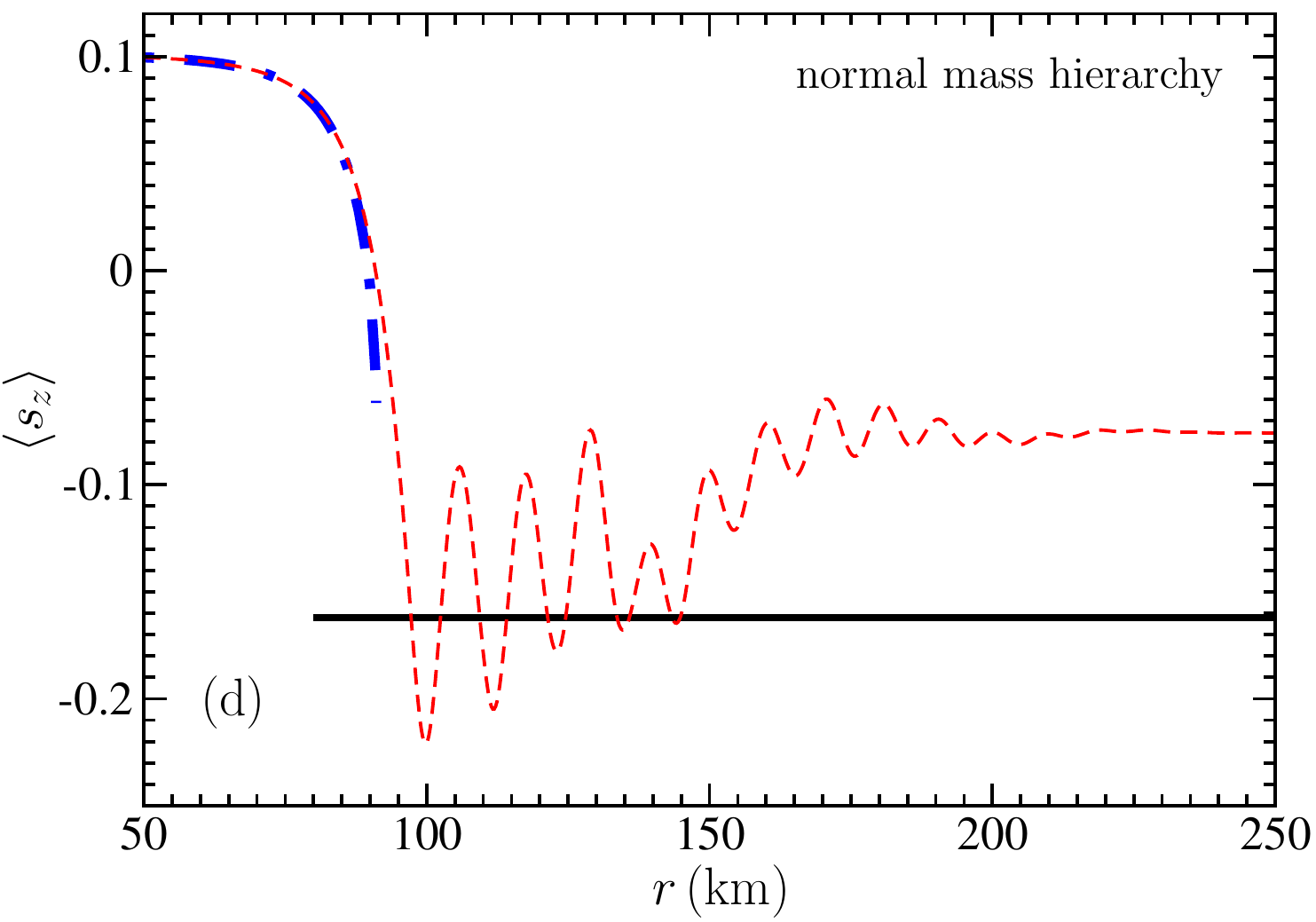}
\end{array}$
%\end{center}
\caption{\label{fig:twosols}
Comparison of the results of the single-angle calculations shown in figure
\ref{fig:P-r} with those in the
adiabatic MSW-like and adiabatic precession solutions.
Figure adapted from \citet{Duan:2007fw}.}
\end{figure}

If neutrinos follow the adiabatic MSW-like flavour evolution, one can
solve equations \eref{eq:avg-sx} and \eref{eq:avg-sz} for $\langle
s_x^\bsv(r)\rangle$ and $\langle s_z^\bsv(r)\rangle$ and then solve
equations \eref{eq:alignment-msw-sx} and \eref{eq:alignment-msw-sz}
for each individual NFIS. 
\citet{Duan:2007fw} have done exactly that with the same physical
settings as those of the single-angle calculations shown in figure
\ref{fig:P-r}. The components of the average NFIS $\langle
s_z^\bsv\rangle$ and
$\langle s_\perp^\bsv\rangle=\sqrt{\langle s_x^\bsv\rangle^2+\langle
  s_y^\bsv\rangle^2}=|\langle s_x^\bsv\rangle|$
as functions of $r$ in this solution are plotted
in figure \ref{fig:twosols}. For comparison the results of the corresponding
 single-angle calculations are also shown in the same figure.

For the inverted mass hierarchy case [figures \ref{fig:twosols}(a,b)]
one observes that at $r\lesssim63$ km $\langle s_\perp^\bsv(r)\rangle$ and 
$\langle s_z^\bsv(r)\rangle$ in the single-angle calculation agree
very well with the ``adiabatic MSW-like solution'' which is solved
from equations \eref{eq:avg-sx} and \eref{eq:avg-sz}.
At $r\gtrsim 63$ km, however, $\langle s_\perp^\bsv(r)\rangle$
abruptly jumps out the track of the adiabatic MSW-like flavour
evolution. For the normal mass hierarchy case [figures
  \ref{fig:twosols}(c,d)] the neutrino system follows the adiabatic
MSW-like solution to a larger radius. \citet{Duan:2007fw} were not able
to solve equations \eref{eq:avg-sx} and \eref{eq:avg-sz} beyond $r\simeq91$
km. Figure \ref{fig:twosols}(d) shows that,
 before the neutrino system deviates from the
adiabatic MSW-like flavour evolution, it seems to experience an MSW
resonance as $\langle s_z^\bsv(r)\rangle$ changes its sign at
$r\simeq90$ km. [According to equations \eref{eq:sz-nu} and
  \eref{eq:sz-anu}
the flavour transformation of a neutrino is
represented by the changing in the orientation of the correspond NFIS.]

Figure \ref{fig:twosols} suggests that configuration of the neutrino
system becomes 
unstable just before it departs from the adiabatic MSW-like flavour
evolution. This 
instability is similar to that of a pendulum near its highest
position which we will look into next.

\subsection{Bipolar systems and flavour pendulum%
\label{sec:pendulum}}

In simulating neutrino flavour transformation in the early Universe
\citet{Kostelecky:1993dm} observed an intriguing phenomenon that with
an inverted mass hierarchy certain neutrino gases can experience ``substantial
flavour oscillation even for extremely small mixing angles''. This
phenomenon can be partly understood by using the concept of the NFIS energy
of a homogeneous, isotropic neutrino gas \citep{Duan:2005cp}:
\begin{equation}
%\fl
\mathcal{E}(t) = 
-\int_{-\infty}^\infty\rmd\omega \ff_\omega\omega\vs_\omega(t)\cdot\vH_\vac
 +\frac{\mu(t)}{2} |\langle\vs(t)\rangle|^2,
\label{eq:nfis-energy}
\end{equation}
where we have assumed $\vH_\matt=0$. Note that the second term on the
right-hand side of equation \eref{eq:nfis-energy} has a positive sign because
the NFIS's are ``antiferromagnetically'' coupled to each other.
Using equation \eref{eq:eom-s} one can show that,
if $n_\nu^\eff$ and, therefore, $\mu$ do not vary with time $t$,
$\mathcal{E}$ must also be constant.

As a simple example, we consider 
a homogeneous and isotropic neutrino gas which initially consists of
equal numbers of pure
$\nu_e$ and $\bar\nu_e$ with the same energy $E_0$.
We assume that $\vH_\matt=0$ and $n_\nu^\eff$ is constant.
In the NFIS notation this system is 
represented by two NFIS's $\vs_{\omega_0}(t)$ and $\vs_{-\omega_0}(t)$
with $\omega_0=\delta m^2/2E_0$. At $t=0$ NFIS's $\vs_{\omega_0}(0)$ and
$\vs_{-\omega_0}(0)$ are aligned and antialigned with $\ve_z^\bsf$,
respectively. 
This is an example of the bipolar system which consists of
two groups of NFIS's approximately opposing each other in direction.
According to equation \eref{eq:nfis-energy}
the NFIS energy of such a neutrino gas, up to a constant, is
\begin{equation}
\fl
\mathcal{E}=-\frac{\delta m^2}{4 E_0}
[\vs_{\omega_0}(t)-\vs_{-\omega_0}(t)]\cdot\ve_z^\bsv
+\frac{\sqrt{2}}{2}\GF n_\nu^\eff 
[\vs_{\omega_0}(t)\cdot\vs_{-\omega_0}(t)]=\mathrm{const}.
\end{equation}
When $n_\nu^\eff$ is large, $\mathcal{E}$ is
dominated by the coupling energy between the NFIS's, and, because of
energy conservation,
$\vs_{\omega_0}(t)$ and $\vs_{-\omega_0}(t)$ must remain in a bipolar
configuration. 

For a normal mass hierarchy ($\delta m^2>0$) the initial configuration
of our simple 
bipolar system becomes absolutely stable
in the limit $\thetav\rightarrow0$ (i.e.\ $\ve_z^\bsf\rightarrow\ve_z^\bsv$). 
This is because in this limit the
bipolar system is initially in the lowest energy configuration where both the
coupling energy between $\vs_{\pm\omega_0}$ and $\vH_\vac$ and that between
$\vs_{\omega_0}$ and $\vs_{-\omega_0}$ are at their minimum
values. This is not the case for an inverted mass hierarchy ($\delta
m^2<0$) with which
these two coupling energies are at their maximum and minimum values,
respectively. Spontaneous collective neutrino
oscillation is, therefore, forbidden to occur in the former case 
but is allowed in the latter. In fact, if $\delta m^2<0$ and 
$\GF n_\nu^\eff\gg|\delta m^2/2E_0|$, 
the bipolar configuration of the NFIS's is energetically
allowed to nearly completely flip its direction, and, therefore,
both neutrinos and antineutrinos can almost entirely change their flavours.

The flavour dynamics of this bipolar system is probably best illustrated
using the pendulum analogy introduced by 
\citet{Hannestad:2006nj}. For this bipolar system
equation \eref{eq:eom-s} can be recast in the form
\begin{eqnarray}
\frac{\rmd}{\rmd t}\vec{J}(t) &= \vec{q}(t)\times M \vec{g},
\label{eq:eom-pendulum-Jdot}\\
\vec{J}(t) &= M \vec{q}(t)\times\frac{\rmd}{\rmd t}\vec{q}(t) 
+ \sigma \vec{q}(t).
\label{eq:eom-pendulum-J}
\end{eqnarray}
Equations \eref{eq:eom-pendulum-Jdot} and \eref{eq:eom-pendulum-J}
describe the motion of a pendulum with total angular momentum
$\vec{J}(t)=\vs_{\omega_0}(t)+\alpha\vs_{-\omega_0}(t)$.
Here, for more generality, we allow the number densities of
antineutrinos and neutrinos to be different with
$\alpha=\ff_{-\omega_0}/\ff_{\omega_0}$.
The mass of the pendulum
$M = (1+\alpha)/\mu$
is located at position
$\vec{q}(t) = \vec{Q}(t)/Q$ 
and it experiences a constant gravity field
$\vec{g} = (Q\omega_0/M) \vH_\vac$,
where vector
\begin{equation}
\vec{Q}(t)=\vs_{\omega_0}(t)-\alpha\vs_{-\omega_0}(t)+
\frac{1+\alpha}{\mu}\omega_0\vH_\vac
\end{equation}
has a constant length $Q$.
For $\alpha\neq1$ the pendulum moves like a gyroscope because it has
a constant, non-vanishing internal spin
$\sigma = \vec{J}(t)\cdot\vec{q}(t)$.

When neutrino number densities are large, the NFIS's in this example 
system maintain a bipolar configuration, and 
$\vec{q}(t)\simeq 2\vs_{\omega_0}(t) \simeq -2\vs_{-\omega_0}(t)$.
For a normal mass hierarchy ($\omega_0>0$), $\vec{g}$ is in the same
direction as 
that of $\vH_\vac$. In the limit $\thetav\rightarrow0$, 
$\vec{q}(t=0)\simeq\vH_\vac\simeq\vec{g}/|\vec{g}|$. In other words,
the flavour pendulum is near its lowest position. 
For an inverted mass hierarchy ($\omega_0<0$), however, $\vec{g}$ is in a
direction opposite to 
that of $\vH_\vac$. In the limit $\thetav\rightarrow0$, one has
$\vec{q}(t=0)\simeq-\vec{g}/|\vec{g}|$, and the flavour pendulum is
near its highest position. 

The inverted mass hierarchy case with $\thetav\ll1$ is interesting.
In this case, if the number densities of neutrinos and antineutrinos
are equal, the flavour pendulum will always swing from near the highest position
through the lowest position and back to its initial height. During
this process bipolar system can experience significant flavour transformation.
If the number densities of neutrinos and antineutrinos are different,
the pendulum will undergo gyroscopic motion because of its
non-vanishing spin, and it will not pass through
the lowest position. 
In particular, 
the pendulum will remain at its highest position like a ``sleeping
top'' if \citep[e.g.][]{Scarborough:1958fk}
\begin{equation}
\frac{\sigma^2}{4M^2g} \geq 1.
\label{eq:sleep-top}
\end{equation}
For the simple bipolar neutrino
system shown above, this is equivalent to the condition that
\citep{Hannestad:2006nj,Duan:2007mv}
\begin{equation}
n_\nu^\eff\geqslant\frac{\sqrt{2}}{2}\frac{1+\alpha}{(1-\sqrt{\alpha})^2}
\frac{|\delta m^2|}{\GF E_0}.
\label{eq:sleep-cond}
\end{equation}
This condition can viewed as the
division between the so-called ``synchronised regime'' and the ``bipolar
regime'' for the following reasons.

According to equation
\eref{eq:eom-s} each individual NFIS
$\vs_\omega(t)$ tends to precess about $\vH_\vac$ with its angular
frequency $\omega$. At the same time, the coupling among NFIS's tend to
make them move collectively. If the latter tendency dominates, all the
NFIS's are locked into one block spin with angular precession frequency
\begin{equation}
\omega_\mathrm{sync}=\int_{-\infty}^\infty\rmd\omega^\prime 
\omega^\prime \ff_{\omega^\prime} 
\frac{\vs_{\omega^\prime}\cdot\langle\vs\rangle}{|\langle\vs\rangle|^2}.
\end{equation}
This is the synchronised neutrino flavour transformation explained by
\citet{Pastor:2001iu}. A necessary condition of synchronisation is that
each NFIS precesses about the block spin much faster than does the
block spin about $\vH_\vac$, i.e.\ 
\begin{equation}
\mu|\langle\vs\rangle|\gtrsim|\omega_\mathrm{sync}|.
\label{eq:sync-cond}
\end{equation}
\citet{Duan:2005cp} pointed that a bipolar system can experience
synchronised or bipolar (pendulum-like) oscillations depending on
whether condition \eref{eq:sync-cond} is satisfied. 
For the simple bipolar system discussed above, the estimates
given by equations \eref{eq:sleep-cond} and \eref{eq:sync-cond} for
 $n_\nu^\eff$ at the boundary between the synchronised and bipolar
regimes are different only by a constant factor.

\subsection{Transition to bipolar oscillations%
\label{sec:bipolar-regime}}

\citet{Pastor:2002we} proposed that 
supernova neutrinos could experience synchronised flavour transformation
when neutrino fluxes are very large. Using equation
\eref{eq:sync-cond} \citet{Duan:2005cp} determined the
synchronised and bipolar regimes in the supernova environment.
The notion of the synchronised regime has since then been
widely adopted. However, figure \ref{fig:twosols} shows that neutrinos
actually 
experience the MSW-like flavour evolution  instead of the synchronised flavour 
transformation in the so-called synchronised
regime when $\vH_\matt\neq0$.
In addition, in the normal mass hierarchy case supernova neutrinos
follow the MSW-like flavour evolution well beyond the synchronised regime
as determined in the inverted mass hierarchy case (figure \ref{fig:twosols}).

\citet{Duan:2007fw} proposed that flavour transformation of
supernova neutrinos in the collective regime,
i.e.\ neutrino self-interaction is not negligible,
 can be explained as the
combination of two adiabatic solutions, i.e.\ the adiabatic MSW-like
solution discussed in section \ref{sec:MSW-like} and the adiabatic
precession solution to be described in section \ref{sec:prec-sol}. 
The flavour pendulum model discussed in section \ref{sec:pendulum} offers
great insights into the 
question why neutrinos deviate from the adiabatic MSW-like
flavour evolution in supernovae. We shall
elaborate on this idea in more detail.
 
 We note that at the
neutrino sphere the most abundant neutrino species are $\nu_e$ and
$\bar\nu_e$, which is similar to the simplistic bipolar system
represented by the flavour pendulum model. In addition, 
as pointed out by \citet{Duan:2005cp}, the effects of
ordinary matter can be ``ignored'' for collective neutrino
oscillations even if the density of ordinary matter is large (but still
low enough to be transparent to neutrinos). The idea of ignoring
ordinary matter is at the core of the collective flavour transformation of
supernova neutrinos and we shall discuss it first.

According to equation \eref{eq:eom-s}, 
all NFIS's tend precess to about the matter field $\vH_\matt$ with the
same frequency and, therefore, $\vH_\matt$ does not break the
collectiveness of neutrino oscillations as $\vH_\vac$ does. In fact, the 
matter field ``disappears'' in the reference frame that
rotates about $\vH_\matt$ with angular frequency $|\vH_\matt|$. This
is similar to the situation where the gravity field ``vanishes'' in a
freely-falling reference frame.
In this corotating frame, however, $\vH_\vac$ is
not stationary but rotates about $-\vH_\matt$ with angular frequency
$|\vH_\matt|$. 
If the matter density is large, $\vH_\vac^\perp$, the component
of $\vH_\vac$ that is perpendicular to $\vH_\matt$, rotates very fast
about $\vH_\matt$. As a result,
its effects on NFIS's average to zero if the NFIS's participate in some
collective motion with a time scale much longer than
$2\pi/|\vH_\matt|$. In other words, when the matter 
density is large,  one can ignore
the effects of ordinary matter for collective neutrino oscillations
by adopting a small effective
mixing angle and an effective mass-squared
difference $\delta m^2_\eff=\delta m^2\cos2\thetav$. 
\citet{Hannestad:2006nj} analysed the effects of ordinary matter
on the flavour pendulum and confirmed that matter density had
little effect on the motion of the pendulum except for a logarithmic
dependence of the period of this motion on the matter density.

Let us now consider the inverted mass hierarchy case with $\thetav\ll1$. 
We can estimate when bipolar oscillations may start using the
analogy of the flavour pendulum. Near the neutrino sphere the effective
total neutrino number 
density $n_\nu^\eff(r)$ is large, and
the mass of the pendulum $M\propto (n_\nu^\eff)^{-1}$ is small. As a
result, the pendulum with nonzero spin $\sigma\simeq(1-\alpha)/2$ is
able to ``defy''  gravity and stays at its highest position.
 In other words, no significant bipolar
oscillation can occur and supernova neutrinos follow the adiabatic
MSW-like flavour evolution in which both neutrinos and antineutrinos
essentially remain in their original flavours. 
As neutrino number densities decrease with $r$ the mass of the flavour pendulum
increases. Eventually, the flavour pendulum becomes so
heavy that its highest position is no longer stable
[see equation \eref{eq:sleep-top}].
When this occurs, supernova neutrinos will break away from the
MSW-like flavour evolution and experience the bipolar oscillation. As
mentioned earlier, unlike in the MSW-like flavour evolution, the
presence of ordinary matter will not affect the bipolar oscillation
once it starts.
%% We assume that
%% supernova neutrinos are represented by a flavour pendulum evolving with
%% ``time'' $r$. Near the neutrino sphere where $n_\nu^\eff(r)$ is large,
%% the flavour pendulum is light [with weight
%% \[M(r)|\vec{g}(r)|\propto1+\frac{2\omega_0}{\mu(r)}
%% =1-\frac{|\delta m^2|}{2\sqrt{2}\GF
%%   n_\nu^\eff(r)E_0}\] 
%% for $\delta m^2<0$ and $\thetav\ll1$] and is stable at its
%% highest position because of its internal spin. As $n_\nu^\eff(r)$
%% decreases with $r$, the flavour pendulum becomes heavier and
%% heavier. Eventually the internal spin of the pendulum can not keep it
%% at its highest position and, as a result, the flavour configuration of
%% the neutrino 
%% system becomes unstable. After this point supernova neutrinos do not
%% follow adiabatic MSW-like flavour evolution but experience collective
%% bipolar oscillations. 

Bipolar neutrino oscillations are represented by
the gyroscopic motion of the flavour pendulum which is a combination of
nutation and precession. The nutation of the flavour pendulum 
signifies simultaneous oscillations of $s_{\omega,z}$, the $z$
components of the NFIS's, or the oscillations neutrino survival
probabilities (see figure \ref{fig:P-r}). Here we do not distinguish
between the vacuum mass basis and the flavour basis because the
effective mixing angle is small.
The precession of the flavour pendulum stands for the
simultaneous precession of all NFIS's about the $z$ axis in flavour space.
From the above discussion it is clear that 
in the inverted mass hierarchy case bipolar oscillations are
insensitive to the exact value of $\thetav$. This explains why 
\citet{Duan:2007bt} obtained similar results for different inverted mass
hierarchy schemes even when $\thetav$ differ by several orders of magnitude.
It is also obvious that bipolar oscillations are not sensitive to
large matter densities in inverted mass hierarchy cases
as confirmed by \citet{Fogli:2007bk}.

For the normal mass hierarchy case, the flavour pendulum is initially
near its lowest position where it is stable even when $n_\nu^\eff(r)$
decreases. As a result, in the single-angle calculations presented in
figure \ref{fig:P-r}, supernova neutrinos follow the adiabatic
MSW-like evolution until they experience significant flavour
transformation at $r\simeq90$ km because of low matter density.  This
flavour transformation is similar to the resonance in the conventional
MSW mechanism and is represented by flipping of the orientation of the
NFIS's or raising of the flavour pendulum. As in the inverted mass
hierarchy case, the flavour pendulum will undergo gyroscopic motion
after the pendulum is raised and supernova neutrinos will, therefore,
experience bipolar oscillations.  But unlike the inverted hierarchy
case, the initiation of bipolar oscillations depends upon the
existence/efficacy of MSW-like resonances. So for the normal hierarchy
collective neutrino flavour transformation is inevitably sensitive to
the value of $\thetav$ and matter densities, an expectation confirmed
by \citet{Duan:2007bt} and \citet{Fogli:2007bk}, respectively.

\subsection{Adiabatic precession solution%
\label{sec:prec-sol}}

Comparing figures \ref{fig:P-r}(c,d) and \ref{fig:twosols}(b) one
notices that, although individual NFIS's can oscillate with significant
amplitudes in the $z$ direction, $\langle s_z(r)\rangle$ roughly stays
constant at $r\lesssim100$ km. This is because, as pointed out by
\citet{Hannestad:2006nj}, neutrino lepton number
\begin{equation}
L=2\int_{-\infty}^\infty\rmd\omega\ff_\omega\vs_\omega(r)\cdot\vH_\vac
=2\langle s_z^\bsv(r)\rangle
\label{eq:L}
\end{equation}
is constant if the matter field can be ``ignored''.
This conservation law arises
because, when the matter field is absent, there is only one external
field $\vH_\vac$ in the NFIS system, and the e.o.m.\ for the
NFIS's are invariant with a simultaneous rotation of all NFIS's about
$\vH_\vac$. However, this symmetry about $\vH_\vac$ is usually broken
because most 
configurations of NFIS systems are not invariant with rotation about
$\vH_\vac$. As a result the NFIS's tend
to rotate about $\vH_\vac$ collectively in order
to restore the symmetry dynamically.

The collective mode of neutrino oscillations discussed above corresponds to the
precession motion of a flavour pendulum. If a pendulum undergoes pure
precession without wobbling, its motion is symmetric about the $z$
axis although its configuration at any instant is not. One can imagine
that such a pendulum will slowly fall but remain in pure precession if
its weight and spin are changed infinitely slowly. 
\citet{Duan:2007mv} 
studied the flavour evolution of the simple bipolar system
discussed in section \ref{sec:pendulum} assuming that the flavour
pendulum stays in pure precession as $n_\nu^\eff(r)$ decreases. They
found qualitative agreement 
between this simple analysis and the numerical simulations of flavour
evolution of supernova neutrinos.

Neutrino gases in the pure collective precession mode satisfy two conditions,
namely, the
pure precession ansatz and the adiabatic ansatz \citep{Duan:2008za}.
The precession ansatz is that at any instant all NFIS's precess about
$\vH_\vac$ with 
the same angular frequency $\omega_\pr(n_\nu^\eff)$ which varies with
$n_\nu^\eff$, or
\begin{equation}
\frac{\rmd}{\rmd r}\vs_\omega(r)=\vs_\omega(r)\times\omega_\pr(n_\nu^\eff)\vH_\vac.
\label{eq:prec-ansatz}
\end{equation}
From equation \eref{eq:prec-ansatz} and the e.o.m.\ for NFIS's
[equation \eref{eq:eom-s}] one concludes that each NFIS $\vs_\omega(r)$ must be
either aligned or antialigned with
\begin{equation}
\fl
\vtH_\omega(r)=\vH_\omega(r)-\omega_\pr(n_\nu^\eff)\vH_\vac
=[\omega-\omega_\pr(n_\nu^\eff)]\vH_\vac-\mu(n_\nu^\eff)\langle\vs(r)\rangle.
\end{equation}
In other words, condition
\begin{equation}
\vs_\omega(r)=\frac{\epsilon_\omega}{2}\frac{\vtH_\omega(r)}{|\vtH_\omega(r)|}
\label{eq:alignment-prec}
\end{equation}
holds, where $\epsilon_\omega=+1$ or $-1$ if $\vs_\omega(r)$ is aligned or
antialigned with $\vtH_\omega(r)$. 
Note that we have ignored $\vH_\matt$ in discussion of collective
neutrino oscillations.

If $n_\nu^\eff(r)$ varies slowly with $r$, the NFIS system transits
from state to state all of which satisfy the precession ansatz. The
adiabatic ansatz is that $\epsilon_\omega$ stays constant during these
transitions: 
\begin{equation}
\frac{\rmd}{\rmd r} \epsilon_\omega = 0.
\label{eq:adiabatic-ansatz}
\end{equation}

\citet{Duan:2006an} showed that
the spectral-swap/split phenomenon 
(figure \ref{fig:P-E-c}) can arise as a result of
collective precession of NFIS's using the following simple argument.
If neutrinos are in the pure collective precession mode, each
NFIS $\vs_\omega(r)$ must stay aligned or antialigned with $\vtH_\omega(r)$
depending on the value of $\epsilon_\omega$. This is true even when
$n_\nu^\eff\rightarrow 0$. In this limit one simply have
\begin{equation}
\vs_\omega|_{n_\nu^\eff\rightarrow0} =
\frac{\epsilon_\omega}{2}\sgn(\omega-\omega_\pr^0)\ve_z^\bsv,
\label{eq:split}
\end{equation}
where $\omega_\pr^0=\omega_\pr(n_\nu^\eff=0)$.
Equation \eref{eq:split} implies that the NFIS orientation
 $\vs_\omega|_{n_\nu^\eff\rightarrow0}$
as a function of $\omega$ flips direction at $\omega_\pr^0$. For
$\thetav\ll1$ this means that the neutrino survival probability as a
function of neutrino energy jumps from 0 to 1 or vice versa at energy
$|\Es|$, where
\begin{equation}
\Es=\frac{\delta m^2}{2\omega_\pr^0}.
\end{equation}
The swapping point is located in the neutrino (antineutrino) sector if
$\Es>0$ ($\Es<0$). 
For $\Es>0$ the whole energy spectra of antineutrinos stay the same or
get swapped depending on whether the neutrino mass hierarchy is normal
or inverted.
\citet{Raffelt:2007cb} pointed out that the value
of $\omega_\pr^0$ or $\Es$ can be determined using equation
\eref{eq:split} and
the conservation of lepton number $L$ defined in equation \eref{eq:L}:
\begin{equation}
L=
\int_{-\infty}^\infty\rmd\omega\ff_\omega\epsilon_\omega\sgn(\omega-\omega_\pr^0).
\end{equation}

\citet{Raffelt:2007cb} showed that,
for homogeneous, isotropic neutrino gases or supernova neutrinos under
the single-angle approximation, the pure collective precession mode of 
neutrino oscillations with any given $n_\nu^\eff$ 
can be solved from a small set of nonlinear equations.
One notes that $\vtH_\omega(r)$ is the total effective field for
$\vs_\omega(r)$ in the frame that rotates about $\vH_\vac$ with
angular frequency $\omega_\pr(n_\nu^\eff)$.
Because a NFIS is only stationary
when it is either aligned or antialigned with its total effective field, the
precession ansatz is equivalent to the assumption that it is possible
to find a corotating frame in which all NFIS's are stationary.
 In this
corotating frame equation \eref{eq:alignment-prec} is equivalent to
\begin{eqnarray}
s_{\omega,\tilde{x}} &=
-\frac{\epsilon_\omega}{2}
\frac{\mu \langle s_{\tilde{x}}\rangle}%
{\sqrt{(\omega-\omega_\pr-\mu  \langle s_z\rangle)^2 +
(\mu \langle s_{\tilde{x}}\rangle)^2}},
\label{eq:alignment-sx-prec}\\
s_{\omega,z} &=
\frac{\epsilon_\omega}{2}
\frac{\omega-\omega_\pr-\mu \langle s_z\rangle}%
{\sqrt{(\omega-\omega_\pr-\mu \langle s_z\rangle)^2 +
(\mu \langle s_{\tilde{x}}\rangle)^2}},
\label{eq:alignment-sz-prec}
\end{eqnarray}
where $s_{\omega,\tilde{x}}$ is the projection of $\vs_\omega$ on $\vte_x$. Here we
define $\vte_x$ to be the unit vector in the same direction of
$\langle\vs_\perp\rangle$, the
vector component of $\langle \vs\rangle$ which is perpendicular to $\ve_z^\bsv$.
Multiplying equations \eref{eq:alignment-sx-prec} and
\eref{eq:alignment-sz-prec} with $\ff_\omega$ and integrating them over
$\omega$ one obtains
\begin{eqnarray}
1 &=
-\frac{1}{2}\int_{-\infty}^\infty\rmd\omega
\frac{\epsilon_\omega\ff_\omega}%
{\sqrt{[(\omega-\omega_\pr)/\mu- \langle s_z\rangle]^2 +
\langle s_{\tilde{x}}\rangle^2}},
\label{eq:unit-cond}\\
\omega_\pr &=
-\frac{1}{2}\int_{-\infty}^\infty\rmd\omega
\frac{\epsilon_\omega\ff_\omega\omega}%
{\sqrt{[(\omega-\omega_\pr)/\mu- \langle s_z\rangle]^2 +
\langle s_{\tilde{x}}\rangle^2}}.
\label{eq:omega-pr}
\end{eqnarray}
Note that in equations \eref{eq:unit-cond} and
\eref{eq:omega-pr} the minus signs arise because of the
antiferromagnetic coupling between NFIS's. 
%% The ostensible differences
%% between these equations and those in \citet{Raffelt:2007cb} are due to
%% the differences in definitions (see section \ref{sec:nfis}).
Given $(L,n_\nu^\eff,\epsilon_\omega)$ one can solve equations
\eref{eq:L}, \eref{eq:unit-cond} and \eref{eq:omega-pr} for 
$(\omega_\pr,\langle s_{\tilde{x}}\rangle,\langle s_z\rangle)$.  One can
then solve equations \eref{eq:alignment-sx-prec} and
\eref{eq:alignment-sz-prec} for each individual NFIS.

\citet{Duan:2007fw} solved the adiabatic, procession solution from
equations \eref{eq:L} and
\eref{eq:alignment-sx-prec}--\eref{eq:omega-pr} that 
corresponds to the single-angle calculations presented
earlier. The results are plotted in figure \ref{fig:twosols}
together with the results from
single-angle calculations and the corresponding adiabatic, MSW-like
solutions. One observes that 
$\langle s_\perp(r)\rangle=|\langle s_{\tilde{x}}(r)\rangle|$ obtained
in single-angle calculations oscillate 
around the adiabatic precession solution after neutrinos stop following
the adiabatic MSW-like flavour evolution. The values of $\langle
s_z(r)\rangle$ are approximately constant when the collection
precession mode first sets in but then changes as the densities of
both ordinary matter and neutrinos continue to decrease with $r$ and
conventional MSW flavour conversion starts to intervene.

\subsection{Collective neutrino
  oscillations in realistic supernova environments%
\label{sec:progress}}

We have so far summarised our understanding of collective neutrino
oscillations under the assumptions that both the two-flavour neutrino
mixing scheme and
the single-angle treatment of a spherical supernova model are
valid. The results obtained using these assumptions shed light on, and
form the basis of, our understanding of three-flavour, collective neutrino 
oscillations in realistic supernova environments.
Here we briefly report the progresses made in these directions.

\citet{Duan:2007sh} carried out the first single-angle
calculations which take into account neutrino self-interaction and
use the full three-flavour neutrino mixing. These calculations are
targeted at a very early epoch of an O-Ne-Mg core-collapse supernova
when the fast deleptonization of the PNS leads to an
intense $\nu_e$ burst. These calculations
show that the neutrino energy spectra emerged from the supernova
surface can also contain step-like features as in two-flavour
calculations. The results of these calculations also suggest that,
like in the conventional MSW flavour transformation, collective
neutrino oscillations in the full three-flavour mixing scheme can 
be factorised into two two-flavour mixing scenarios that occur in
sequence.

To generalise the spin-precession analogy to three flavour
oscillations, 
\citet{Dasgupta:2007ws} defined eight-dimensional Bloch vector
$\vP_\omega$, the three-flavour version of the polarisation vector, as
\begin{equation}
\varrho_\omega=\frac{n_\nu^\eff}{3}+\frac{1}{2}\vP_\omega\cdot\vec{\Lambda},
\end{equation} 
where $\Lambda_i$ ($i=1,2,\ldots,8$) are the Gell-Mann
matrices. One can define the cross product of eight-dimensional 
vectors by replacing the structure constant $\epsilon_{ijk}$ of the
SU(2) group 
with the structure constant $f_{ijk}$ of SU(3) \citep{Kim:1987bv}.
The e.o.m.\ for $\vP_\omega$ can then be written in a form similar to equation
\eref{eq:eom-P}. With Bloch vectors, three-flavour collective neutrino
oscillations can be visualised by the ``$\ve_3$--$\ve_8$'' diagram,
where $\ve_i$ ($i=1,2,\ldots,8$) are the unit basis vectors for the
eight-dimensional flavour space.
Using the concept of Bloch vectors and a simplified supernova model
\citet{Dasgupta:2008cd} showed that the step-like
features in the energy spectra of supernova neutrinos in \citet{Duan:2007sh} can
indeed be understood as two spectral swaps/splits each of which
corresponds to the conservation of a neutrino lepton number.

Independently, \citet{Duan:2008za} attacked the problem of
three-flavour, collective neutrino oscillations using a different
approach. Using the correspondence between the operations for NFIS's and
flavour matrices:
\begin{equation}
\vs_\omega\times\vs_{\omega^\prime} \sim
   [\varrho_\omega,\varrho_{\omega^\prime}]
\quad\mathrm{and}\quad
\vs_\omega\cdot\vs_{\omega^\prime} \sim
   \Tr(\varrho_\omega\varrho_{\omega^\prime}),
\end{equation}
one can translate the analyses for two-flavour neutrino mixing from the
spin representation to matrix representation.
In this translation a corotating frame 
is equivalent to a unitary transformation. Similarly, the statement that a
NFIS $\vs_\omega$ is stationary in such a corotating frame means that
$\varrho_\omega$ commutes with its Hamiltonian after an appropriate
unitary transformation. Expressing
the precession ansatz [equation \eref{eq:prec-ansatz}]
and the adiabatic ansatz [equation \eref{eq:adiabatic-ansatz}] in
terms of flavour matrices and generalising them to the three-flavour
mixing scheme,
\citet{Duan:2008za} showed that the e.o.m.~for
$\varrho_\omega$ in the three-flavour scheme
are invariant under independent rotations of the
whole system about $\Lambda_3$ and $\Lambda_8$ when the matter field
is absent. Two collective precession modes of neutrino oscillations
can arise because of these symmetries which would naturally lead to
 spectral swaps/splits as in three-flavour mixing scenarios.

While all of the above analyses on collective neutrino oscillations
are based on the homogeneity and isotropy of the neutrino gas 
or the validity of the single-angle
approximation of supernova neutrinos, a real supernova
environment is clearly inhomogeneous and anisotropic. 
Some of the conclusions that are correct for homogeneous, isotropic
neutrino gases may not be so for inhomogeneous, anisotropic ones.
For example, \citet{EstebanPretel:2008ni} recently
showed that very dense ordinary matter can play a bigger role in
suppressing collective neutrino oscillations in anisotropic environments
than in isotropic environments.

\citet{Raffelt:2007yz} studied some simple anisotropic
neutrino gases initially consisting of  neutrinos
and antineutrinos with equal numbers. They found that collective
oscillations in such 
systems quickly break down and flavour equipartition is always
achieved. We note that such a system correspond to a flavour pendulums
with no internal spin which does not undergo precession at all.  
This study, therefore, suggests that
bipolar neutrino oscillations that correspond to the nutation motion of
the flavour pendulum are not collective in anisotropic
environments. However, the collective neutrino oscillation mode that
corresponds to the precession motion of the flavour pendulum
arises because of a symmetry in the equations that govern flavour
evolution of neutrinos. This symmetry also exists in inhomogeneous,
anisotropic neutrino gases \citep{Duan:2008fd}.
Protected by this symmetry, the collective precession mode of
neutrino flavour transformation may not break down even in inhomogeneous,
anisotropic environments. % in the same way that the nutation mode does. 
Indeed,
\citet{EstebanPretel:2007ec} showed that single-angle-like collective
neutrino oscillations appear as the neutrino-antineutrino symmetry is dropped
which permits the precession motion of the flavour pendulum.

Using the symmetry of the equations that govern neutrino flavour evolution
\citet{Duan:2008fd} prescribed the adiabatic, precessional
flavour evolution in inhomogeneous, anisotropic neutrino
gases, which is similar to the simpler cases with homogeneous and isotropic
neutrino gases. The key idea is that, assuming all NFIS's to precess
collectively about $\vH_\vac$, at any  
\textit{space-time point} (instead of any instant in homogeneous,
isotropic cases) one can find a corotating frame in which all the
NFIS's at this point are stationary. Such a collective neutrino
oscillation mode is described by a wave-like solution.
As neutrinos
travel from one ``wavefront'' to another, the corresponding NFIS's rotate
about $\vH_\vac$ with a common angle which is equal to the phase
difference between the two ``wavefronts'' of the solution.

%%%%%%%%%%%%%%%%%%%%%%%%%%%%%%%%%%%%%%%%%%%%%%%%%%%%%%%%%%%%%%%%%%%%%%%%%%%%%%%%%%%
%%%%%%%%%%%%%%%%%%%%%%%%%%%%%%%%%%%%%%%%%%%%%%%%%%%%%%%%%%%%%%%%%%%%%%%%%%%%%%%%%%%
%%%%%%%%%%%%%%%%%%%%%%%%%%%%%%%%%%%%%%%%%%%%%%%%%%%%%%%%%%%%%%%%%%%%%%%%%%%%%%%%%%%
\section{Summary and outlook\label{sec:summary}}

While the neutrino telescopes around the globe are waiting for the
neutrino shower from the next Galactic supernova, 
our rapidly-growing knowledge of neutrino flavour transformation in
supernovae suggests that these cosmic neutrinos will reveal
much valuable information about the supernova as well as about
themselves. Through these neutrinos the dynamical features of the
supernova such as shocks and turbulence may be observed and
analysed. Undetermined fundamental properties of the neutrino such as
its mass hierarchy may be disclosed through the 
swapped/split neutrino energy spectra due to neutrino
self-interaction. However, much still need to be learned about this
subject. 

For example, collective neutrino oscillations in highly
inhomogeneous, anisotropic environments are so far not well
understood. Most of our understanding of collective neutrino
oscillations is based on the single-angle approximation which ignores
the anisotropic nature of the supernova environment. Meanwhile,
direct numerical simulation of neutrino oscillations
with neutrino self-interaction in such environments (using 2D and/or
3D matter profiles) remains to be done. The subject of turbulence 
in supernova and its impact is also uncertain.
One of the imminent tasks is to combine the effect of the dynamic
matter profile and that of neutrino self-interaction, which has
been studied largely in parallel so far. The matter contribution is
important near the region where $\GF\rho(r)\sim\delta m^2/(2E)$ and
collective neutrino oscillations can occur if 
$\GF n_\nu(r)\gtrsim\delta m^2/(2E)$. 
If the two regions are well separated, one can conduct two
calculations in series with the results from the computation of collective
oscillations as inputs to that on the MSW effect. This is the 
approximation made by \citet{Kneller:2008PhRvD..77d5023K,Chakraborty:2008zp,Gava:2009pj,Lunardini:2007vn}. In the case where the two
regions overlap with each other, in principle one should carry out a
comprehensive calculation with both effects included which could be
a time-consuming task.

In summary, we have reviewed the field of neutrino oscillations in
supernovae focusing upon the recent discoveries of the dynamism of the
pure-MSW effect and the emergence of neutrino self-interactions near
the neutrino sphere.
We have attempted to give those readers interested in
pursuing research in this area sufficient detail that he or she can
now begin to digest the ever-growing literature on the subject and, at
the same time, give the casual observer a overarching understanding of
these two phenomena.  But perhaps the most general summary we can make
is that the evolution of this field over the past few years has been
both fruitful and frenetic and we fully expect that this trend will
continue for some time yet.

%%%%%%%%%%%%%%%%%%%%%%%%%%%%%%%%%%%%%%%%%%%%%%%%%%%%%%%%%%%%%%%%%%%%%%%%%%%%%%%%%%%

\ack 
The authors are very grateful to the editor for giving us the
opportunity to write this review and to G M Fuller, J Gava, R
Humphreys, G McLaughlin, A Mirizzi, Y-Z Qian, and C Volpe for the
discussions during the drafting process. This work was supported in
part by US DOE grants DE-FG02-00ER41132 at INT and DE-FG02-87ER40328
at UMN, and by ``Non standard neutrino properties and their impact in
astrophysics and cosmology'', Project No.\ ANR-05-JCJC-0023 at IPN
Orsay.

%%%%%%%%%%%%%%%%%%%%%%%%%%%%%%%%%%%%%%%%%%%%%%%%%%%%%%%%%%%%%%%%%%%%%%%%%%%%%%%%%%%

%\section*{References}
%\bibliographystyle{iopart-num}
\bibliographystyle{iopart-alpha}
\bibliography{ref}
%\bibliography{ref.old}

\end{document}